\documentclass[twocolumn,preprintnumbers,amsmath,amssymb]{revtex4}
\usepackage[utf8]{inputenc}
\usepackage{epsfig}
\usepackage{graphicx}
\usepackage{dcolumn}
\usepackage{bm}
\usepackage{amsmath, amsfonts}
\usepackage{amssymb}
\usepackage{amsthm}
\usepackage{physics}
\usepackage{comment}
\usepackage{color} 
\usepackage[colorlinks,linkcolor=blue,anchorcolor=blue,urlcolor=blue, citecolor=blue]{hyperref}
\usepackage{caption}
\usepackage{subcaption}
\usepackage{siunitx,booktabs}
\usepackage{epstopdf}
\newcommand{\unit}{1\!\!1}
\usepackage{lipsum}
\begin{document}
	\title{Secure quantum key distribution with a subset of malicious devices}
	\author{Ví­ctor Zapatero}
	\email{vzapatero@com.uvigo.es}  
	\author{Marcos Curty}
	\affiliation{Escuela de Ingeniería de Telecomunicación, Department of Signal Theory and Communications, ­University of Vigo, Vigo E-36310, Spain}
\begin{abstract}
	The malicious manipulation of quantum key distribution (QKD) hardware is a serious threat to its security, as, typically, neither end users nor QKD manufacturers can validate the integrity of every component of their QKD system in practice. One possible approach to re-establish the security of QKD is to use a redundant number of devices. Following this idea, we address various corruption models of the possibly malicious devices and show that, compared to the most conservative model of active and collaborative corrupted devices, natural assumptions allow to significantly enhance the secret key rate or considerably reduce the necessary resources. Furthermore, we show that, for most practical situations, the resulting finite-size secret key rate is similar to that of the standard scenario assuming trusted devices.
\end{abstract}
\maketitle
\section{Introduction}\label{Introduction}
Quantum key distribution~\cite{BB84,review1,review2,Feihu} (QKD) allows for information-theoretically secure communications, unaffected by the long-term security weakening inherent to public-key cryptography~\cite{DHS,RSA}. Its security relies on fundamental physical principles and various assumptions, a crucial one being that the legitimate QKD users, say Alice and Bob, hold honest devices that stick to the QKD protocol and do not intentionally leak their private information to an eavesdropper (Eve). However, this strong assumption is probably unjustified, considering the amount of hardware and software Trojan horse attacks (THAs) against conventional cryptographic systems reported in the last years~\cite{Gligor,Zander,Prevelakis,Yang,Robertson}. After all, likewise conventional security hardware, QKD devices incorporate many sophisticated components typically provided by specialised companies, and neither QKD vendors nor users are capable of validating the security of all these components in practice~\cite{Adee}. However, a malicious component can totally compromise the security of QKD. Indeed, the fabrication process of QKD systems might provide Eve with plenty of opportunities to meddle with the QKD hardware, including both the optical equipment and the classical post-processing (CP) units. Moreover, Eve could even sidestep post-fabrication tests by arranging attack triggers that depend on a sequence of unlikely events~\cite{Yang,Becker}. 

Remarkably, not even device-independent (DI) QKD \cite{Mayers,Acin,Vazirani,Rotem,Miller} can provide security against malicious devices, as shown in~\cite{Barrett}. It is the classical nature of the secret keys that makes QKD systems vulnerable to classical hacking in both the DI and the non-DI scenarios, because classical keys are susceptible to copying.

A possible solution to foil malicious hardware and software in QKD was recently presented in~\cite{Curty}, and then experimentally demonstrated in~\cite{Wei}. The triggering idea is that it might be more difficult for Eve to corrupt various devices than a single device, for example, if they originate from different providers. Therefore, one can use a redundant number of devices for both the raw key generation and the post-processing of QKD. As shown in~\cite{Curty}, under the assumption that the number of devices controlled by Eve is restricted, secure QKD is possible by combining verifiable secret sharing (VSS)~\cite{VSS,MPC,Ben-Or,Chaum,Maurer}---whose essential building block is secret sharing~\cite{Shamir,Blakley}, a standard technique in secure hardware design~\cite{Mitra}--- and privacy amplification (PA)~\cite{PA,Tomamichel1}. Of course, both tools operate on top of DI and non-DI security analyses, which determine the secret key length that one can extract from the honest optical apparatuses.

However, a major limitation of the proposal in~\cite{Curty} is that it is conceived for the case where all the corrupted devices fully obey a single Eve who can access their internal information and make them arbitrarily misbehave from the protocol. This scenario, which we refer to as the active collaborative (AC) model, might be over-conservative in many practical situations. For instance, if Alice and Bob purchase devices from different vendors, it might be reasonable to expect that, even if they are corrupted, they do not collaborate, meaning that they do not share their private information with each other or cooperate in any way. Also, if the information delivered by a certain device is different from the one prescribed by the protocol, it might be detected by Alice and Bob a posteriori. In this sense, some QKD users might only request security against non-collaborative (rather than collaborative) or passive (rather than active) corrupted devices.

Crucially, when applied in more sensible corruption models like these, the proposal in~\cite{Curty} provides no advantage at all with respect to the AC model. One major contribution of this work is to prove that some of these models actually enable a significant enhancement of the secret key rate, require fewer honest devices and classical communications than the AC model, or allow to remarkably diminish the post-processing time, a severe bottleneck in QKD. In particular, we introduce conditional VSS, a weaker version of VSS that is more suitable for the task of QKD. In addition, we present a general distributed QKD post-processing protocol appropriate for all the corruption models. Lastly, we evaluate the performance of two well-known QKD schemes in the presence of malicious devices. The simulations corroborate that notably improved non-asymptotic key rates can be reached by replacing the AC model by less conservative and probably more realistic models. Furthermore, in all the considered models, we find that the increased authentication cost of our protocol (compared to that of standard QKD post-processing) is negligible with respect to the secret key length for practical data block-sizes and moderate numbers of corrupted devices.
\section{Results}\label{Results}
We start by describing the general formalism we consider. Without loss of generality, a standard QKD setup can be divided into two parts with separate roles: a QKD module and a classical post-processing (CP) unit. Alice's and Bob's QKD modules form a so-called QKD pair, whose role is to generate raw correlated data between the parties via quantum communication. Each module transfers its raw data to its local CP unit, and the two distant CP units distill a pair of secret keys from the raw data via coordinated classical post-processing and authenticated classical communication.

The focus of this work is the general scenario where not all the devices are trusted, thus forcing the parties to use a redundant number of them~\cite{Curty}. Throughout the paper, we shall consider that Alice and Bob share $n_{\rm q}$ QKD pairs (or simply ``pairs"), and that each of them holds $n_{\rm c}$ CP units (or simply ``units").
Similarly, we assume that up to $t_{\rm q}$ QKD pairs are corrupted (a QKD pair is corrupted when at least one of its modules is) and up to $t_{\rm c}$ CP units are corrupted per lab. Nevertheless, our results could be easily adapted to contemplate different numbers of honest and corrupted units in each lab. For $j=1,\ldots,n_{\rm q}$, Alice's (Bob's) module $\textrm{QKD}_{\textrm{A}_{j}}$ ($\textrm{QKD}_{\textrm{B}_{j}}$) is connected to all her (his) units $\{\textrm{CP}_{\textrm{A}_{l}}\}_{l=1}^{n_{\rm c}}$ ($\{\textrm{CP}_{\textrm{B}_{l'}}\}_{l'=1}^{n_{\rm c}}$) via secure channels, \textit{i.e.}, channels that provide both privacy and authentication. Also, all of Alice's (Bob's) units are pairwise connected by secure channels too. Since all these links take place within Alice's (Bob's) lab, in practice security could be enforced by using, say, physically protected cables. Similarly, the $\textrm{CP}_{\textrm{A}_{l}}$ are connected to the $\textrm{CP}_{\textrm{B}_{l'}}$ by authenticated classical channels. And lastly, as usual, a quantum channel fully accessible to an eavesdropper links $\textrm{QKD}_{\textrm{A}_{j}}$ to its partner $\textrm{QKD}_{\textrm{B}_{j}}$. A schematic of this QKD setup is given in Fig.~1.
\begin{figure}[!htbp]
	\centering 
	\includegraphics[width=9.7cm,height=4.2cm]{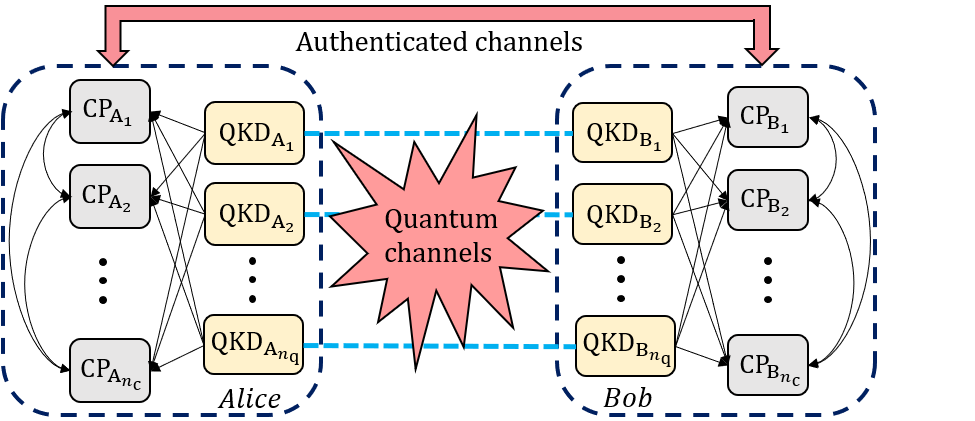}\\
	\caption{Proposal of a QKD setup with redundant devices suggested in~\cite{Curty}. The areas surrounded by dashed lines define Alice's and Bob's labs. Alice's (Bob's) lab contains $n_{\rm q}$ QKD modules (yellow boxes). Each module of Alice is linked to a single module of Bob through a quantum channel (dashed blue lines), forming a so-called QKD pair, and $t_{\rm q}$ pairs are possibly malicious at most. In addition, Alice (Bob) holds $n_{\rm c}$ CP units (grey boxes), $t_{\rm c}$ of them being possibly malicious at most. In each lab, all the CP units are connected to each other and to all $n_{\rm q}$ local QKD modules via secure channels that provide both privacy and authentication (black solid arrows). Also, every unit of Alice is linked to every unit of Bob through an authenticated classical channel (all of them together symbolised by the red double-end arrow).}
	\label{fig:setting}
\end{figure}
\subsection{AC corruption}\label{AC}
In the first place, let us briefly summarize the proposal in~\cite{Curty}, which establishes the security of QKD in the AC model using the setup of Fig.~1. On the one hand, given that $n_{\rm q}>t_{\rm q}$, PA allows to ``remove" not only the information Eve gains through her intervention in the quantum channel (as it is done in standard QKD post-processing), but also the information she learns from the corrupted QKD pairs. On the other hand, given that $n_{\rm c}>3t_{\rm c}$~\cite{Maurer}, VSS enables an honest QKD module to split a raw key into shares and redundantly allocate them among its local CP units for distributed post-processing. Crucially, the properties of VSS may guarantee the secrecy and the correctness of the final keys reconstructible by Alice and Bob at the end of this post-processing.

Before we analyse alternative corruption models, it is convenient to tight up some few loose ends affecting the proposal in~\cite{Curty}. In the first place, it requires the execution of $n_{\rm q}+1$ separate PA steps to distill a secret key. On the contrary, in Sec.~\ref{alternative models} we show that a single PA step suffices, which actually applies to all possible corruption models (see also Sec.~\ref{protocol} for a distributed post-processing protocol that implements PA in a single step).

In the second place, the use of standard VSS assures that the post-processing is resilient to the misbehaving of the CP units at the price of relying on simulated broadcast, better known as byzantine agreement~\cite{Lamport}. However, this is a very stringent task: it requires the exchange of an exponentially increasing number of classical messages, say $C\sim{O(n_{\rm c}^{t_{\rm c}})}$, among the units that want to reach the agreement~\cite{Lamport}. What is more, the achieved resiliency is probably not relevant for QKD. After all, Eve has unrestricted access to the quantum channel and thus may induce the abortion of the QKD protocol at will. For these reasons, in all corruption models we replace VSS by a weaker cryptographic primitive, namely, conditional VSS (defined in Sec.~\ref{Methods}), which circumvents simulated broadcast by simply allowing the CP units to abort the protocol. As seen in Sec.~\ref{alternative models} below, this replacement is not only advantageous in the AC model, but whenever actively corrupted CP units are considered, whether they collaborate or not.
\subsection{Alternative corruption models}\label{alternative models}
In what follows, we address various adversarial scenarios alternative to the AC model. In particular, three looser non-mixed corruption models exist: passive and collaborative (PC), active and non-collaborative (AN) and passive and non-collaborative (PN), where non-collaboration is obviously only defined if multiple corrupted devices exist. Importantly, we decouple the analysis of the different corruption models for the QKD modules and the CP units, such that the results we present for the QKD modules do not assume a specific model for the CP units and vice versa. In addition, we maintain the general QKD setup presented in Fig.~1.\\

Let us discuss the QKD modules first. In virtue of the \textit{privacy} of conditional/standard VSS (see the Methods section), a distributed QKD post-processing protocol using VSS guarantees that the extractable secret key length does not depend on the corruption model of the CP units, but only on that of the QKD pairs. What is more, let us assume for now that the parties select the AC model as their preferred model for the QKD pairs. For $j=1,\ldots,n_{\rm q}$, the $j$-th QKD pair runs an independent QKD session. As shown in Supplementary Note 1, for $n_{\rm q}=t_{\rm q}+1$ (minimum valid choice of $n_{\rm q}$ for a given $t_{\rm q}$), the $\epsilon_{\rm cor}$-correct, $\epsilon_{\rm sec}$-secret key length $l$ extractable via one-step PA from all these sessions is given by
\begin{equation}\label{key length AC}
l=\left\lfloor{\min_{j}\left\{h^{j}_{\varepsilon}-\lambda_{j}\right\}-\log_{2}\left(\frac{1}{\hat{\epsilon}_{\rm cor}\epsilon_{\rm PA}^{2}\delta}\right)}\right\rfloor
\end{equation}
where $h_{\varepsilon}^{j}$ is a hypothetical lower bound on the $\varepsilon$-smooth min-entropy of Bob's $j$-th raw key conditioned on the information held by Eve up to the parameter estimation (PE) step, and the smooth parameter $\varepsilon$ depends on the PE procedure. As explained in Supplementary Note 1, the term ``hypothetical" here refers to the fact that the information delivered by corrupted QKD modules cannot be trusted. Similarly, $\lambda_{j}$ is the public syndrome information required for the reconciliation of the $j$-th pair of raw keys, and $\hat{\epsilon}_{\rm cor}=\epsilon_{\rm cor}-\epsilon_{\rm AU}$ for a pre-agreed authentication error $\epsilon_{\rm AU}$, such that $\epsilon_{\rm AU}<\epsilon_{\rm cor}$ and $\epsilon_{\rm AU}<\epsilon_{\rm sec}$. Lastly, $\epsilon_{\rm PA}$ is the error probability of the PA step and $\delta>0$, such that
\begin{equation}\label{secrecy AC}
\epsilon_{\rm sec}\geq{2\varepsilon+\delta+\epsilon_{\rm PA}+\epsilon_{\rm AU}}.
\end{equation}
As one would expect, from Eq.~(\ref{key length AC}) we see that, if a single honest QKD pair exists, ``a single key" can be extracted from all $n_{\rm q}$ raw keys in the AC model. Notably, the generalization of Eq.~(\ref{key length AC}) to $n_{\rm q}-t_{\rm q}>1$ is straightforward.

Now, let us address the alternative models, PC, AN and PN. As long as the malicious QKD pairs are collaborative, an omniscient Eve could learn all the information they hold about the keys, and as long as they are active, they can deliver untrustworthy protocol information unsuitable for correct PE. Hence, although for different reasons, the intermediate scenarios PC and AN cannot lead to an enhancement of the secret key length with respect to the AC model: they also require to remove all the key material that comes from corrupted QKD pairs via PA, thus demanding $n_{\rm q}>t_{\rm q}$ as well. In particular, the extractable key length for $n_{\rm q}=t_{\rm q}+1$ in the PC (AN) corruption model is given by Eq.~(\ref{key length AC}) too.

In the PN corruption model, one assumes an independent Eve per malicious QKD pair who does not collaborate with the eavesdroppers possibly controlling the other pairs. Moreover, passivity implies that corrupted pairs deliver trustworthy protocol information which allows to quantify the ``ignorance" (in secret bits) that the Eves possibly corrupting other pairs have about their raw data. Thus, it suffices to remove the information held by the most knowledgeable eavesdropper via PA in order to provide security against all of them. As a consequence, secure QKD is possible even if all the QKD pairs are corrupted in the PN model, \textit{i.e.,} even if $n_{\rm q}=t_{\rm q}$. In this setting, one can show that the $\epsilon_{\rm cor}$-correct, $\epsilon_{\rm sec}$-secret key length $l$ extractable via one-step PA in the PN model (see Supplementary Note 2) is given by
\begin{equation}\label{key length PN}
l=\Biggl\lfloor\min_{v}\sum_{j\neq{v}}^{n_{\rm q}}\left\{H_{\rm min}^{\varepsilon}(s_{\rm B}^{j}|E_{v})-\lambda_{j}\right\}-\log_{2}\left(\frac{1}{\hat{\epsilon}_{\rm cor}\epsilon_{\rm PA}^{2}\delta^{n_{\rm q}-1}}\right)\Biggr\rfloor,
\end{equation}
where $H_{\rm min}^{\varepsilon}(s_{\rm B}^{j}|E_{v})$ denotes the $\varepsilon$-smooth min-entropy of Bob's $j$-th raw key, $s_{\rm B}^{j}$, conditioned on the information $E_{v}$ held by the $v$-th eavesdropper (\textit{i.e.}, the one that corrupts the $v$-th QKD pair, with $v=1,\ldots{},n_{\rm q}$). The remaining parameters were introduced in Eq.~(\ref{key length AC}), and the secrecy parameter now satisfies
\begin{equation}\label{secrecy PN}
\epsilon_{\rm sec}\geq{(n_{\rm q}-1)(2\varepsilon+\delta)+\epsilon_{\rm PA}+\epsilon_{\rm AU}}.
\end{equation}
Remarkably, Eq.~(\ref{key length PN}) trivially outperforms Eq.~(\ref{key length AC}) for any given $t_{\rm q}>1$ (and we recall that non-collaboration is only defined in this case).\\

In what follows, we discuss the CP units. Although the corruption model of the CP units does not affect the extractable key length, $l$, it determines the necessary resources to securely implement a distributed post-processing using conditional VSS: the number of units per party, $n_{\rm c}$, the number $R$ of copies per share of raw key to be delivered by any given QKD module, and the total number of raw key shares managed per CP unit, say $r$, originating from a given QKD module. On the one hand, $n_{\rm c}$ and $R$ determine the necessary classical communications both between labs and inside each lab, and the total authentication cost of the former, say $l_{\rm AU}$. On the other hand, $r$ strongly affects the post-processing time, a usual concern in the performance of QKD. In Table 1 we list the minimum values of $n_{\rm c}$, $R$ and $r$ required for distributed QKD post-processing, depending on the corruption model of the CP units.

\begin{table}[htb]
	\centering
	\begin{tabular}{|c||c|c|}\hline
		$n_{\rm c}$, $R$, $r$ & active & passive \\ \hline\hline
		collaborative &
		\begin{tabular}{c} $n_{\rm c}=3t_{\rm c}+1$ \\ $R=2t_{\rm c}+1$ \\ $r=\binom{n_{\rm c}-1}{t_{\rm c}}$ \\
		\end{tabular} &
		\begin{tabular}{c} $n_{\rm c}=t_{\rm c}+1$ \\ $R=1$ \\ $r=1$ \\
		\end{tabular} \\ \hline
		non-collaborative &
		\begin{tabular}{c} $n_{\rm c}=2t_{\rm c}+2$ \\ $R=2t_{\rm c}+1$ \\ $r=n_{\rm c}-1$ \\
		\end{tabular} &
		\begin{tabular}{c} $n_{\rm c}=2$ \\ $R=1$ \\ $r=1$ \\
		\end{tabular} \\ \hline	
	\end{tabular}
	\caption{Minimum resources of a distributed QKD post-processing protocol based on conditional VSS, depending on the corruption model of the CP units. While $n_{\rm c}$ is the total number of units per party, $R$ is the redundancy of each raw key share and $r$ is the number of key shares managed per CP unit from each of its local QKD modules. The number $t_{\rm c}$ of possibly corrupted units per lab is at least two for the non-collaborative models (AN and PN), as non-collaboration is only defined in this case.}
	\label{table:1}
\end{table}
The entries of the table follow from the requirements of conditional VSS and are established in Proposition 1 of Sec.~\ref{Methods} (see Supplementary Note 3 for a proof of this proposition). As we observe, all the restricted models allow to reduce the resources with respect to the AC model. For instance, note that the number $r$ of shares per unit grows exponentially with $n_{\rm c}$ for a fixed fraction of corrupted units in the AC model. This might lead to prohibitively long post-processing times even for small values of $n_{\rm c}$. Nevertheless, this problem disappears if one assumes that the possibly corrupted units are non-collaborative, thus moving to the AN model. Also in this model, it is worth noting that conditional VSS tolerates $n_{\rm c}=2t_{\rm c}+2$, while standard VSS would still require $n_{\rm c}=3t_{\rm c}+1$, a constraint imposed by the necessity to allow for simulated broadcast.

Within the passive models (PC and PN), the distributed post-processing has the extra advantage that the PE and the lab-to-lab classical communications can be conducted by a single CP unit per lab. On the contrary, the active models require the participation of $R=2t_{\rm c}+1$ units per lab for these tasks, in order to assure the presence of a majority of honest units.

Remarkably, in Sec.~\ref{Methods}, we formulate a distributed QKD post-processing protocol adequate for all the corruption models, matching the entries of Table 1 in each case. The security of this protocol, established in Proposition 3 (see Sec.~\ref{protocol}), is proven in Supplementary Note 4 combining conditional VSS with a standard QKD security analysis.

Lastly, as stated above, the corruption model of the CP units also determines the authentication cost, $l_{\rm AU}$, of the distributed post-processing. The classical communications require to select $R$ distinct $\textrm{CP}_{\textrm{A}_{l}}$ and $R$ distinct $\textrm{CP}_{\textrm{B}_{l'}}$, such that each of the former pre-shares a dedicated pool of secret key bits with each of the latter for authentication purposes. Thus, denoting the common size of every \textit{key pool} by $\abs{k}$, it follows that
\begin{equation}\label{authentication cost}
l_{\rm AU}=R^{2}\times{}\abs{k},
\end{equation}
where $R$ is given in Table 1 for each model. A possible estimation of $\abs{k}$ using a typical authentication scheme~\cite{LFSR} is presented in Supplementary Note 5. Within this scheme, the authentication cost of a message scales logarithmically with its length, meaning that for most practical situations $l_{\rm AU}<<l$, as we shall corroborate in the next section.
\subsection{Performance evaluation}\label{Performance evaluation}
To complete the analysis, we calculate explicit secret key rates in various significant corruption models, and in the finite key regime. The secret key rate is defined as
\begin{equation}\label{key rate}
K=\frac{l-l_{\rm AU}}{n_{\rm q}N},
\end{equation}
where we recall that $l$ ($l_{\rm AU}$) is the extractable secret key length (authentication cost) and $n_{\rm q}$ ($N$) is the number of QKD pairs (number of signals transmitted per pair). For illustration purposes, $l_{\rm AU}$ is computed according to the classical communications of the distributed post-processing presented in the Methods section.

For concreteness, we assume the same corruption model for the QKD modules and the CP units, a natural supposition in practice. Moreover, we restrict ourselves to the extreme corruption models, AC and PN, as the intermediate scenarios (AN and PC) do not allow to enhance the secret key rate, disregarding the authentication cost (see Sec.~\ref{alternative models}). We also assume that Alice and Bob use the minimum number of devices that allows for $K>0$, which depends on the corruption model they consider. For AC corruption, this means that they agree on the number $t_{\rm q}\geq{}0$ ($t_{\rm c}\geq{}0$) of malicious QKD pairs (CP units per lab) they want to be protected against, and use $n_{\rm q}=t_{\rm q}+1$ pairs ($n_{\rm c}=3t_{\rm c}+1$ units per lab). Alternatively, for PN corruption, they use $n_{\rm q}=2$ QKD pairs and $n_{\rm c}=2$ CP units per party, which suffices to achieve $K>0$ even if all the devices are possibly malicious (see Sec.~\ref{alternative models}).

We consider two practical QKD protocols with decoy states: an efficient MDI-QKD scheme~\cite{X.B.Wang} with three decoy intensities in the basis X (devoted to PE) and one signal intensity in the basis Z (devoted to  key distillation), and the standard decoy-state BB84 scheme~\cite{Lim} with three decoy intensities per basis. Detailed analyses of these protocols are provided in Supplementary Notes 6 and 7, respectively. For each protocol, we compute estimates of $l$ (given by Eq.~(\ref{key length AC}) for the AC model and by Eq.~(\ref{key length PN}) for the PN model) and $l_{\rm AU}$ (given by Eq.~(\ref{authentication cost})), by setting the observables to their expected values according to respective channel models described in the cited Supplementary Notes. These channel models depend on various common experimental parameters: the efficiency of the photo-detectors, set to $\eta_{\rm det}=65\%$, their dark count probability, set to $p_{\rm d}=7.2\times{}10^{-8}$ (both values matching the recent MDI-QKD experiment reported in~\cite{Yin}), and the polarization misalignment, set to, say $\delta_{\rm mis}=0.08$ for illustration purposes. Moreover, in both the MDI-QKD and the BB84 schemes, the weakest decoy intensity is set to $\omega=10^{-3}$ for the numerics. In each case, we optimise the remaining protocol inputs (\textit{i.e.}, intensity settings, and basis and decoy probabilities) to maximize $K$ as a function of the channel loss between Alice and Bob.

For the finite key analysis, we select a post-processing block-size of $M$ bits. Then, for every value of the channel loss, we choose the smallest number of transmission rounds per QKD pair, $N$, that assures that all $n_{\rm q}$ sifted keys reach this block-size except with a probability of, say $\gamma_{\rm sift}=5\times{}10^{-3}$, according to the channel model.

Regarding the EC leakage, we assume the typical model $|{sy}\bigl(s_{\rm B}^{j}\bigr)\bigr|={M}f_{\rm EC}h(E_{\rm tol})$ for every EC syndrome, where $f_{\rm EC}=1.16$ is the efficiency of the EC protocol, $h(\cdot)$ is the binary entropy function, and $E_{\rm tol}$ is a pre-fixed threshold QBER. In particular, $E_{\rm tol}$ is an upper bound on the QBER that any pair of sifted keys can reach according to the channel model, except with an error probability of $\gamma_{\rm EC}=5\times{}10^{-3}$.

Finally, the security parameters are set to $\epsilon_{\rm cor}=\epsilon_{\rm sec}=10^{-8}$ and $\epsilon_{\rm AU}=5\times{}10^{-9}$. As shown in Supplementary Note 5, $\epsilon_{\rm AU}$ determines the individual authentication error probability $\gamma_{\rm AU}$ via $\epsilon_{\rm AU}=(t_{\rm c}+1)^{2}(n_{\rm q}+1)\gamma_{\rm AU}$ ($\epsilon_{\rm AU}=(n_{\rm q}+1)\gamma_{\rm AU}$) in the presence of actively (passively) corrupted CP units. Given $\epsilon_{\rm sec}$ and $\epsilon_{\rm AU}$, the remaining parameters, $\epsilon_{\rm PA}$ and $\delta$, entering the extractable key length, $l$, are determined by imposing a common value, $\gamma_{\rm sec}$, for every error term that contributes to $\hat{\epsilon}_{\rm sec}=\epsilon_{\rm sec}-\epsilon_{\rm AU}$ (given by Eq.~(\ref{secrecy AC}) and Eq.~(\ref{secrecy PN})). In particular, from the PE procedure presented in Supplementary Note 6 (Supplementary Note 7), it follows that $\gamma_{\rm sec}=\hat{\epsilon}_{\rm sec}/48$ ($\gamma_{\rm sec}=\hat{\epsilon}_{\rm sec}/20$) in the MDI-QKD (BB84) scheme within both the AC and the PN scenarios, where we used the fact that $n_{\rm q}=2$ in the latter case.

\begin{figure}[!htbp]
	\centering 
	\includegraphics[width=10.5cm,height=11cm]{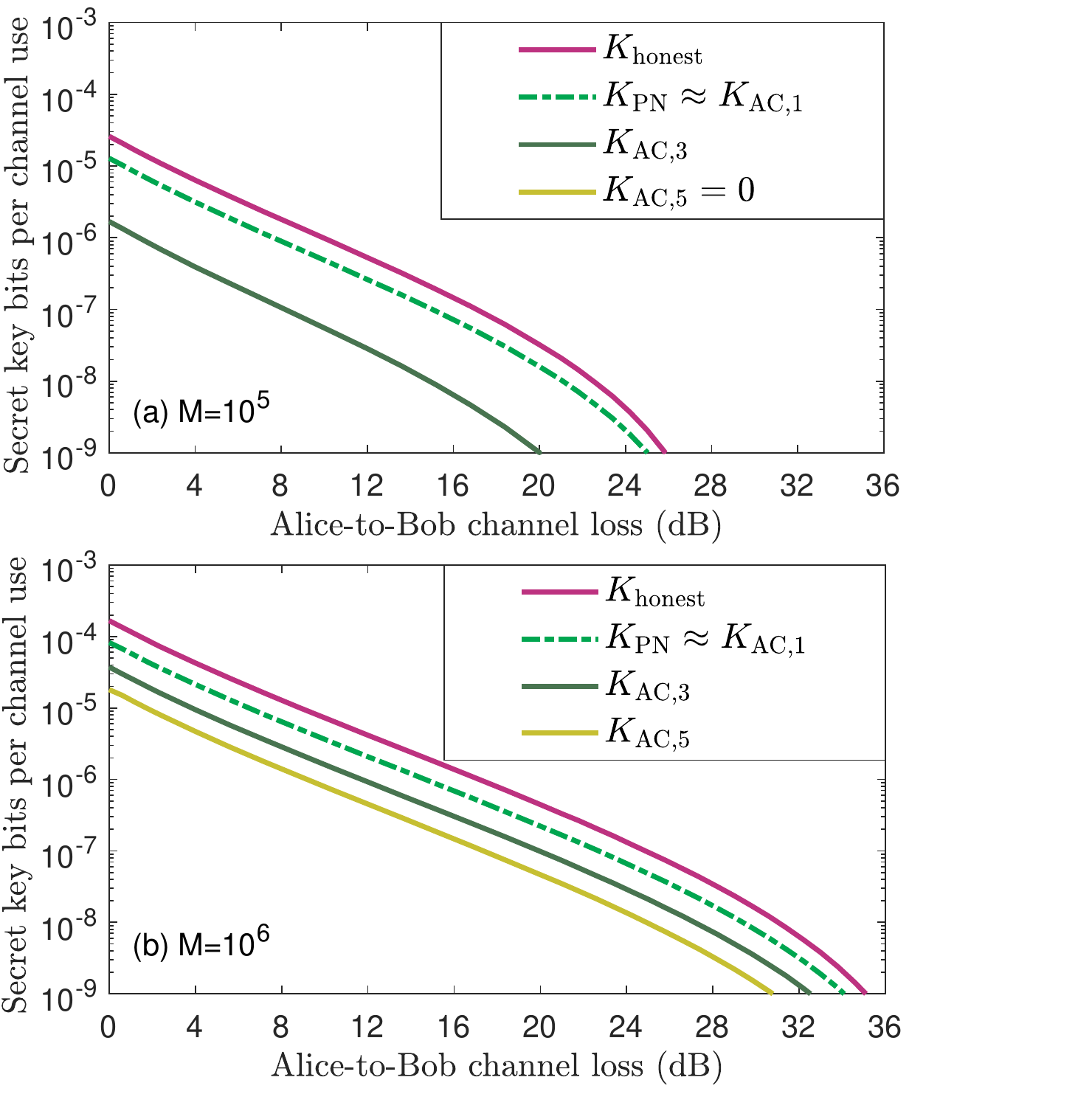}
	\caption{Secret key rate, $K$ of a decoy-state MDI-QKD scheme~\cite{X.B.Wang} in various adversarial scenarios with malicious devices, as a function of the total channel loss between Alice and Bob (assumed to be at the same distance of the untrusted measurement node). Two finite block-sizes are considered, \textbf{(a)} $M=10^{5}$ and \textbf{(b)} $M=10^{6}$, and the authentication cost is computed according to the distributed post-processing protocol of the Methods section. In both figures, the purple line is the secret key rate in the standard scenario ---where each party holds one QKD module and one classical post-processing (CP) unit, both of them trusted--- and green lines denote different corruption models. In particular, the dashed-dotted phosphorescent line is the secret key rate assuming passive and non-collaborative corrupted devices, which requires the use of two QKD pairs and two CP units per lab (all of them being possibly malicious) to provide security. A more conservative scenario is represented by the solid non-phosphorescent green lines, which assume active and collaborative corrupted devices. These lines further assume the same number, say $t$, of malicious QKD pairs and malicious CP units per lab, which requires the use of at least $n_{\rm q}=t+1$ QKD pairs and $n_{\rm c}=3t+1$ CP units per party to provide security. Specifically, the dark (light) green line corresponds to $t=3$ ($t=5$).}
	\label{fig:performance}
\end{figure}

Adhering to all the above, in Fig.~2, we plot the secret key rate as a function of the total channel loss for the MDI-QKD scheme, considering that Alice and Bob are at the same distance of the central untrusted node. Similarly, the secret key rate of the BB84 scheme is plotted in Supplementary Figure 4. In both cases, for illustration purposes two different block-sizes are considered, $M\in\{10^{5},10^{6}\}$. Within the AC corruption model, for concreteness we only address the symmetric case $t_{\rm q}=t_{\rm c}=t$, such that $n_{\rm q}=t+1$ and $n_{\rm c}=3t+1$. Hence, we use the notation $K_{\mathrm{AC},t}$ ($l_{\mathrm{AC},t}$) for the secret key rate (length) secure against $t$ corrupted devices of each kind in this model. Similarly, $K_{\rm PN}$ ($l_{\mathrm{PN}}$) denotes the secret key rate (length) in the PN model, which, as explained above, unambiguously requires $n_{\rm q}=n_{\rm c}=2$. Lastly, $K_{\rm honest}$ ($l_{\rm honest}$) denotes the secret key rate (length) in the standard situation where
each party holds one trusted QKD module and one trusted CP unit, \textit{i.e.,} $K_{\rm honest}=K_{\mathrm{AC},0}$ ($l_{\rm honest}=l_{\mathrm{AC},0}$).

The conclusions gathered from Fig.~2 are readily understood in view of the results of Sec.~\ref{Results}. In the first place, for both $M=10^{5}$ and $M=10^{6}$, we find that $K_{\rm PN}\approx{}K_{\mathrm{AC},1}$ to a precision that cannot be distinguished in the figure. This follows from the fact that, in both cases, two raw keys are generated (as $n_{\rm q}=2$) and the parties need to remove the information from one of them via PA. Indeed, comparing Eq.~(\ref{key length AC}) and Eq.~(\ref{secrecy AC}) with Eq.~(\ref{key length PN}) and Eq.~(\ref{secrecy PN}), one observes that
\begin{equation}
l_{\rm PN}=l_{\mathrm{AC},1},
\end{equation}
\textit{i.e.,} the secret key lengths coincide exactly for fixed security parameters, fixed experimental inputs ($N$ and $E_{\rm tol}$), and average observables. Thus, the minuscule difference between $K_{\rm PN}$ and $K_{\mathrm{AC},1}$ comes from the authentication cost, as $l_{\rm AU}\propto{R^{2}}$ with $R=2t+1$ ($R=1$) in the AC (PN) model.

The same argument relates $K_{\mathrm{AC},t}$ and $K_{\rm honest}/(t+1)$ for all $t$. On the one hand, for the specifications above,
\begin{equation}\label{match}
l_{\mathrm{AC},t}=l_{\rm honest}
\end{equation}
for all $t$, which corresponds to the key material coming from the honest QKD pair. On the other hand, in the presence of $t$ malicious QKD pairs, the extraction of the above key length requires the generation of $t+1$ raw keys in the AC model. Thus, from Eq.~(\ref{key rate}), it follows that
\begin{equation}\label{delta_AU}
\frac{K_{\rm honest}}{t+1}-K_{\mathrm{AC},t}=\frac{{\delta{l}}_{\rm AU}}{(t+1)N}
\end{equation}
in the simulations, where ${\delta{l}}_{\rm AU}$ denotes the extra authentication cost of the AC model with $t_{\rm q}=t_{\rm c}=t$, compared to the honest scenario. Due to the factor $N^{-1}$ in the right-hand side of Eq.~(\ref{delta_AU}), larger block sizes lead to smaller differences $K_{\rm honest}/(t+1)-K_{\mathrm{AC},t}$.

Finally, since $K_{\mathrm{AC},t}\propto{}(l_{\rm honest}-l_{\mathrm{AU}})$ and $l_{\mathrm{AU}}\propto(2t+1)^{2}$ in the AC model, $K_{\mathrm{AC},t}$ vanishes for any given block size if a large enough number of CP units is considered, as eventually $l_{\mathrm{AU}}>l_{\rm honest}$. This is the case for $M=10^{5}$ and $t=5$ in Fig.~2.
\section{Discussion}\label{Discussion}
QKD security today requires every QKD component to be honest and follow the protocol steps. Nevertheless, our experience in classical cryptography indicates that this might be very hard to certify in practice. Even in the DI setting, where the QKD devices are often referred to as uncharacterised black boxes, it is mandatory to assure that, beyond the reception of quantum signals from an untrusted source, the only interaction these boxes have with the outside world is the exchange of inputs and outputs with the legitimate parties. This assumption, despite weak, is still very hard to verify. Fortunately, as pointed out in~\cite{Curty}, one can protect QKD against malicious equipment by using redundant devices to combine VSS with PA, an approach that we follow in this work.

However, a major limitation of the proposal in~\cite{Curty} is that it relies on simulated broadcast, a very high-priced task in terms of total communication, especially for large numbers of CP units. What is more, the scheme presented in~\cite{Curty} requires the execution of $n_{\rm q}+1$ PA steps, where $n_{\rm q}$ is the total number of QKD pairs. In this work, we eliminate the limitation of simulated broadcast and show that a single PA step suffices, thus turning the approach in~\cite{Curty} into practical.

Moreover, the proposal in~\cite{Curty} assumes that the malicious devices may actively deviate from the protocol and collaborate with each other, which is probably over-pessimistic. For instance, an archetypical security breach consists of a malicious item implanted by an eavesdropper in an honest apparatus, leading to a passively corrupted device that may leak private information but sticks to the protocol prescriptions. Likewise, if the devices originate from different vendors, it is reasonable to expect that possibly corrupted apparatuses do not collaborate. In this work, we show that very natural assumptions like these allow to achieve a better performance than the active-collaborative model, both in terms of secret key rate and necessary resources.

Also, it is often stated in the QKD community that one could simply bitwise XOR the final keys generated by different QKD systems to defeat malicious equipment. Although this alternative may assure the privacy of the output, it has the major problem of generally requiring more devices than actually necessary to establish security, due to the non-distributed post-processing. For instance, note that not only the QKD module but also the CP unit in any given QKD system learns the raw key in the XOR approach, leading to a double-trouble situation where one must contemplate the worst possible combination of modules and units to guarantee the privacy of the raw key material. Similarly, the XOR approach does not prevent an actively malicious unit from jeopardizing the post-processing of the raw key generated by its module.

Furthermore, we would like to note that secret-sharing-type techniques are in fact the standard tool to guarantee security against untrusted devices. For instance, it is the adopted solution in modern hardware secure modules~\cite{Thales,Gemalto,Amazon}. Likewise, similar ideas to those we present here may be deployed in QKD to relax the security assumptions in trusted node network architectures, such that one can establish the security of the final keys even if some intermediate nodes are compromised~\cite{Salvail}.

Another contribution of this work is to evaluate the finite secret key rate of practical QKD schemes in the presence of malicious devices, for different corruption models and accounting for the authentication cost of the redundant classical communications. Particularly, based on our theoretical results, we devise an efficient distributed QKD post-processing protocol adequate for all the corruption models we examine. The simulations confirm that our techniques may achieve finite secret key rates comparable to those of standard QKD with trusted devices. Putting it all together, this work is a fundamental step towards the development of practical QKD systems secure against malicious devices possibly sabotaged by a third party, a major threat against classical cryptography today that cannot be put aside in the quantum-safe era.
\section{Methods}\label{Methods}
\subsection{Conditional verifiable secret sharing}\label{VSS}
Here, we introduce a modified version of the VSS scheme presented in~\cite{Maurer} that contemplates the possibility of aborting, thus providing a weaker cryptographic primitive than standard VSS. For this reason, we refer to it as conditional VSS.

We consider a scenario with one possibly dishonest dealer, $D$, and a set of $n$ parties, $\mathbb{P}=\{P_{1},\ldots,P_{n}\}$, $t$ of which are possibly corrupted. In this setting, a conditional VSS scheme is a pair of protocols, (Share, Reconstruct), satisfying three properties: \textit{privacy}, \textit{conditional commitment} and \textit{conditional correctness} (defined below). In full generality, Share and Reconstruct run as follows. During Share, $D$ \textit{distributes} an input $m$ among the $n$ parties, which pairwise perform consistency tests on their common information via secure channels and possibly abort. Upon non-abortion of Share, during Reconstruct the parties collaborate to retrieve $m$. The defining properties of conditional VSS are given below:
\begin{enumerate}
	\item{\textbf{Privacy.} If $D$ is honest, the information obtained by any set of $t$ or less parties prior to Reconstruct is independent of $m$.}
	\item{\textbf{Conditional commitment.} Upon non-abortion of Share, Reconstruct yields the same output for all non-actively corrupted parties.}
	\item{\textbf{Conditional correctness.} Upon non-abortion of Share, if $D$ is honest the common output of all non-actively corrupted parties is the input $m$.}
\end{enumerate}
Regarding the parties, all four non-mixed corruption models presented in the main text shall be addressed: AC, AN, PC and PN. However, we do not restrict to any of them yet. Also, note that the set of non-actively corrupted parties includes all the parties (and not only the honest ones) in the passive models.
As for the dealer, D is said to be dishonest if it may distribute incorrect/inconsistent information about his input to the parties or directly reveal it to them. 
In particular, this means that if the QKD modules belong to the PN model, even corrupted modules are honest dealers.

In what follows, we describe a pair of protocols, (Share, Reconstruct), that depend on various settings, and such that adequate choices of these settings confer the pair the category of a conditional VSS scheme. We remark that the adequacy of some given settings depends on the corruption model one assumes for the parties. As in the main text, the protocol definitions below assume that the parties and the dealer do not misbehave, whether or not these protocols are robust against active corruption. Also, the dealer's input $m$ is assumed to be a binary string, and we recall that the symbol ``$\oplus$" denotes bitwise XOR. In addition, this operation is generalised to a pair of strings with different lengths by padding the shortest one with as many zeros as necessary for the lengths to match. This said, Share runs as follows.
%
\begin{enumerate}
	\item{$D$ uses a $q$-out-of-$q$ SS scheme to split a message $m$ into $q$ random shares, by selecting the first $q-1$ shares $m_{i}$ at random and then choosing $m_{q}=m\oplus{m_{1}\oplus\ldots\oplus{m_{q-1}}}$.}
	\item{For $i=1,\ldots,q$, $D$ sends $m_{i}$ to all the parties in a certain subset, say $\sigma_{i}\subsetneq{\mathbb{P}}$, via secure channels. If any of these parties does not receive the share, she takes a zero bit string as default share.}
	\item{If $|\sigma_{i}|>1$, all pairs of parties in $\sigma_{i}$ perform a consistency test: they send each other their copies of $m_{i}$ over secure channels to check if they are equal. If any party finds an inconsistency, she aborts the protocol.}
\end{enumerate}
Importantly, abortion proceeds in two steps: the aborting party sends an abortion order to all other parties, and each receiving party resends the order to all the rest. Upon reception of an abortion order, the parties abort. Step two assures that the non-actively corrupted parties always abort collectively. Upon non-abortion of Share, Reconstruct runs as follows.

\begin{enumerate}
	\item{All pairs of parties send each other their shares through authenticated channels.}
	\item{For $i=1,\ldots,q$, each party uses MV to reconstruct the share $m_{i}$, and then obtains $m=\oplus_{i=1}^{q}m_{i}$.}
\end{enumerate}
In general, in order for MV to be well-defined, the output must be set to a default value in case of a tie. Nevertheless, ties never occur for the adequate choices of the parameters $n$ and $q$ and the subsets $\sigma_{i}$ we present next.

\textbf{Proposition 1.} \textit{Let $t$ be the maximum number of corrupted parties, and let $\{T_{1},\ldots,T_{\binom{n}{t}}\}$ be any ordered list of all possible combinations of $t$ parties. Under the following settings, (Share, Reconstruct) defines a conditional VSS scheme:
\begin{enumerate}
	\item{$n=3t+1$, $q=\binom{n}{t}$ and $\sigma_{i}=\mathbb{P}/T_{i}$ (AC corruption).}
	\item{$n=2t+2$, $q=n$ and $\sigma_{i}=\mathbb{P}/P_{i}$ (AN corruption).}
	\item{$n=t+1$, $q=n$ and $\sigma_{i}=P_{i}$ (PC corruption).}
	\item{$n=2$, $q=n$ and $\sigma_{i}=P_{i}$ (PN corruption).}
\end{enumerate}
What is more, the above settings are optimal in the number of parties.}\\

The reader is referred to Supplementary Note 3 for a proof of Proposition 1. Also, note that, by definition of $R$ (see Sec.~(\ref{alternative models})), we have that $R=|\sigma_{i}|$ for all $i$.

Finally, we remark that the above conditional VSS scheme enables secure multiparty computation of linear functions of the shared private input in a very simple way. Let $L(\cdot)$ be the linear function to be computed on $m$. Upon non-abortion of Share, each party applies $L$ to its shares of $m$, in so obtaining shares of $L(m)$. Since this step requires null communication, \textit{privacy}, \textit{conditional commitment} and \textit{conditional correctness} are trivially maintained.

\subsection{Generation of random bit strings}\label{RBS}
Distributed QKD post-processing also relies on the possibility to generate unbiased random bit strings (RBS) of a pre-fixed length $L$ among $n$ parties, when up to $t$ of them are possibly corrupted. Here, we describe a RBS generation protocol suitable for the active corruption models, AC and AN, that builds on conditional VSS to safeguard the randomness of its output string (the passive models shall be addressed afterwards).

Let us set the total number of parties, $n$, the total number of shares, $q$, and the subsets of parties, $\sigma_{i}$, as specified in Proposition 1 for the considered model (AC or AN). The RBS generation protocol runs as follows.

\begin{enumerate}
\item{For $k=1,\ldots,t+1$, $P_{k}$ creates a random $L$-bit string, $R_{k}$, and distributes it among all $n$ parties (including itself) using Share. If, for some $k$, Share aborts, the RBS generation protocol aborts. If a party receives any share whose length differs from $L$, she aborts.}
\item{Upon non-abortion of step 1, the parties use Reconstruct to obtain $R_{k}$ for all $k=1\ldots,t+1$. Then, each of them individually calculates $R=\oplus_{k=1}^{t+1}R_{k}$.}
\end{enumerate}
\textbf{Proposition 2.} \textit{The RBS generation protocol outputs a common random $L$-bits string for all non-actively corrupted parties.}\\

The reader is referred to Supplementary Note 3 for a proof of Proposition 2.

Finally, using the standard notion of passivity given in the main text, one can avoid the use of conditional VSS for RBS generation in the passive models (PC and PN). Instead, any given unit can generate the strings directly, and such strings are truly random by assumption.
\subsection{Distributed QKD post-processing protocol}\label{protocol}
Making use of our theoretical results, here we present a distributed QKD post-processing protocol based on conditional VSS that is appropriate for all non-mixed corruption models introduced in Sec.~\ref{Results}. We refer to it simply as Protocol.

In the first place, the parties agree on the corruption models they assume for the QKD modules and the CP units (which might be different in general), and also select the numbers $t_{\rm q}$ and $t_{\rm c}$ of corrupted devices they want to be protected against. In case they choose AC, AN or PC corruption (PN corruption) for the modules, they must hold $n_{\rm q}=t_{\rm q}+1$ ($n_{\rm q}=2$) QKD pairs in total ---given that they stick to the rule of using the minimum valid amount of devices--- and the secret key length $l$ is given by Eq.~(\ref{key length AC}) (Eq.~(\ref{key length PN})). Similarly, they provide themselves with as many CP units as specified in Table 1 for their preferred model. Coming next, they agree on a correctness (secrecy) parameter, $\epsilon_{\rm cor}$ ($\epsilon_{\rm sec}$), and a total authentication error $\epsilon_{\rm AU}$, such that $\epsilon_{\rm AU}<\epsilon_{\rm cor}$ and $\epsilon_{\rm AU}<\epsilon_{\rm sec}$.

For $j=1,\ldots,n_{\rm q}$, the pair $(\textrm{QKD}_{\textrm{A}_{j}},\textrm{QKD}_{\textrm{B}_{j}})$ runs a QKD session to generate the basis Z raw key strings, $(r_{\rm A}^{j},r_{\rm B}^{j})$, to be kept private, and some non-private protocol information, $(\textrm{info}_{\textrm{A}}^{j},\textrm{info}_{\textrm{B}}^{j})$, typically including the basis and intensity settings, detection events, etc. Crucially, $(\textrm{info}_{\textrm{A}}^{j},\textrm{info}_{\textrm{B}}^{j})$ includes all the raw key material required for parameter estimation. The post-processing procedure (namely, Protocol) starts next and is described below. Although the description assumes that the possibly corrupted devices do not deviate from the protocol prescriptions, Protocol is indeed secure against active eavesdroppers, as established in Proposition 3 below. Finally, although not explicitly stated, in case of abortion, the aborting party must notify the other party. This said, Protocol runs as follows.

Let us focus on, say, the $j$-th QKD pair.
\begin{enumerate}
	\item[1.]{Distribution of data.} $\textrm{QKD}_{\textrm{A}_{j}}$ ($\textrm{QKD}_{\textrm{B}_{j}}$) distributes shares of its raw key $r_{\rm A}^{j}$ ($r_{\rm B}^{j}$) among the $\textrm{CP}_{\textrm{A}_{l}}$ following the Share protocol of a conditional VSS scheme (see Sec.~\ref{VSS}) for the selected corruption model of the CP units. We denote the set of units that receive the $i$-th share of $r_{\rm A}^{j}$ ($r_{\rm B}^{j}$) by $\sigma_{i}^{\rm A}$ ($\sigma_{i}^{\rm B}$), which without loss of generality is common for all $j=1,\ldots{},n_{\rm q}$. In addition, $\textrm{QKD}_{\textrm{A}_{j}}$ ($\textrm{QKD}_{\textrm{B}_{j}}$) sends the protocol information $\textrm{info}_{\textrm{A}}^{j}$ ($\textrm{info}_{\textrm{B}}^{j}$) to all $\textrm{CP}_{\textrm{A}_{l}}\in\sigma_{1}^{\rm A}$ ($\textrm{CP}_{\textrm{B}_{l'}}\in\sigma_{1}^{\rm B}$), and the latter perform a consistency test on this data: they pairwise check that their copies of $\textrm{info}_{\textrm{A}}^{j}$ ($\textrm{info}_{\textrm{B}}^{j}$) match via authenticated channels. If a $\textrm{CP}_{\textrm{A}_{l}}$ ($\textrm{CP}_{\textrm{B}_{l'}}$) finds an inconsistency, it aborts the protocol (see the Share protocol in the section devoted to conditional VSS for the two-step abortion procedure we consider).
	\item[2.]{Sifting.} Each $\textrm{CP}_{\textrm{A}_{l}}\in{}\sigma_{1}^{\rm A}$ sends its copy of $\textrm{info}_{\textrm{A}}^{j}$ to all $\textrm{CP}_{\textrm{B}_{l'}}\in{}\sigma_{1}^{\rm B}$, which individually apply majority voting (MV) to decide on a single copy. Then, the $\textrm{CP}_{\textrm{B}_{l'}}\in{}\sigma_{1}^{\rm B}$ forward some sifting information, $\textrm{sift}^{j}$, computable from the pair $(\textrm{info}_{\textrm{A}}^{j},\textrm{info}_{\textrm{B}}^{j})$, to the $\textrm{CP}_{\textrm{B}_{l'}}\notin{}\sigma_{1}^{\rm B}$, which apply MV too. Using $\textrm{sift}^{j}$, every $\textrm{CP}_{\textrm{B}_{l'}}$ discards some key bits from their shares of $r_{\rm B}^{j}$ to obtain shares of the sifted key, $s_{\rm B}^{j}$. Alternative sifting schemes that require to discard random subsets of the data could easily be adapted by including a random bit string (RBS) generation protocol (see Sec.~\ref{RBS}).
	\item[3.]{Parameter estimation.}
	Using $(\textrm{info}_{\textrm{A}}^{j},\textrm{info}_{\textrm{B}}^{j})$, each $\textrm{CP}_{\textrm{B}_{l'}}\in\sigma_{1}^{\rm B}$ computes a hypothetical lower bound $h_{\varepsilon}^{j}$ (see Supplementary Notes 1 and 2 for the details) on the $\varepsilon$-smooth min-entropy of $s_{\rm B}^{j}$ conditioned on the information held by an eavesdropper up to the parameter estimation (PE) step, for a certain $\varepsilon$ that depends on the PE procedure.
\end{enumerate}
Once steps 1 to 3 are implemented for $j=1,\ldots{}n_{\rm q}$, all $\textrm{CP}_{\textrm{B}_{l'}}$ construct their shares of the concatenated sifted key $s_{\rm B}=[s_{\rm B}^{1},\ldots{},s_{\rm B}^{n_{\rm q}}]$, such that the $k$-th share of $s_{\rm B}$ is simply given by the concatenation of the $k$-th share of $s_{\rm B}^{1}$, the $k$-th share of $s_{\rm B}^{2}$, and so on. In addition, from all $n_{\rm q}$ values $h_{\varepsilon}^{j}$, every $\textrm{CP}_{\textrm{B}_{l'}}\in{}\sigma_{1}^{\rm B}$ computes a lower bound $l$ on the secret key length extractable from $s_{\rm B}$ via PA. If a $\textrm{CP}_{\textrm{B}_{l'}}\in{}\sigma_{1}^{\rm B}$ finds $l\leq{0}$, it aborts the protocol. Otherwise, the post-processing proceeds as follows.
\begin{enumerate}
\item[4.]{RBS generation.} Every $\textrm{CP}_{\textrm{B}_{l'}}\in{}\sigma_{1}^{\rm B}$ forwards $l$ to the $\textrm{CP}_{\textrm{B}_{l'}}\notin{}\sigma_{1}^{\rm B}$, which apply MV. All $\textrm{CP}_{\textrm{B}_{l'}}$ perform a RBS generation protocol to select two random 2-universal hash functions, $h_{\rm EV}$ and $h_{\rm PA}$, respectively devoted to error verification (EV) and PA.
\item[5.]{Information reconciliation.} Every $\textrm{CP}_{\textrm{B}_{l'}}$ computes its shares of the string of concatenated syndromes, $sy_{\rm B}=[sy(s_{\rm B}^{1}),\ldots{},sy(s_{\rm B}^{n_{\rm q}})]$, and the EV tag $h_{\rm EV,B}=h_{\rm EV}(s_{\rm B})$. Here, $sy(\cdot)$ is a linear function specified by an error correction (EC) protocol for a pre-agreed quantum bit error rate (QBER). All together, the $\textrm{CP}_{\textrm{B}_{l'}}$ reconstruct $sy_{\rm B}$ and $h_{\rm EV,B}$ via the Reconstruct protocol of a conditional VSS scheme (see Sec.~\ref{VSS}). Each $\textrm{CP}_{\textrm{B}_{l'}}\in{}\sigma_{1}^{\rm B}$ sends the following items to every $\textrm{CP}_{\textrm{A}_{l}}\in{}\sigma_{1}^{\rm A}$:
\begin{enumerate}
	\item{The total sifting information, $\{\textrm{sift}^{j}\}_{j=1}^{n_{\rm q}}$.}
	\item{The syndrome information, $sy_{\rm B}$, a description of $h_{\rm EV}$ and the EV tag, $h_{\rm EV,B}$.}
	\item{A description of $h_{\rm PA}$.}
\end{enumerate}
For all 3 items, each $\textrm{CP}_{\textrm{A}_{l}}\in{}\sigma_{1}^{\rm A}$ applies MV to decide on a single copy. Then, it forwards $\{\textrm{sift} ^{j}\}_{j=1}^{n_{\rm q}}$, $h_{\rm EV}$ and $h_{\rm PA}$ to the $\textrm{CP}_{\textrm{A}_{l}}\notin{}\sigma_{1}^{\rm A}$ (which apply MV too), and every $\textrm{CP}_{\textrm{A}_{l}}$ sifts its shares of the raw keys $r_{\rm A}^{j}$ to obtain shares of the concatenated sifted key $s_{\rm A}=[s_{\rm A}^{1},\cdots{},s_{\rm A}^{n_{\rm q}}]$. Following the EC protocol, all $\textrm{CP}_{\textrm{A}_{l}}$ compute their shares of the concatenated syndrome string, $sy_{\rm A}=[sy(s_{\rm A}^{1}),\ldots{},sy(s_{\rm A}^{n_{\rm q}})]$, and jointly reconstruct it via the Reconstruct protocol of a conditional VSS scheme. From $sy_{\rm B}$ and $sy_{\rm A}$, each $\textrm{CP}_{\textrm{A}_{l}}\in{}\sigma_{1}^{\rm A}$ computes the error pattern $\hat{e}$ and updates its copy of the first share of $s_{\rm A}$ by XOR-ing it with $\hat{e}$. Thus, by construction, Alice's corrected key is $\hat{s}_{\rm A}=s_{\rm A}\oplus\hat{e}$, ``$\oplus$" denoting bitwise XOR. Then, all $\textrm{CP}_{\textrm{A}_{l}}$ compute their shares of the EV tag $h_{\rm EV,A}=h_{\rm EV}(\hat{s}_{\rm A})$ and jointly reconstruct it via the Reconstruct protocol of a conditional VSS scheme. Finally, every $\textrm{CP}_{\textrm{A}_{l}}\in{}\sigma_{1}^{\rm A}$ checks that $h_{\rm EV,A}=h_{\rm EV,B}$. Otherwise, it aborts the protocol.
\item[6.]{Privacy amplification.} In case of not aborting, every $\textrm{CP}_{\textrm{A}_{l}}$ ($\textrm{CP}_{\textrm{B}_{l'}}$) computes its shares of the final key $k_{\rm A}=h_{\rm PA}(\hat{s}_{\rm A})$ ($k_{\rm B}=h_{\rm PA}(s_{\rm B})$).
\end{enumerate}
In Supplementary Note 4, we prove that the following security claim holds for all (non-mixed) corruption models of the QKD modules and the CP units.

\textbf{Proposition 3.} \textit{Suppose that Protocol does not abort. Then, Alice and Bob can unambiguously determine unique $\epsilon_{\rm cor}$-correct and $\epsilon_{\rm sec}$-secret final keys.}\\

Importantly, the determination of such final keys by Alice and Bob can be done by simply applying MV on the key shares held by their respective CP units, followed by an XOR operation. More generally, in the presence of actively corrupted units, the $\textrm{CP}_{\textrm{A}_{l}}$ ($\textrm{CP}_{\textrm{B}_{l'}}$) can forward their final shares to a local key management layer~\cite{Peev,Sasaki}. There, they could be stored in distributed memories or employed for applications such as message encryption, which in turn can be performed share-wise too.
\section{Data availability}
No datasets were generated or analysed during the current study.
\section{Acknowledgements}
We thank Liu Zhang Chen-Da for useful discussions on verifiable secret sharing and secure multiparty computation. We acknowledge support from the Spanish Ministry of Economy and Competitiveness (MINECO), the Fondo Europeo de Desarrollo Regional (FEDER) through the grant TEC2017-88243-R, and the European Union's Horizon 2020 research and innovation programme under the Marie Sklodowska-Curie grant agreement No 675662 (project QCALL) for financial support. VZ gratefully acknowledges support from a FPU scholarship from the Spanish Ministry of Education.
\section{Author contributions}
M.C. conceived the initial idea and triggered the consideration of
this research project. V.Z. made the theoretical analysis and
performed the numerical simulations, with inputs from both authors.
M.C. and V.Z. analysed the results and prepared the manuscript. 
\section{Competing interests}
The authors declare no competing interests.
\onecolumngrid
\section{Supplementary Information}
\section*{Supplementary Note 1: secret key length in the AC, AN and PN models for the QKD modules}\label{AC}
In this section, we derive the extractable secret key length under the assumption that the QKD modules belong to the AC model and $n_{\rm q}=t_{\rm q}+1$, and its validity for the intermediate models AN and PC is therefore trivial.

In what follows, asterisks will be used to denote the well-defined versions of certain quantities to which the QKD modules are committed with respect to the honest CP units in the distributed QKD post-processing, in virtue of the properties of standard/conditional VSS and the redundancy of the classical communications. For a detailed proof of the well-definiteness of many of these quantities in a specific distributed post-processing scheme, the reader is referred to Protocol in the Methods section of the main text and Supplementary Note 4.

In any case, despite the misbehaving of the possibly corrupted QKD modules and CP units, the distributed post-processing can guarantee the existence of a well-defined sifted key at Bob's lab, $s_{\rm B}^{*}$, reconstructible through the Reconstruct protocol of a standard/conditional VSS scheme (see the Methods section in the main text), and given by the concatenation of all well-defined sifted keys from the different QKD pairs. Namely, $s_{\rm B}^{*}=[s_{\rm B}^{1*},\ldots{},s_{\rm B}^{n_{\rm q}*}]$. By applying PA with 2-universal hashing~\cite{Tomamichel1}, a $\hat{\epsilon}_{\rm sec}$-secret key can be extracted from $s_{\rm B}^{*}$ as long as its length $l^{*}$ verifies~\cite{Renner}
\begin{equation}\label{LHL}
l^{*}\leq{}\left\lfloor{H_{\rm min}^{\epsilon}(s_{\rm B}^{*}|E')-2\log_{2}}\left(\frac{1}{2\epsilon_{\rm PA}}\right)\right\rfloor,
\end{equation}
for all $\hat{\epsilon}_{\rm sec}\geq{\epsilon+\epsilon_{\rm PA}}$, where $H_{\rm min}^{\epsilon}(s_{\rm B}^{*}|E')$ is the $\epsilon$-smooth min-entropy of $s_{\rm B}^{*}$ conditioned on the (possibly quantum) information $E'$ held by Eve ---the omniscient eavesdropper controlling all corrupted QKD modules---, and $\epsilon_{\rm PA}$ is the error probability of PA.

Crucially, note that no adversary may have access to more information about the final keys than the omniscient Eve just presented, so it suffices to refer to this Eve. To be precise, such an Eve potentially knows all the raw key material coming from corrupted QKD pairs, and all the information about the key of the honest pair revealed by the public discussion and her interaction with the quantum channel. In particular, possible adversaries corrupting the CP units do not have access to any more information about the honest pair's keys than the Eve above, because a distributed post-processing ---say, Protocol in the main text--- may assure that, since these keys are delivered by two honest dealers, \textit{i.e.,} the honest QKD modules, they are kept private to the CP units. The reader is referred to the Methods section of the main text for a definition of the \textit{privacy} property of conditional VSS. This said, the derivation goes as follows.

Without loss of generality, $E'$ can be decomposed as $E'=CE$, where $C$ denotes the information gained by Eve when she learns the syndrome, $sy^{*}_{\rm B}$, and the EV tag, $h^{*}_{\rm EV,B}$, and $E$ denotes the information she holds in advance of that. Assuming that EC is applied individually on each $s_{\rm A}^{j*}$ to reconcile it with the corresponding $s_{\rm B}^{j*}$, the well-defined syndrome information sent to Bob in the information reconciliation (IR) step (see for instance Protocol in the main text) splits as $sy^{*}_{\rm B}=[{sy}^{*}(s_{\rm B}^{1*}),\ldots{},{sy}^{*}(s_{\rm B}^{n_{\rm q}*})]$. Clearly, all $n_{\rm q}$ items in $sy_{\rm B}$ but the one that comes from the honest QKD pair are possibly known to Eve a priori. If we denote the pair index of the honest QKD pair by $``\mathrm{h}"$, this implies that only ${sy}^{*}(s_{\rm B}^{\mathrm{h}*})$ contributes to $C$, together with the error verification tag $h^{*}_{\rm EV,B}$, whose size is $|h^{*}_{\rm EV,B}|=\lceil\log_{2}(2/\hat{\epsilon}_{\rm cor})\rceil$ bits. Then, from a chain inequality for smooth entropies~\cite{Renner}, $H_{\rm min}^{\epsilon}(s_{\rm B}^{*}|E')\geq{}H_{\rm min}^{\epsilon}(s_{\rm B}^{*}|E)-|C|$ and therefore
\begin{equation}\label{chain}
H_{\rm min}^{\epsilon}(s_{\rm B}^{*}|E')\geq{}H_{\rm min}^{\epsilon}(s_{\rm B}^{*}|E)-\left|{sy}^{*}(s_{\rm B}^{\mathrm{h}*})\right|-\left\lceil{\log_{2}\left(\frac{2}{\hat{\epsilon}_{\rm{cor}}}\right)}\right\rceil.
\end{equation}
If we use the decomposition $s_{\rm B}^{*}=s_{\rm B}^{\mathrm{h}*}s_{\rm B}^{\mathrm{d}*}$ (where $s_{\rm B}^{\mathrm{d}*}$ includes all the substrings of $s_{\rm B}^{*}$ that come from dishonest QKD pairs), the following chain rule holds~\cite{Vitanov}. For all $\varepsilon,\varepsilon'\geq{0}$ and for all $\epsilon$ such that  $\epsilon>2\varepsilon+\varepsilon'$,
\begin{equation}\label{Vitanov}
H_{\rm min}^{\epsilon}(s_{\rm B}^{\mathrm{h}*}s_{\rm B}^{\mathrm{d}*}|E)\geq{}H_{\rm min}^{\varepsilon}(s_{\rm B}^{\mathrm{h}*}|s_{\rm B}^{\mathrm{d}*}E)+H_{\rm min}^{\varepsilon'}(s_{\rm B}^{\mathrm{d}*}|E)-\log_{2}\left(\frac{1}{\epsilon-2\varepsilon-\varepsilon'}\right),
\end{equation}
where $\varepsilon$ and $\varepsilon'$ are the smoothing parameters of the corresponding smooth min-entropies~\cite{Renner}. We recall that $\varepsilon$ depends on the parameter estimation (PE) procedure followed by Alice and Bob. Also, one can set $\varepsilon'=0$ and use the trivial bound $H_{\rm min}^{\varepsilon'}(s_{\rm B}^{\mathrm{d}*}|E)\geq{0}$ valid for all $\varepsilon'\geq{}0$, as $s_{\rm B}^{\mathrm{d}*}$ could be entirely known to Eve. This amounts to say that $s_{\rm B}^{\mathrm{d}*}$ is included in $E$, which further implies that $H_{\rm min}^{\varepsilon}(s_{\rm B}^{\mathrm{h}*}|s_{\rm B}^{\mathrm{d}*}E)=H_{\rm min}^{\varepsilon}(s_{\rm B}^{\mathrm{h}*}|E)$. From these two results, inserting Eq.~(\ref{Vitanov}) in Eq.~(\ref{chain}) one finds
\begin{equation}
H_{\rm min}^{2\varepsilon+\delta}(s_{\rm B}^{*}|E')\geq{}H_{\rm min}^{\varepsilon}(s_{\rm B}^{\mathrm{h}*}|E)-\left|{sy}^{*}(s_{\rm B}^{\mathrm{h}*})\right|-\log_{2}\left(\frac{4}{\hat{\epsilon}_{\rm cor}\delta}\right),
\end{equation}
where we use the fact that $\lceil{\log_{2}(2/\epsilon_{\rm{cor}})}\rceil\leq{\log_{2}\left({4}/\epsilon_{\rm{cor}}\right)}$ and also define the slack variable $\delta=\epsilon-2\varepsilon$, such that $\delta>0$. Further inserting the previous equation in Eq.~(\ref{LHL}), it follows that one can extract
\begin{equation}\label{key length}
l^{*}\leq{}\left\lfloor{H_{\rm min}^{\varepsilon}(s_{\rm B}^{\mathrm{h}*}|E)-\left|{sy}^{*}(s_{\rm B}^{\mathrm{h}*})\right|-\log_{2}\left(\frac{1}{\hat{\epsilon}_{\rm cor}\epsilon_{\rm PA}^{2}\delta}\right)}\right\rfloor
\end{equation}
$\hat{\epsilon}_{\rm sec}$-secret key bits for all
\begin{equation}\label{secrecy parameter}
\hat{\epsilon}_{\rm sec}\geq{2\varepsilon+\delta+\epsilon_{\rm PA}},
\end{equation}
and $\delta>0$.

Notably, the analysis above is conditioned on the successful authentication of all the classical communications. Thus, for a given total authentication error $\epsilon_{\rm AU}$, the overall secrecy parameter is given by
\begin{equation}\label{secrecy}
\epsilon_{\rm sec}=\hat{\epsilon}_{\rm sec}+\epsilon_{\rm AU}.
\end{equation}
Crucially, the honest QKD pair is unknown and thus Eq.~(\ref{key length}) cannot be evaluated in practice. However, it implies a looser but more convenient bound that does not rely on the knowledge of the honest pair by assuming a worst case scenario.
Precisely, let $h^{j*}_{\varepsilon}$ denote the hypothetical lower bound on $H_{\rm min}^{\varepsilon}(s_{\rm B}^{j*}|E)$ determined by 
the well-defined protocol information, say $(\textrm{info}_{\textrm{A}}^{j*},\textrm{info}_{\textrm{B}}^{j*})$, delivered by the $j$-th QKD module. We use the term \textit{hypothetical} here because, even though the distributed QKD post-processing can assure that the $j$-th QKD pair is committed to a single value $h^{j*}_{\varepsilon}$ via $(\textrm{info}_{\textrm{A}}^{j*},\textrm{info}_{\textrm{B}}^{j*})$, one cannot assure that such $h^{j*}_{\varepsilon}$ is a valid lower bound on $H_{\rm min}^{\varepsilon}(s_{\rm B}^{j*}|E)$ unless $j=\rm h$. Let us further explain this point. On the one hand,
if $j\neq{}\rm h$, $(\textrm{info}_{\textrm{A}}^{j*},\textrm{info}_{\textrm{B}}^{j*})$ might be unfaithful information ---thus, unsuitable for correct PE--- and all one can guarantee is that the trivial bound $H_{\rm min}^{\epsilon}(s_{\rm B}^{j*}|E)=0$ holds for all $\epsilon$. On the other hand, let us focus on the case $j=\mathrm{h}$. $\mathrm{QKD}_{A_{\mathrm{h}}}$ and $\mathrm{QKD}_{B_{\mathrm{h}}}$ indeed create a pair of raw keys via quantum communication to be delivered in the post processing (see the \textit{correctness} of conditional VSS in the Methods section of the main text), and indeed generate the related faithful protocol information. Then, distributed QKD post-processing (for instance, the protocol based on conditional VSS presented in the main text, whose security is addressed in Supplementary Note 4) allows to assure that the pair of keys coming from $\mathrm{QKD}_{A_{\mathrm{h}}}$ and $\mathrm{QKD}_{B_{\mathrm{h}}}$ is sifted, reconciled and subjected to PA correctly by an honest majority of CP units in each lab. In particular, $h^{\rm h*}_{\varepsilon}$ is a valid lower bound on $H_{\rm min}^{\varepsilon}(s_{\rm B}^{\rm h*}|E)$, such that the more convenient lower bound on $H_{\rm min}^{\varepsilon}(s_{\rm B}^{\mathrm{h}*}|E)-\left|{sy}^{*}(s_{\rm B}^{\mathrm{h}*})\right|$ that we referred to above is the straightforward bound $\min_{j}\{h^{j*}_{\varepsilon}-\bigl|{sy}^{*}(s_{\rm B}^{j*})\bigr|\}$. Defining $\lambda_{j}^{*}=\bigl|{sy}^{*}(s_{\rm B}^{j*})\bigr|$ to match the notation in the main text, we have that the well defined $l^{*}$ reached by all honest CP units reads
\begin{equation}\label{looser}
l^{*}=\left\lfloor{\min_{j}\biggl\{h^{j*}_{\varepsilon}-\lambda_{j}^{*}\biggr\}-\log_{2}\left(\frac{1}{\hat{\epsilon}_{\rm cor}\epsilon_{\rm PA}^{2}\delta}\right)}\right\rfloor.
\end{equation}
Also note that, for simplicity of the notation, the asterisks are omitted in the main text.
%
\section*{Supplementary Note 2: secret key length in the PN corruption model for the QKD modules}\label{PN}
In this section we derive the secret key length that one can extract via distributed QKD post-processing (say, Protocol in the main text) under the assumption that the possibly corrupted QKD pairs belong to the PN corruption model.

As explained in the main text, in this scenario we can assume $n_{\rm q}=t_{\rm q}$. This choice allows to fairly compare the performance of the AC and the PN corruption models in terms of the secret key rate, and it means that every QKD pair might be corrupted by an independent eavesdropper, say $\mathrm{Eve}_{j}$ (see Supplementary Figure~\ref{fig:multiple_Eves}), with $j=1,\ldots{}n_{\rm q}$. Let us focus on one of them, say $\mathrm{Eve}_{v}$. We denote by $E_{v}$ the information held by $\mathrm{Eve}_{v}$ prior to the IR step. Defining, for instance, $Z_{1}=s_{\rm B}^{v*}$, $Z_{j}=s_{\rm B}^{(j-1)*}$ for $j=2,\ldots,v$ and $Z_{j}=s_{\rm B}^{j*}$ for $j=v+1,\ldots,n_{\rm q}$, the next holds:
\begin{enumerate}
	\item{$H_{\rm min}^{\epsilon_{1}}(Z_{1}|E_{v})=0$ for all $\epsilon_{1}$.}
	\item{$H_{\rm min}^{\epsilon'_{j}}(Z_{j}|Z_{j-1}\ldots{}Z_{1}E_{v})=H_{\rm min}^{\epsilon'_{j}}(Z_{j}|E_{v})$ for all $\epsilon'_{j}$ and $j=2,\ldots{},n_{\rm q}$.}
\end{enumerate}
Therefore, one can apply the simplified version, Eq.~(\ref{simplified}), of the generalised chain rule for conditional smooth min-entropies presented in Supplementary Note 9. This yields,
\begin{equation}
H_{\rm min}^{(n_{\rm q}-1)(2\varepsilon+\delta)}(s^{*}_{\rm B}|E_{v})\geq{\sum_{j\neq{v}}^{n_{\rm q}}H_{\rm min}^{\varepsilon}(s_{\rm B}^{j*}|E_{v})-\log_{2}\left(\frac{1}{\delta^{n_{\rm q}-1}}\right)},
\end{equation}
with $\varepsilon,\delta>0$ and $n_{\rm q}\geq{}2$. Coming next, we account for the information that $\mathrm{Eve}_{v}$ gains at the IR step. The total information held by $\mathrm{Eve}_{v}$ a posteriori of IR can be decomposed as ${E}_{v}^{'}=C_{v}{E}_{v}$, where $C_{v}$ denotes the information she learns during IR. Precisely, $C_{v}$ contemplates all the syndromes, $sy^{*}(s_{\rm B}^{j*})$, with $j\neq{v}$, and the EV tag $h^{*}_{\rm EV,B}$, such that $|h^{*}_{\rm EV,B}|=\lceil\log_{2}(2/\hat{\epsilon}_{\rm cor})\rceil$. Note that we are assuming that EC (but not EV) is implemented separately for each $j=1,\ldots{}n_{\rm q}$ in the post-processing. From a chain inequality for smooth entropies~\cite{Renner} previously used in Supplementary Note 1, we have that
\begin{equation}
H_{\rm min}^{(n_{\rm q}-1)(2\varepsilon+\delta)}(s_{\rm B}^{*}|{E}_{v}^{'})\geq{\sum_{j\neq{v}}^{n_{\rm q}}\biggl\{H_{\rm min}^{\varepsilon}(s_{\rm B}^{j*}|E_{v})-|sy^{*}(s_{\rm B}^{j*})|\biggr\}-\log_{2}\left(\frac{4}{\hat{\epsilon}_{\rm cor}\delta^{n_{\rm q}-1}}\right)}.
\end{equation}
By applying PA with 2-universal hashing~\cite{Tomamichel1}, a key that is $\hat{\epsilon}_{\rm sec}$-secret with respect to ${E}_{v}^{'}$ can be extracted from $s_{\rm B}^{*}$, as long as the output length satisfies~\cite{Renner}
\begin{equation}\label{looser PN}
l^{*}\leq{}\Biggl\lfloor\sum_{j\neq{v}}^{n_{\rm q}}\biggl\{H_{\rm min}^{\varepsilon}(s_{\rm B}^{j*}|E_{v})-\bigl|{sy}^{*}\bigl(s_{\rm B}^{j*}\bigr)\bigr|\biggr\}-\log_{2}\left(\frac{1}{\hat{\epsilon}_{\rm cor}\epsilon_{\rm PA}^{2}\delta^{n_{\rm q}-1}}\right)\Biggr\rfloor,
\end{equation}
for all
\begin{equation}
\hat{\epsilon}_{\rm sec}\geq{}(n_{\rm q}-1)(2\varepsilon+\delta)+\epsilon_{\rm PA}.
\end{equation}
Note that all the parameters above are defined as in Supplementary Note 1. Lastly, composing the total authentication error $\epsilon_{\rm AU}$ (pre-agreed by the parties), the overall secrecy parameter reads
\begin{equation}\label{secrecy_2}
\epsilon_{\rm sec}=\hat{\epsilon}_{\rm sec}+\epsilon_{\rm AU}.
\end{equation}
\begin{figure}[!htbp]
	\centering 
	\includegraphics[width=5.6cm,height=3.6cm]{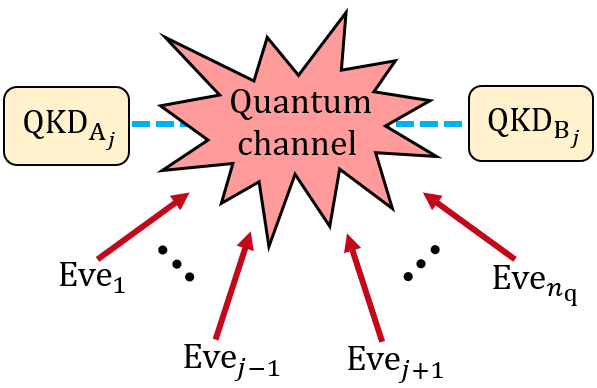}\\
	\caption{Supplementary Figure 1. Depiction of a setting where $n_{\rm q}-1$ non-collaborative eavesdroppers, $\{\mathrm{Eve}_{v}\}_{v\neq{j}}$ (where $\mathrm{Eve}_{v}$ is the eavesdropper controlling the $v$-th QKD pair), attack the quantum communication between $\mathrm{QKD}_{\mathrm{A}_{j}}$ and $\mathrm{QKD}_{\mathrm{B}_{j}}$. The assumption that the possibly corrupted QKD pairs are passive implies that $\mathrm{Eve}_{j}$ has total access to the internal information of $\mathrm{QKD}_{\mathrm{A}_{j}}$ and $\mathrm{QKD}_{\mathrm{B}_{j}}$, but the latter deliver faithful key material and protocol information.}
	\label{fig:multiple_Eves}
\end{figure}

Finally, note that Eq.~(\ref{looser PN}) determines the extractable key length that provides security with respect to the information ${E}_{v}^{'}$ held by $\mathrm{Eve}_{v}$. Nevertheless, one can provide security against all $\{\mathrm{Eve}_{v}\}_{v=1}^{n_{\rm q}}$ by taking
\begin{equation}\label{secrecy PN 3}
l^{*}=\Biggl\lfloor\min_{v}\sum_{j\neq{v}}^{n_{\rm q}}\biggl\{H_{\rm min}^{\varepsilon}(s_{\rm B}^{j*}|E_{v})-\lambda_{j}^{*}\biggr\}-\log_{2}\left(\frac{1}{\hat{\epsilon}_{\rm cor}\epsilon_{\rm PA}^{2}\delta^{n_{\rm q}-1}}\right)\Biggr\rfloor,
\end{equation}
where we defined $\lambda_{j}^{*}=\bigl|{sy}^{*}\bigl(s_{\rm B}^{j*}\bigr)\bigr|$ to match the notation in the main text.

To conclude this part, we remark that, in the AC corruption model, $h_{\varepsilon}^{j*}$ does not necessarily pose a lower bound on $H_{\rm min}^{\varepsilon}(s_{\rm B}^{j*}|E)$. However, by assumption, in the PN model the $j$-th QKD pair indeed creates a pair of raw keys via quantum communication to be delivered in the post-processing (see the \textit{correctness} of conditional VSS in the Methods section of the main text), and is committed to a value $h_{\varepsilon}^{j*}$ that poses a lower bound on $H_{\rm min}^{\varepsilon}(s_{\rm B}^{j*}|E_{v})$ for all $v\neq{}j$ (in the main text the asterisks are omitted for simplicity). Finally, we recall that these bounds also hold against those ``Eves" possibly corrupting the CP units, as long as one implements a distributed post-processing (say, Protocol in the main text). This is so because, since in the PN model all the QKD modules are honest dealers, in particular they keep their secret key material private to the CP units during the distribution stage (see the \textit{privacy} of conditional VSS in the Methods section of the main text).
\section*{Supplementary Note 3: proof of propositions 1 and 2}\label{props}
Here, we give detailed proofs of propositions 1 and 2 in the Methods section of the main text.\\

\textbf{Proposition 1.} Proposition 1 establishes adequate settings under which the pair of protocols (Share, Reconstruct) presented in the Methods section of the main text defines a conditional VSS scheme for every non-mixed corruption model of the parties. Here, we address all four scenarios one by one.
\begin{enumerate}
	
	\item{\textbf{AC corruption ($t>0$).}} The considered settings are $n=3t+1$, $q=\binom{n}{t}$ and $\sigma_{i}=\mathbb{P}/T_{i}$ for $i=1,\ldots,q$. Let $\{T_{1},\ldots,T_{\binom{n}{t}}\}$ be an ordered list of all possible combinations of $t$ parties. Since an honest $D$ distributes $m$ according to the previous settings, every combination of $t$ parties is missing exactly one distinct share. Thus, \textit{privacy} follows. Let us now assume that, for some $i$, two honest parties in $\sigma_{i}$ receive different copies of $m_{i}$ ---note that such parties are guaranteed to exist for all $i$, because $|\sigma_{i}|=n-t=2t+1\geq{}t+2$ for all $t>0$---. Then, Share certainly aborts. Conversely, upon non-abortion of Share, all honest parties in each $\sigma_{i}$ hold identical copies of $m_{i}$ (possibly, a default zero string). What is more, $|\sigma_{i}|=2t+1$ implies that every $\sigma_{i}$ contains a majority of honest parties, such that \textit{conditional commitment} follows from the use of MV in Reconstruct. \textit{Conditional commitment} implies that, upon non-abortion of Share, $D$ is committed to an input with respect to the honest parties. \textit{Conditional correctness} follows identically as \textit{conditional commitment}, given the fact that an honest $D$ commits to his actual input value $m$. This completes the proof.
	
	Note that, in the AC model, $n>3t$ is necessary to assure \textit{conditional commitment} by enforcing the success of MV during Reconstruct. In fact, it is known to be a general necessary condition for secure MPC~\cite{MPC,Ben-Or,Chaum} in the AC model, such that setting $n=3t+1$ is optimal. What is more, within our conditional VSS scheme, any attempt to reduce the total number of shares, $q$, comes at the price of increasing the number of parties, $n$. To see this, let us assume that such improved settings exist, satisfying all three properties of conditional VSS while keeping $q<\binom{n}{t}$ for a given number of parties, $n$. On the one hand, \textit{privacy} implies that every combination of $t$ parties is missing one share at least. On the other hand, by the pigeonhole principle, $q<\binom{n}{t}$ implies that at least two distinct combinations of $t$ parties, say $T_{k}$ and $T_{l}$, have one common missing share, say $m_{s}$, for some $s=1,\ldots,q$. Since $|T_{k}\cup{}T_{l}|\geq{t+1}$, it follows that $|\sigma_{s}|\leq{}n-t-1$, and thus \textit{conditional commitment} requires $n\geq{}3t+2$ at least, in order for MV to certainly succeed when applied to all copies of $m_{s}$.
	
	\item{\textbf{AN corruption ($t>1$).}} The considered settings are $n=2t+2$, $q=n$ and $\sigma_{i}=\mathbb{P}/P_{i}$ for $i=1,\ldots,q$. Since an honest $D$ distributes $m$ according to the previous settings, every party is missing exactly one distinct share. This suffices to establish \textit{privacy} in a non-collaborative setting. Let us now assume that, for some $i$, two honest parties in $\sigma_{i}$ receive different copies of $m_{i}$ ---note that such parties are guaranteed to exist for all $i$, because $|\sigma_{i}|=n-1=2t+1>t+2$ for all $t>1$---. Then, Share certainly aborts. Conversely, upon non-abortion of Share, all honest parties in each $\sigma_{i}$ hold identical copies of $m_{i}$ (possibly, a default zero string), and since every $\sigma_{i}$ contains a majority of honest parties, \textit{conditional commitment} follows from the use of MV in Reconstruct. \textit{Conditional correctness} follows identically as in the AC model.
	
	The optimality of the setting $n=2t+2$ for the pair of protocols (Share, Reconstruct) in the AN model follows from the next lemma.
	
	\textbf{Lemma.} \textit{If, for some $i=1,\ldots,q$, $|\sigma_{i}|<2t+1$, the pair of protocols (Share, Reconstruct) does not provide a conditional VSS scheme in the AN model.}
	
	For the AC model, such an assertion is straightforward. However, at a first glance, it seems reasonable that non-collaboration of the corrupted parties may allow to overcome the restriction that each share is held by an honest majority of parties. This is so because, for any given share $m_{i}$, the values declared by any two corrupted parties in $\sigma_{i}$ that misbehave during Reconstruct are not expected to coincide, except with the minuscule probability of a random match. Nevertheless, Lemma states that this is not the case, and we prove it in what follows. For this purpose, let us consider that $D$ is corrupted, and let us assume the worst-case scenario where, for some $i$, $\sigma_{i}$ contains all $t$ corrupted parties. With a non-negligible probability of success, $D$ could, for instance, select two distinct versions of the share $m_{i}$, say $m_{i}^{\rm h}$ and $m_{i}^{\rm d}$, and deliver $m_{i}^{\rm h}$ ($m_{i}^{\rm d}$) to all honest (dishonest) parties in $\sigma_{i}$. Note that this does not necessarily imply the abortion of Share, as the dishonest parties in $\sigma_{i}$ can simply declare the copy $m_{i}^{\rm h}$ they receive from the honest ones during the consistency test of $m_{i}$. If, in addition, $|\sigma_{i}|<2t+1$, $\sigma_{i}$ does not contain a majority of honest parties and thus \textit{conditional commitment} is compromised, because one cannot assure the consistency of the copies of $m_{i}$ reached by all honest parties via MV. This completes the proof.
	
	From the lemma, $|\sigma_{i}|\geq{}2t+1$ is necessary for (Share, Reconstruct) to define a conditional VSS scheme in the AN model. Since, in addition, $\sigma_{i}\subsetneq{\mathbb{P}}$, the requirement $n\geq{}2t+2$ follows, which means that our setting $n=2t+2$ is optimal. Indeed, our setting $\sigma_{i}=\mathbb{P}/P_{i}$ is such that $|\sigma_{i}|=n-1=2t+1$, which is optimal too according to the lemma. Lastly, as in the AC model, direct application of the pigeonhole principle implies that any attempt to reduce the total number of shares, $q$, comes at the price of increasing the number of parties, $n$, in order to maintain the defining properties of conditional VSS.
	
	\item{\textbf{PC corruption ($t>0$).}} The considered settings are $n=t+1$, $q=n$ and $\sigma_{i}=P_{i}$ for $i=1,\ldots,q$. Since an honest $D$ distributes $m$ according to the previous settings, every combination of $t=n-1$ parties is missing exactly one distinct share. Thus, \textit{privacy} follows. In addition, \textit{conditional commitment} holds due to passive corruption of the parties and the fact that $|\sigma_{i}|=1$ for all $i$ (which implies that MV trivially succeeds). \textit{Conditional correctness} follows identically as in the previous models.
	
	Note that the optimality of $n=t+1$ in the PC model is obvious in full generality, and not only within our specific protocols Share and Reconstruct. This is so because setting $n=t$ would compromise \textit{privacy} in the presence of collaborative corrupted parties. Also, as in the previous models, any attempt to reduce the total number of shares, $q$, comes at the price of increasing the number of parties (if one aims to preserve conditional VSS).
	Remarkably, as a consequence of considering passive corruption, $|\sigma_{i}|=1$ suffices for all $i=1,\ldots,q$, in which case step 3 of Share vanishes and thus Share never aborts. This being the case, in the PC model, (Share, Reconstruct) with the above settings not only provides a conditional VSS scheme, but also a standard VSS scheme. Moreover, in the absence of step 3 of Share, no consistency test occurs, which means that VSS reduces to secret sharing (SS) by definition.
	%
	\item{\textbf{PN corruption ($t>1$).}} The considered settings are $n=2$, $q=n$ and $\sigma_{i}=P_{i}$ for $i=1,2$. We clarify that $n=2$ for all $t$ means that it suffices to select two parties out of all corrupted parties in order for (Share, Reconstruct) to define a conditional VSS scheme. An honest $D$ splits $m$ into two random shares and delivers each of them to a different party. \textit{Privacy} holds because each party is missing one share and they do not collaborate. \textit{Conditional commitment} follows due to passivity and the fact that $|\sigma_{i}|=1$ for $i=1,2$. \textit{Conditional correctness} follows identically as in the previous models.
	
	The optimality of $n=2$ and $q=2$ is trivial, and it is not restricted to our pair of protocols (Share, Reconstruct).
\end{enumerate}

\textbf{Proposition 2.} Proposition 2 asserts that the RBS generation protocol yields a common random $L$-bits string for all non-actively corrupted parties. The proposition refers to the active corruption models, and since non-mixed corruption is assumed, all non-actively corrupted parties are honest.

The reasoning is identical for both the AC and the AN model. Let the settings be selected as prescribed by Proposition 1 and assume that the RBS generation protocol does not abort. This implies that Share terminated successfully for all $k=1,\ldots,t+1$. In virtue of \textit{conditional commitment}, non-abortion of Share for, say $P_{k}$, means that all honest parties reach a common string $R_{k}$ via Reconstruct. Thus, all of them output a common final string $R=\oplus_{k=1}^{t+1}R_{k}$. What is more, non-abortion implies that $|R_{k}|=L$ bits for all $k$, such that $|R|=L$ too. Then, Proposition 2 follows if we prove the randomness of $R$. On the one hand, since at least one dealer party is honest, say $P_{\rm h}$, for some $\mathrm{h}\in\{1,\ldots,t+1\}$, \textit{conditional correctness} assures that $R_{\rm h}$ is random. What is more, in virtue of \textit{privacy} and disregarding the honest dealer party $P_{\rm h}$ itself, the information obtained by any set of $t$ or less parties prior to Reconstruct is statistically uncorrelated to $R_{\rm h}$. In particular, the string to which every dealer different from $P_{\rm h}$ is committed upon non-abortion of its Share protocol is uncorrelated to $R_{\rm h}$. Therefore, $R=\oplus_{k=1}^{t+1}R_{k}$ is indeed random.\\

Notably, since the parties do not collaborate in the AN model, one could feel tempted to select two dealers instead of $t+1$, as the strings they would generate would be uncorrelated to each other. Nevertheless, their bitwise XOR would not be necessarily random due to the active character of the two dealers.
\section*{Supplementary Note 4: proof of proposition 3}\label{Claim}
In what follows, we give a detailed proof of Proposition 3, which establishes the security of Protocol (see the Methods section in the main text) within all non-mixed corruption models of the QKD modules and the CP units.

For ease of understanding, we shall refer to the AC model for both kinds of devices hereafter, and the security of Protocol follows identically for all the alternative models we consider, as it builds on (1) the defining properties of conditional VSS (established for each corruption model of the CP units in the main text), (2) the extractable secret key length (established for each corruption model of the QKD pairs in the main text), and (3) the redundancy of the classical communications (set to the adequate value for each corruption model of the CP units in the protocol description).

Below, as we did in Supplementary Note 1, we use asterisks to denote the well-defined versions of quantities to which the QKD modules are committed with respect to the honest CP units in the distributed QKD post-processing, in virtue of the properties of conditional VSS and the redundancy of the classical communications. These quantities include the raw keys, the sifted keys, the EC syndromes, the EV and PA hash functions, the corrected keys, the EV tags and the final keys, together with some other quantities which are not divided into shares: the protocol information, the sifting information, the hypothetical lower bounds computed in the PE step, the error pattern and the secret key length.\\

\textbf{Correctness.} We first prove the correctness established in Proposition 3. Precisely, Proposition 3 asserts the $\epsilon_{\rm cor}$-correctness of the output keys upon non-abortion of Protocol (if Protocol aborts, correctness follows trivially). Hence, let us assume Protocol does not abort and refer to \textit{conditional commitment} (\textit{conditional correctness}) simply as \textit{commitment} (\textit{correctness}) for conciseness.

In the first place, the Share protocol in step 1 guarantees the \textit{commitment} of the raw keys. In the second place, given the \textit{commitment} of the raw keys, the \textit{commitment} of the sifted keys follows from the uniqueness of the sifting information, say $\{\textrm{sift}^{*j}\}_{j=1}^{n_{\rm q}}$, used by the honest CP units to sift their shares of the raw keys. And, in particular, the uniqueness of $\{\textrm{sift}^{j*}\}_{j=1}^{n_{\rm q}}$ is trivially enforced by the consistency tests and the redundancy of the communications in step 2. In the third place, given the \textit{commitment} of the sifted keys, the \textit{commitment} of the corrected keys is enforced by the uniqueness of the error pattern $\hat{e}^{*}$ that all honest $\textrm{CP}_{\textrm{A}_{l}}\in{}\sigma_{1}^{\rm A}$ apply on their copies of the first share of the sifted key. In turn, the uniqueness of $\hat{e}^{*}$ follows from the \textit{commitment} of the syndrome strings $sy_{\rm A}$ and $sy_{\rm B}$, ensured by the \textit{commitment} of the sifted keys and the redundancy of the communications. Fourthly, the \textit{commitment} of the EV tags $h_{\rm EV,A}$ ($h_{\rm EV,B}$) follows from the \textit{commitment} of the corrected keys and that of the function $h_{\rm EV}$ (assured by the RBS generation protocol). In particular, due to the redundancy of the classical communications in step 5, all honest $\textrm{CP}_{\textrm{B}_{l'}}\in{}\sigma_{1}^{\rm B}$ reach the well-defined copies $h_{\rm EV,A}^{*}$ and $h_{\rm EV,B}^{*}$, where $h^{*}_{\rm EV,A}=h^{*}_{\rm EV}(\hat{s}_{\rm A}^{*})$ ($h^{*}_{\rm EV,B}=h^{*}_{\rm EV}({s}_{\rm B}^{*})$) is the well-defined EV tag reached by all honest $\textrm{CP}_{\textrm{A}_{l}}$ ($\textrm{CP}_{\textrm{B}_{l'}}$), computed on the well-defined corrected key $\hat{s}_{\rm A}^{*}$ (sifted key $s_{\rm B}^{*}$). Furthermore, from step 5 of Protocol it follows that EV aborts if $h_{\rm EV,A}^{*}\neq{}h_{\rm EV,B}^{*}$. Conversely, no abortion of the EV step guarantees that $h_{\rm EV,A}^{*}=h_{\rm EV,B}^{*}$. The $\hat{\epsilon}_{\rm cor}$-correctness follows from this fact as long as $h_{\rm EV}^{*}$ (well-defined EV function reached by all honest $\textrm{CP}_{\textrm{B}_{l}}$ at the RBS generation protocol) is a random 2-universal hash function with output length $\lceil{\log_{2}(2/\hat{\epsilon}_{\rm cor})}\rceil$ at least~\cite{Renner}. But this is indeed ensured by the \textit{correctness} of conditional VSS within the RBS generation protocol. To finish with, the \textit{commitment} of the final keys, $k_{\rm A}$ and $k_{\rm B}$, follows from the \textit{commitment} of the corrected keys and that of the function $h_{\rm PA}$. In turn, the latter follows from the \textit{commitment} of conditional VSS within the RBS generation protocol, in which all honest $\textrm{CP}_{\textrm{B}_{l'}}\in{\sigma_{1}^{\rm B}}$ select a unique length $l^{*}$ due to the consistency tests in step 1 and the redundancy of the classical communications. Also note that the \textit{commitment} of $h_{\rm PA}$ (plus the redundancy of the communications) guarantees that correctness is not compromised in the final PA step.

Notably, one should not confuse the \textit{correctness} of conditional VSS (see the Methods section in the main text) with the correctness of the output keys of Protocol. In fact, except from the implicit use of \textit{correctness} in the RBS generation protocol, only the \textit{commitment} (but not the \textit{correctness} or the \textit{privacy}) of conditional VSS is required to establish the correctness of the final keys.

Lastly, we remark that an authentication error may allow a corrupted CP unit to impersonate an honest one, thus possibly compromising the correctness. Therefore, one must compose the error probability,  $\hat{\epsilon}_{\rm cor}$, of the EC ---which presumes the successful authentication of all the classical communications--- with the total error probability of the authentication, $\epsilon_{\rm AU}$, pre-selected by Alice and Bob. In this way, the overall correctness parameter is
\begin{equation}\label{correctness parameter}
\epsilon_{\rm cor}=\hat{\epsilon}_{\rm cor}+\epsilon_{\rm AU}.
\end{equation}\\

\textbf{Secrecy.} In what follows, we prove the secrecy of Protocol, established in Proposition 3. Importantly, the reasoning we present below does not assume a specific QKD scheme, but it applies to a wide variety of them.

Let us assume again that Protocol does not abort. Given the redundancy of the communications and the uniqueness of $l^{*}$ (established in the previous section), the RBS generation protocol guarantees that all honest CP units reach a well-defined function $h_{\rm PA}^{*}$ (\textit{conditional commitment}) and that this function is indeed a 2-universal hash function of input length $l^{*}$ selected at random, as required for PA~\cite{Tomamichel1} (\textit{conditional correctness}). Under these circumstances, the $\epsilon_{\rm sec}$-secrecy of $k_{\rm A}^{*}$ and $k_{\rm B}^{*}$ asserted in Proposition 3 follows as long as $l^{*}$ is a valid lower bound on the extractable secret key length. But this is indeed the case if $l^{*}$ is selected as established in Supplementary Note 1 (Supplementary Note 2) for AC, AN or PC (PN) corruption of the QKD modules.
\section*{Supplementary Note 5: authentication cost}\label{Authentication cost}
We consider the authentication scheme presented in~\cite{LFSR} and described in~\cite{Fung}, based on the construction of Toeplitz matrices using a linear feedback seed register (LFSR).

The sender and the receiver must pre-share a key pool of secret bits, and for every classical message $m$ to be authenticated, they draw bits from this key pool to construct a LFSR-based Toeplitz matrix $T$. Let $\gamma_{\rm AU}$ ($|m|$) be a pre-fixed error probability (the length of the message $m$), and let $k=\left\lceil{\log_{2}{\left(2|m|/\gamma_{\rm AU}\right)}}\right\rceil$.
The construction of $T$ uses $2k$ secret bits and the size of the resulting matrix is $k\times{|m|}$. The sender multiplies the matrix $T$ by the message $m$ to generate an authentication tag $t=T\times{m}$, of $k$ bits. Then, he encrypts the tag using the one-time-pad, thus consuming another $k$ secret bits from the key pool. Nevertheless, the encryption of the tag guarantees that the first $2k$ bits used for the construction of $T$ remain secure and can be reallocated in the key pool, in such a way that the net secret key cost of the authentication is $k$ bits. Finally, the sender transmits both $m$ and its encrypted tag through the public channel. The receiver calculates its own tag using $T$ and the received message, and authentication succeeds if this tag matches the one sent by the sender after decrypting it.\\ 

Coming next, we quantify the amount of secret bits that Protocol consumes for the authentication of the classical communications between labs. Following the protocol description, this task requires every $\textrm{CP}_{\textrm{A}_{l}}\in{}\sigma_{1}^{\rm A}$ to pre-share a so-called \textit{key pool} of secret bits with every $\textrm{CP}_{\textrm{B}_{l'}}\in{}\sigma_{1}^{\rm B}$. Assuming a pre-fixed size $|k|$ for all the key pools, the overall authentication cost reads
\begin{equation}\label{authentication cost}
l_{\rm AU}=R^{2}\times|k|
\end{equation}
bits, where we recall that $R=|\sigma_{1}^{\rm A}|=|\sigma_{1}^{\rm B}|$. Indeed, $R$ is the common size of all $\sigma_{i}^{\rm A}$ ($\sigma_{i}^{\rm B}$), given by $R=2t_{\rm c}+1$ ($R=1$) for the active (passive) corruption models of the CP units.\\

Let us now estimate $|k|$ for the authentication scheme based on Toeplitz matrices presented above. According to Protocol, the sifting step requires the authentication of $n_{\rm q}$ messages from Alice to Bob, say $\{m_{\rm A}^{j}\}_{j=1}^{n_{\rm q}}$ with $m_{\rm A}^{j}=\textrm{info}_{\textrm{A}}^{j}$, and the information reconciliation step requires the authentication of a single message from Bob to Alice, say $m_{\rm B}$ (consisting of various items). Therefore, although $\{\bigl|m_{\rm A}^{j}\bigr|\}_{j=1}^{n_{\rm q}}$ and $|m_{\rm B}|$ are not known a priori, it is required that
\begin{equation}\label{key pool}
|k|\geq{}\sum_{j=1}^{n_{\rm q}}\left\lceil{\log_{2}{\left(\frac{2\bigl|m_{\rm A}^{j}\bigr|}{\gamma_{\rm AU}}\right)}}\right\rceil+\left\lceil{\log_{2}{\left(\frac{2|m_{\rm B}|}{\gamma_{\rm AU}}\right)}}\right\rceil
\end{equation}
secret bits, setting a common error probability, $\gamma_{\rm AU}$, for every communication. In fact, since only an authentication error between two honest units may compromise the security, we find that $\gamma_{\rm AU}$ and $\epsilon_{\rm AU}$ must be related as
\begin{equation}\label{gamma_AU_active}
\epsilon_{\rm AU}\geq{}(t_{\rm c}+1)^{2}(n_{\rm q}+1)\gamma_{\rm AU}
\end{equation}
for active corruption of the CP units, and as
\begin{equation}\label{gamma_AU_passive}
\epsilon_{\rm AU}\geq{}(n_{\rm q}+1)\gamma_{\rm AU}
\end{equation}
for passive corruption. Thus, in order to fulfill a pre-selected total authentication error $\epsilon_{\rm AU}$, one can set $\gamma_{\rm AU}$ to the largest value compatible with Eq.~(\ref{gamma_AU_active}) (Eq.~(\ref{gamma_AU_passive})) for active (passive) corruption of the CP units.
\section*{Supplementary Note 6: decoy-state MDI-QKD}\label{MDI-QKD}
Here, we combine Protocol (in the main text) with the efficient MDI-QKD scheme proposed in~\cite{X.B.Wang}. In this scheme, Alice and Bob use a single intensity for the basis Z, devoted to key extraction, and perform PE with the basis X alone, for which they use three different intensities. Asterisks are omitted for readability throughout this note.\\

\textbf{QKD protocol.} The description below assumes that the possibly corrupted devices of any kind do not misbehave from the protocol description. Nevertheless, the scheme is secure against active eavesdroppers (see Proposition 3 in the main text).\\

For $j=1,\ldots,n_{\rm q}$, $\textrm{QKD}_{\textrm{A}_{j}}$ and $\textrm{QKD}_{\textrm{B}_{j}}$ create the pairs of strings $(r_{\rm A}^{j},a^{j})$ and $(r_{\rm B}^{j},b^{j})$, respectively. While $r_{\rm A}^{j}$ and $r_{\rm B}^{j}\in\left\{0,1\right\}^{N}$ are fully random polarization bit strings, $a^{j}$ and $b^{j}\in\left\{\lambda,\mu,\nu,\omega\right\}^{N}$ are strings of intensities that verify $P\bigl[a_{i}^{j}=\lambda\bigr]=P\bigl[b_{i}^{j}=\lambda\bigr]=q_{\rm Z}$ and $P\bigl[a_{i}^{j}=a\bigr]=P\bigl[b_{i}^{j}=a\bigr]=q_{\rm X}p_{a}$, for $a\in{\rm A}=\{\mu,\nu,\omega\}$ and $i=1,\ldots,N$. On each side, the intensity $\lambda$ determines the use of the basis Z, and the basis X is used otherwise.\\

Let us now focus on, say, the $j$-th QKD pair. For $i$ ranging from $1$ to $N$, steps (i) to (iii) are repeated.
\begin{enumerate}
	\item[(i)]\textit{State preparation.} $\textrm{QKD}_{\textrm{A}_{j}}$ ($\textrm{QKD}_{\textrm{B}_{j}}$) prepares a phase-randomized weak coherent pulse (PR-WCP) with intensity $a_{i}^{j}$ ($b_{i}^{j}$) in the BB84 state defined by $a_{i}^{j}$ and $r_{{\rm A}_{i}}^{j}$ ($b_{i}^{j}$ and $r_{{\rm B}_{i}}^{j}$).
	\item[(ii)]\textit{Transmission.} $\textrm{QKD}_{\textrm{A}_{j}}$ and $\textrm{QKD}_{\textrm{B}_{j}}$ send the states to Charles via the quantum channel. 
	\item[(iii)]\textit{Measurement.} If Charles is honest, he measures the received signals with a Bell state measurement (BSM).
\end{enumerate}
After the above quantum communication phase, the distributed QKD post-processing starts. Again, we focus on a single QKD pair (the $j$-th one).
\begin{enumerate}
	\item[1.]\textit{Distribution of data.}
	Charles sends a $N$-trit string $c^{j}$ to both modules. If he is honest, this is the string of successes, such that $c_{i}^{j}=1$ ($c_{i}^{j}=2$) if a successful BSM associated to the Bell state $\ket{\psi^{+}}=1/\sqrt{2}(\ket{01}+\ket{10})$ ($\ket{\psi^{-}}=1/\sqrt{2}(\ket{01}-\ket{10})$) occurred at the $i$-th round, and $c_{i}^{j}=0$ otherwise. Let $a^{j}|_{c^{j}}$ ($b^{j}|_{c^{j}}$) be the restriction of the intensities string $a^{j}$ ($b^{j}$) to the non-zero entries of $c^{j}$. Also, let $r_{\rm A}^{j}|_{c^{j},\rm X}$ ($r_{\rm B}^{j}|_{c^{j},\rm X}$) and $r_{\rm A}^{j}|_{c^{j},\rm Z}$ (${r}_{\rm B}^{j}|_{c^{j},\rm Z}$) be the restrictions of $r_{\rm A}^{j}$ ($r_{\rm B}^{j}$) to the non-zero entries of $c^{j}$ where Alice (Bob) uses basis X and basis Z, respectively. 
	
	$\textrm{QKD}_{\textrm{A}_{j}}$ ($\textrm{QKD}_{\textrm{B}_{j}}$) uses the Share protocol of a conditional VSS scheme to distribute shares of $r_{\rm A}^{j}|_{c^{j},\rm Z}$ ($r_{\rm B}^{j}|_{c^{j},\rm Z}$) among all $\textrm{CP}_{\textrm{A}_{l}}$ ($\textrm{CP}_{\textrm{B}_{l}}$), where we recall that the details of the Share protocol depend on the corruption model of the CP units as specified in Proposition 1 in the main text. In particular, let $\sigma_{i}^{\rm A}$ ($\sigma_{i}^{\rm B}$) denote the set of units that receives the $i$-th share of $r_{\rm A}^{j}|_{c^{j},\rm Z}$ ($r_{\rm B}^{j}|_{c^{j},\rm Z}$), which without loss of generality is common for all $j=1,\ldots,n_{\rm q}$. $\textrm{QKD}_{\textrm{A}_{j}}$ ($\textrm{QKD}_{\textrm{B}_{j}}$) communicates $a^{j}|_{c^{j}}$ ($c^{j}$, $b^{j}|_{c^{j}}$) and $r_{\rm A}^{j}|_{c^{j},\rm X}$ ($r_{\rm B}^{j}|_{c^{j},\rm X}$) to every $\textrm{CP}_{\textrm{A}_{l}}\in\sigma_{1}^{\rm A}$ ($\textrm{CP}_{\textrm{B}_{l'}}\in\sigma_{1}^{\rm B}$). Then, all the $\textrm{CP}_{\textrm{A}_{l}}\in\sigma_{1}^{\rm A}$ ($\textrm{CP}_{\textrm{B}_{l'}}\in\sigma_{1}^{\rm B}$) perform a consistency test on $a^{j}|_{c^{j}}$ ($c^{j}$, $b^{j}|_{c^{j}}$) and $r_{\rm A}^{j}|_{c^{j},\rm X}$ ($r_{\rm B}^{j}|_{c^{j},\rm X}$).
	\item[2.]\textit{Sifting.}
	Every $\textrm{CP}_{\textrm{A}_{l}}\in{}\sigma_{1}^{\rm A}$ sends $a^{j}|_{c^{j}}$ and $r_{\rm A}^{j}|_{c^{j},\rm X}$ to every $\textrm{CP}_{\textrm{B}_{l'}}\in{}\sigma_{1}^{\rm B}$ and each of the latter applies MV. Using $a^{j}|_{c^{j}}$ and $b^{j}|_{c^{j}}$, each $\textrm{CP}_{\textrm{B}_{l'}}\in{}\sigma_{1}^{\rm B}$ unit builds the index sets
	\begin{eqnarray}
	&&\mathcal{Z}_{j}=\left\{i|c_{i}^{j}\neq{0},a_{i}^{j}=b_{i}^{j}=\lambda\right\}\hspace{.2cm}\textrm{and} \nonumber \\
	&&\mathcal{X}_{j}^{a,b}=\left\{i|c_{i}^{j}\neq{0},a_{i}^{j}=a,b_{i}^{j}=b\right\}
	\end{eqnarray}
	for all $a,b\in{\rm A}$, and checks if the sifting condition $\bigl|\mathcal{Z}_{j}\bigr|\geq{}M$ is met for a pre-established threshold value $M$. If it is not met, the $\textrm{CP}_{\textrm{B}_{l'}}\in{}\sigma_{1}^{\rm B}$ abort the protocol. In case of not aborting, the $\textrm{CP}_{\textrm{B}_{l'}}\in{}\sigma_{1}^{\rm B}$ forward the set $\mathcal{Z}_{j}$ to the rest of Bob's units, which apply MV. All together, the $\textrm{CP}_{\textrm{B}_{l'}}$ perform a RBS generation protocol to select a random subset $\mathcal{Z'}_{j}\subseteq{\mathcal{Z}_{j}}$, of size $M$. Then, Bob's units proceed to the sifting. Precisely, every $\textrm{CP}_{\textrm{B}_{l'}}$ builds its shares of the sifted key $s_{\rm B}^{j}=r_{\rm B}^{j}|_{\mathcal{Z'}_{j}}$ from those of $r_{\rm B}^{j}|_{c^{j},\rm Z}$ (discarding the data external to $\mathcal{Z'}_{j}$).
	\item[3.]\textit{Parameter estimation.}
	For each pair $a,b\in{\rm A}$, every $\textrm{CP}_{\textrm{B}_{l'}}\in{}\sigma_{1}^{\rm B}$ builds the PE strings $r_{\rm B}^{j}|_{\mathcal{X}_{j}^{a,b}}$ and $r_{\rm A}^{j}|_{\mathcal{X}_{j}^{a,b}}$ from the respective strings $r_{\rm A}^{j}|_{c^{j},\rm X}$ and $r_{\rm B}^{j}|_{c^{j},\rm X}$, discarding the data external to $\mathcal{X}_{j}^{a,b}$. Also, every $\textrm{CP}_{\textrm{B}_{l'}}\in{}\sigma_{1}^{\rm B}$ performs the required bit flips on the strings $r_{\mathrm{B}}^{j}|_{\mathcal{X}_{j}^{a,b}}$ depending on the list of successes, $c^{j}$, declared by $\textrm{QKD}_{\textrm{B}_{j}}$~(see \cite{Curty1,Curty2}). In this way, $r_{\rm A}^{j}|_{\mathcal{X}_{j}^{a,b}}$ and $r_{\rm B}^{j}|_{\mathcal{X}_{j}^{a,b}}$ are properly correlated for all $a,b$. Then, each of them computes the numbers of bit errors
	\begin{equation}\label{error_numbers_MDI-QKD}
	e_{a,b}^{j}=\sum_{k=1}^{\bigl|\mathcal{X}_{j}^{a,b}\bigr|}r_{{\rm A}_{k}}^{j}\bigr|_{\mathcal{X}_{j}^{a,b}}\oplus{r_{{\rm B}_{k}}^{j}\bigr|_{\mathcal{X}_{j}^{a,b}}},
	\end{equation}
	where $r_{{\rm A}_{k}}^{j}\bigr|_{\mathcal{X}_{j}^{a,b}}$ ($r_{{\rm B}_{k}}^{j}\bigr|_{\mathcal{X}_{j}^{a,b}}$) denotes the $k$-th bit of the corresponding string. Using $\left|\mathcal{Z}_{j}\right|$ and the different $\bigl|\mathcal{X}_{j}^{a,b}\bigr|$ and $e_{a,b}^{j}$, every $\textrm{CP}_{\textrm{B}_{l'}}\in{}\sigma_{1}^{\rm B}$ computes a lower bound on the number $n_{11,\rm Z}^{j}$ of single-photon successes in $\mathcal{Z'}_{j}$ and an upper bound on the single-photon phase-error rate $\phi_{11,\rm Z}^{j}$ associated to the single-photon successes in $\mathcal{Z'}_{j}$.
\end{enumerate}
The above steps 1 to 3 are performed for all $j=1,\ldots,n_{\rm q}$. At this stage, every $\textrm{CP}_{\textrm{B}_{l'}}\in{}\sigma_{1}^{\rm B}$ derives a lower bound $l$ (given in the next section) on the secret key length that can be extracted from the concatenated sifted key $s_{\rm B}=s_{\rm B}^{1}\ldots{}s_{\rm B}^{n_{\rm q}}$ via PA. If a $\textrm{CP}_{\textrm{B}_{l'}}\in{}\sigma_{1}^{\rm B}$ finds $l\leq{0}$, it aborts the protocol. Importantly, all $\textrm{CP}_{\textrm{B}_{l'}}\in{}\sigma_{1}^{\rm B}$ hold copies of the first share of $s_{\rm B}$, so each of them performs the relevant bit flips (see~\cite{Curty1,Curty2}) on its copy of this share to correctly correlate $s_{\rm A}$ (defined below) and $s_{\rm B}$.
\begin{enumerate}
	\item[4.]\textit{RBS generation.}
	If the protocol does not abort, every $\textrm{CP}_{\textrm{B}_{l'}}\in{}\sigma_{1}^{\rm B}$ forwards $l$ to the rest of Bob's units, which apply MV. All $\textrm{CP}_{\textrm{B}_{l'}}$ perform a RBS generation protocol to randomly select two 2-universal hash functions $h_{\rm EV}$ and $h_{\rm PA}$, respectively devoted to error verification (EV) and PA. Following~\cite{Fung}, if Toeplitz matrices are used for this purpose, $2\lceil\log_{2}(2/\hat{\epsilon}_{\rm cor})\rceil$  ($Mn_{\rm q}+l-1$) bits are required to specify $h_{\rm EV}$ ($h_{\rm PA}$).
	\item[5.]\textit{Information reconciliation.} Every $\textrm{CP}_{\textrm{B}_{l'}}$ computes its shares of (1) the concatenated syndromes string $sy_{\rm B}=sy(s_{\rm B}^{1})\ldots{}sy(s_{\rm B}^{n_{\rm q}})$ and (2) the EV tag $h_{\rm EV,B}=h_{\rm EV}(s_{\rm B})$. All together, the $\textrm{CP}_{\textrm{B}_{l'}}$ reconstruct $sy_{\rm B}$ and $h_{\rm EV,B}$ via the Reconstruct protocol of a conditional VSS scheme (see the Methods section in the main text). Each $\textrm{CP}_{\textrm{B}_{l'}}\in{}\sigma_{1}^{\rm B}$ sends the following items to every $\textrm{CP}_{\textrm{A}_{l}}\in{}\sigma_{1}^{\rm A}$:
	\begin{enumerate}
		\item[1.]{The string $s_{\mathcal{Z'}}=s_{\mathcal{Z'}_{1}}\ldots{}s_{\mathcal{Z'}_{n_{\rm q}}}$, where $s_{\mathcal{Z'}_{j}}$ specifies, say, the positions in $r_{\rm A}^{j}|_{c^{j},\rm Z}$ that contribute to $\mathcal{Z'}_{j}$.}
		\item[2.]{The syndrome information $sy(s_{\rm B})$, together with the description of $h_{\rm EV}$ and the EV tag $h_{\rm EV}(s_{\rm B})$.}
		\item[3.]{The description of $h_{\rm PA}$.}
	\end{enumerate}
	Each $\textrm{CP}_{\textrm{A}_{l}}\in{}\sigma_{1}^{\rm A}$ decides on all three items via MV and communicate $s_{\mathcal{Z'}}$, $h_{\rm EV}$ and $h_{\rm PA}$ to the rest of Alice's units, which apply MV too. Then, they proceed as follows. Using $s_{\mathcal{Z'}}$, all $\textrm{CP}_{\textrm{A}_{l}}$ shrink their shares of $r_{\rm A}|_{c,\rm Z}=r_{\rm A}^{1}|_{c^{1},\rm Z}\ldots{}r_{\rm A}^{n_{\rm q}}|_{c^{n_{\rm q}},\rm Z}$ into shares of $s_{\rm A}=s_{\rm A}^{1}\ldots{}s_{\rm A}^{n_{\rm q}}$, where $s_{\rm A}^{j}=r_{\rm A}^{j}|_{\mathcal{Z'}_{j}}$. All the $\textrm{CP}_{\textrm{A}_{l}}$ compute shares of $sy(s_{\rm A})$ from those of $s_{\rm A}$ and then perform the Reconstruct protocol of a conditional VSS scheme to agree on $sy(s_{\rm A})$. Coming next, the $\textrm{CP}_{\textrm{A}_{l}}\in{}\sigma_{1}^{\rm A}$ compute the error pattern $\hat{e}$ from $sy(s_{\rm B})$ and $sy(s_{\rm A})$ and update the first share of $s_{\rm A}$ XOR-ing it with $\hat{e}$ (\textit{i.e.}, key reconciliation is achieved by acting on a single share). We denote the corrected key by $\hat{s}_{\rm A}=s_{\rm A}\oplus{\hat{e}}$. Using $h_{\rm EV}$, all the $\textrm{CP}_{\textrm{A}_{l}}$ compute their shares of $h_{\rm EV}(\hat{s}_{\rm A})$ and reconstruct it via the Reconstruct protocol of a conditional VSS scheme. Then, each $\textrm{CP}_{\textrm{A}_{l}}\in{}\sigma_{1}^{\rm A}$ checks that $h_{\rm EV}(\hat{s}_{\rm A})=h_{\rm EV}(s_{\rm B})$. Otherwise, it aborts the protocol.
	\item[6.]\textit{Privacy amplification.}
	In case of not aborting, all the $\textrm{CP}_{\textrm{A}_{l}}$ compute their shares of Alice's final key $S_{\rm A}=h_{\rm PA}(\hat{s}_{\rm A})$. Similarly, if no abortion is notified, all the $\textrm{CP}_{\textrm{B}_{l'}}$ compute their shares of Bob's final key $S_{\rm B}=h_{\rm PA}(s_{\rm B})$.\\
\end{enumerate}

\textbf{Secret key length formula in the AC, AN and PC corruption models for the QKD modules.} In this section, we particularize the extractable key length (Eq.~(\ref{looser})) for the decoy-state MDI-QKD protocol presented above. As seen in the main text, this formula is tight within the AC, AN and PC corruption models for the QKD modules, and to evaluate it, it suffices to derive the explicit formula of $h^{j*}_{\varepsilon}$. Assuming perfect state preparation, the entropic uncertainty relation~\cite{Tomamichel2} gives
\begin{equation}\label{uncertainty}
H_{\rm min}^{\varepsilon}(s_{\rm B}^{\mathrm{h}*}|E)\geq{}n_{11,\rm Z}^{\mathrm{h},\rm L*}\left[1-h\left(\phi_{11,\rm Z}^{\mathrm{h},\rm U*}\right)\right],
\end{equation}
where $h(\cdot)$ is the binary entropy function, $n_{11,\rm Z}^{\mathrm{h},\rm L*}$ stands for a lower bound on $n_{11,\rm Z}^{\mathrm{h}*}$ and $\phi_{11,\rm Z}^{\mathrm{h},\rm U*}$ stands for an upper bound on $\phi_{11,\rm Z}^{\mathrm{h}*}$, $n_{11,\rm Z}^{j}$ and $\phi_{11,\rm Z}^{j}$ being defined in the QKD protocol description at the beginning of this note. From the definition of the smooth min-entropies, it follows that $\varepsilon$ is upper-bounded by the sum of the error probabilities of the estimates of $n_{11,\rm Z}^{\mathrm{h},\rm L*}$ and $\phi_{11,\rm Z}^{\mathrm{h},\rm U*}$.

Eq.~(\ref{uncertainty}) implies that, for all $j=1,\ldots,n_{\rm q}$, one should define
\begin{equation}\label{hypothetical}
{h^{j*}_{\varepsilon}}=n_{11,\rm Z}^{j,\rm L*}\left[1-h\left(\phi_{11,\rm Z}^{j,\rm U*}\right)\right],
\end{equation}
which indeed determines a lower bound on $H_{\rm min}^{\varepsilon}(s_{\rm B}^{j*}|E)$ if the $j$-th QKD pair delivers faithful protocol information. Putting it all together, the extractable key length of the protocol reads
\begin{equation}\label{MDI-QKD AC}
l^{*}=\left\lfloor{\min_{j}\biggl\{n_{11,\rm Z}^{j,\rm L*}\left[1-h(\phi_{11,\rm Z}^{j,\rm U*})\right]-\bigl|{sy}^{*}(s_{\rm B}^{j*})\bigr|\biggr\}-\log_{2}\left(\frac{1}{\hat{\epsilon}_{\rm cor}\epsilon_{\rm PA}^{2}\delta}\right)}\right\rfloor,
\end{equation}
where we recall that $|{sy}^{*}(s_{\rm B}^{j*})|$ is the size of the $j$-th EC syndrome, $\hat{\epsilon}_{\rm cor}$ is the correctness parameter, $\epsilon_{\rm PA}$ is the error probability of the privacy amplification and $\delta>0$. Also, as shown in Supplementary Note 1, the above key length is $\epsilon_{\rm sec}$-secret for all $\epsilon_{\rm sec}=\hat{\epsilon}_{\rm sec}+\epsilon_{\rm AU}$, where $\hat{\epsilon}_{\rm sec}\geq{2\varepsilon+\delta+\epsilon_{\rm PA}}$ and $\epsilon_{\rm AU}$ is the total error probability of the authentication, which is selected by the parties a priori.

Explicit expressions of $n_{11,\rm Z}^{j,\rm L*}$ and $\phi_{11,\rm Z}^{j,\rm U*}$ in terms of the observables of the protocol are given in the next section, together with an upper bound on the smooth-parameter $\varepsilon$.\\

\textbf{Secret key length formula in the PN corruption model for the QKD modules.} In Supplementary Note 2, we derived a tighter secret key length formula valid for the PN corruption model, given by Eq.~(\ref{looser PN}). When particularized in our MDI-QKD scheme, this formula reads
\begin{equation}\label{MDI-QKD PN}
l^{*}=\Biggl\lfloor\min_{v}\sum_{j\neq{v}}^{n_{\rm q}}\biggl\{n_{11,\rm Z}^{j,\rm L*}\left[1-h(\phi_{11,\rm Z}^{j,\rm U*})\right]-\bigl|{sy}\bigl(s_{\rm B}^{j*}\bigr)\bigr|\biggr\}-\log_{2}\left(\frac{1}{\hat{\epsilon}_{\rm cor}\epsilon_{\rm PA}^{2}\delta^{n_{\rm q}-1}}\right)\Biggr\rfloor,
\end{equation}
with $\epsilon_{\rm sec}=\hat{\epsilon}_{\rm sec}+\epsilon_{\rm AU}$ and $\hat{\epsilon}_{\rm sec}\geq{(n_{\rm q}-1)(2\varepsilon+\delta)+\epsilon_{\rm PA}}$. Also, we recall that $\epsilon_{\rm AU}$ is pre-determined by Alice and Bob.\\

\textbf{Parameter estimation.} Here, we compute the bounds $n_{11,\rm Z}^{j,\rm L*}$ ($n_{11,\rm Z}^{j,\rm L}$) and $\phi_{11,\rm Z}^{j,\rm U*}$ ($\phi_{11,\rm Z}^{j,\rm U}$) that enter the secret key length, Eq.~(\ref{MDI-QKD AC}) (Eq.~(\ref{MDI-QKD PN})). Since the analysis below is common for every $j=1,\ldots{},n_{\rm q}$, for simplicity of notation we drop the QKD pair index $j$ and refer to any of the QKD pairs.\\

PE is divided into two steps. In a first step, we use the observables of the protocol to calculate bounds on the number $S_{11,\rm X}$ ($E_{11,\rm X}$) of single-photon successes (errors) in $\mathcal{X}=\cup_{a,b}\mathcal{X}_{a,b}$. For this purpose, we apply the decoy-state bounds presented in~\cite{Curty2}, although a slightly simpler technique is used to estimate the expected sizes of the sets ${\mathcal{X}}^{a,b}$ given their realisations (see Supplementary Note 8). In a second step, since PE is only performed with the basis X data in the protocol (see the protocol description at the beginning of this note), we use basis-indistinguishability arguments for the single-photon contributions and standard results from large deviation theory to compute a lower bound on $n_{11,\rm Z}$ and an upper bound on $\phi_{11,\rm Z}$ given the former bounds on $S_{11,\rm X}$ and $E_{11,\rm X}$.\\

In the first place, let us write down the relevant bounds on $S_{11,\rm X}$ and $E_{11,\rm X}$, respectively denoted by $S_{11,\rm X}^{\rm L}$ and $E_{11,\rm X}^{\rm U}$. Let $\mathrm{A}=\left\{\mu,\nu,\omega\right\}$ be the set of intensities that the parties use when they select the basis X, such that $\mu>\nu>\omega$, and let $p_{\mu}$, $p_{\nu}$ and $p_{\omega}$ be the corresponding probabilities. Also, let us introduce a list $\mathcal{V}=\left\{(v_{i},v'_{i})\right\}_{i=1}^{9}$ of pairs of vectors given by:
\begin{eqnarray}
&&(v_{1},v'_{1})=\left([\mu,\nu,\mu,\nu],[\mu,\omega,\mu,\omega]\right),\hspace{.3cm}(v_{2},v'_{2})=\left([\mu,\nu,\mu,\nu],[\mu,\omega,\nu,\omega]\right),\hspace{.3cm}(v_{3},v'_{3})=\left([\mu,\nu,\mu,\nu],[\nu,\omega,\mu,\omega]\right), \nonumber \\
&&(v_{4},v'_{4})=\left([\mu,\nu,\mu,\nu],[\nu,\omega,\nu,\omega]\right),\hspace{.3cm}(v_{5},v'_{5})=\left([\mu,\nu,\mu,\omega],[\mu,\omega,\nu,\omega]\right),\hspace{.3cm}(v_{6},v'_{6})=\left([\mu,\nu,\mu,\omega],[\nu,\omega,\nu,\omega]\right), \nonumber \\
&&(v_{7},v'_{7})=\left([\mu,\omega,\mu,\nu],[\nu,\omega,\mu,\omega]\right),\hspace{.3cm}(v_{8},v'_{8})=\left([\mu,\omega,\mu,\nu],[\nu,\omega,\nu,\omega]\right),\hspace{.3cm}(v_{9},v'_{9})=\left([\mu,\omega,\mu,\omega],[\nu,\omega,\nu,\omega]\right). \nonumber \\
\end{eqnarray}
Then, the lower bound $S_{11,\rm X}^{\rm L}$ is given by~\cite{Curty2}
\begin{equation}\label{lower_bound}
S_{11,\rm X}^{\rm L}=\biggl\lfloor{\max_{(v_{i},v'_{i})\in\mathcal{V}}\left\{\frac{\tau_{11}}{c_{11}}\left(J_{v_{i}v'_{i}}-\Gamma_{v_{i}v'_{i}}\right)\right\}\biggr\rfloor}
\end{equation}
except with probability at most $\epsilon_{11,\rm X}=\sum_{a,b}\epsilon_{a,b}$, for a series of error terms $\left\{\epsilon_{a,b}\right\}_{a,b\in{\rm A}}$ specified by the parties, and some specific quantities $\tau_{11}$, $c_{11}$, $J_{vv'}$, and $\Gamma_{vv'}$ that we define in what follows. First,
\begin{equation}\label{tau_11}
\tau_{nm}=\frac{1}{n!m!}\sum_{a,b\in{\rm A}}e^{-(a+b)}a^{n}b^{m}\hspace{.05cm}p_{a,b,\rm X},
\end{equation}
where $p_{a,b,\rm X}$ stands for the probability of a basis X coincidence with intensity settings $a\in{}A$ for Alice and $b\in{}A$ for Bob. That is, $p_{a,b,\rm X}=p_{a}p_{b}q_{\rm X}^{2}$. Regarding $c_{11}$, $J_{vv'}$, and $\Gamma_{vv'}$, we distinguish two cases depending on the sign of $(a_{0}+a_{1})/(a'_{0}+a'_{1})-(b_{0}+b_{1})/(b'_{0}+b'_{1})$, where for convenience we use the generic notation $v=[a_{0},a_{1},b_{0},b_{1}]$ and $v'=[a'_{0},a'_{1},b'_{0},b'_{1}]$ for the pairs of vectors in the list $\mathcal{V}$.\\

\textit{Case 1: $(a_{0}+a_{1})/(a'_{0}+a'_{1})>(b_{0}+b_{1})/(b'_{0}+b'_{1})$.}\\

In this case, the definitions are
\begin{equation}\label{c_11} c_{nm}=(b_{0}^2-b_{1}^{2})(a_{0}-a_{1})({a'_{0}}^{n}-{a'_{1}}^{n})({b'_{0}}^{m}-{b'_{1}}^{m})-({b'_{0}}^2-{b'_{1}}^{2})({a'_{0}}-{a'_{1}})(a_{0}^{n}-a_{1}^{n})(b_{0}^{m}-b_{1}^{m}),
\end{equation}
\begin{equation}\label{J_vvp}
J_{vv'}=(b_{0}^{2}-b_{1}^2)(a_{0}-a_{1})G_{v'}-({b'_{0}}^{2}-{b'_{1}}^2)({a'_{0}}-{a'_{1}})G_{v},
\end{equation}
with
\begin{equation}\label{G_v}
G_{v}=|\tilde{\mathcal{X}}^{a_{0},b_{0}}|+|\tilde{\mathcal{X}}^{a_{1},b_{1}}|-|\tilde{\mathcal{X}}^{a_{0},b_{1}}|-|\tilde{\mathcal{X}}^{a_{1},b_{0}}|,
\end{equation},
\begin{equation}\label{G_vp}
G_{v'}=|\tilde{\mathcal{X}}^{a'_{0},b'_{0}}|+|\tilde{\mathcal{X}}^{a'_{1},b'_{1}}|-|\tilde{\mathcal{X}}^{a'_{0},b'_{1}}|-|\tilde{\mathcal{X}}^{a'_{1},b'_{0}}|
\end{equation}
and $|\tilde{\mathcal{X}}^{a,b}|=e^{a+b}|{\mathcal{X}}^{a,b}|/p_{a,b,\rm X}$.\\

Lastly,
\begin{equation}\label{Gamma_vvp} \Gamma_{vv'}=(b_{0}^{2}-b_{1}^{2})(a_{0}-a_{1})(\hat{\Gamma}_{a'_{0},b'_{0}}+\hat{\Gamma}_{a'_{1},b'_{1}}+\hat{\Gamma}_{a'_{0},b'_{1}}+\hat{\Gamma}_{a'_{1},b'_{0}})+({b'_{0}}^{2}-{b'_{1}}^{2})(a'_{0}-a'_{1})({\Gamma}_{a_{0},b_{0}}+{\Gamma}_{a_{1},b_{1}}+{\Gamma}_{a_{0},b_{1}}+{\Gamma}_{a_{1},b_{0}}),
\end{equation}
where $\hat{\Gamma}_{a,b}=e^{a+b}\hat{\Delta}(|{\mathcal{X}}^{a,b}|,\epsilon_{a,b})/p_{a,b,\rm X}$ and ${\Gamma}_{a,b}=e^{a+b}{\Delta}(|{\mathcal{X}}^{a,b}|,\epsilon_{a,b})/p_{a,b,\rm X}$. The functions $\hat{\Delta}(x,y)$ and ${\Delta}(x,y)$ are defined in Supplementary Note 8. There, we explain the technique we use to relate the observed set sizes, $|{\mathcal{X}}^{a,b}|$, with their expected values, in order to set statistical bounds on the latter.\\

\textit{Case 2: $(a_{0}+a_{1})/(a'_{0}+a'_{1})\leq{}(b_{0}+b_{1})/(b'_{0}+b'_{1})$}.\\

For this case,
\begin{equation}\label{c_11_2}
c_{nm}=(a_{0}-a_{1})(b_{0}-b_{1})({a'_{0}}-{a'_{1}})({b'_{0}}-{b'_{1}})(a_{0}+a_{1}-{a'_{0}}-{a'_{1}}),
\end{equation}
\begin{equation}\label{J_vvp_2}
J_{vv'}=(a_{0}^{2}-a_{1}^2)(b_{0}-b_{1})G_{v'}-({a'_{0}}^{2}-{a'_{1}}^2)({b'_{0}}-{b'_{1}})G_{v},
\end{equation}
and
\begin{equation}\label{Gamma_vvp_2} \Gamma_{vv'}=(a_{0}^{2}-a_{1}^{2})(b_{0}-b_{1})(\hat{\Gamma}_{a'_{0},b'_{0}}+\hat{\Gamma}_{a'_{1},b'_{1}}+\hat{\Gamma}_{a'_{0},b'_{1}}+\hat{\Gamma}_{a'_{1},b'_{0}})+({a'_{0}}^{2}-{a'_{1}}^{2})(b'_{0}-b'_{1})({\Gamma}_{a_{0},b_{0}}+{\Gamma}_{a_{1},b_{1}}+{\Gamma}_{a_{0},b_{1}}+{\Gamma}_{a_{1},b_{0}}),
\end{equation}
where the definitions of $G_{v}$, $G_{v'}$, $\hat{\Gamma}_{a,b}$ and ${\Gamma}_{a,b}$ are the same as in Case 1.\\

Coming next, we compute an upper bound $E_{11,\rm X}^{\rm U}$ on $E_{11,\rm X}$. For this, let us introduce the list of vectors $\mathcal{W}=\left\{[a_{0},a_{1},b_{0},b_{1}]\hspace{.1cm}|\hspace{.1cm}a_{0}>a_{1},b_{0}>b_{1},\hspace{.1cm}a_{0},a_{1},b_{0},b_{1}\in{\rm A}\right\}$. Then, the upper bound $E_{11,\rm X}^{\rm U}$ is given by~\cite{Curty2}
\begin{equation}\label{upper_bound}
E_{11,\rm X}^{\rm U}=\biggl\lceil{\max_{v\in\mathcal{W}}\left\{\frac{\tau_{11}(F_{v}-\Gamma_{v})}{(a_{0}-a_{1})(b_{0}-b_{1})}\right\}\biggr\rceil},
\end{equation}
except with probability at most ${\epsilon'}_{11,\rm X}=\sum_{a,b}\epsilon'_{a,b}$, for a series of error terms $\left\{\epsilon'_{a,b}\right\}_{a,b\in{\rm A}}$ specified by the parties and some specific quantities $F_{v}$, and $\Gamma_{v}$ that we define in what follows:
\begin{equation}\label{F_v}
F_{v}=\tilde{e}^{a_{0},b_{0}}+\tilde{e}^{a_{1},b_{1}}-\tilde{e}^{a_{0},b_{1}}-\tilde{e}^{a_{1},b_{0}}
\end{equation}
with $\tilde{e}^{a,b}=e^{a+b}e_{a,b}/p_{a,b,\rm X}$, and
\begin{equation}\label{Gamma_v} \Gamma_{v}=-{\Gamma'}_{a_{0},b_{0}}-{\Gamma'}_{a_{1},b_{1}}-{\hat{\Gamma}'}_{a_{0},b_{1}}-{\hat{\Gamma}'}_{a_{1},b_{0}}
\end{equation}
with ${\Gamma'}_{a,b}=e^{a+b}\Delta(e_{a,b},\epsilon'_{a,b})/p_{a,b,\rm X}$ and ${\hat{\Gamma}'}_{a,b}=e^{a+b}\hat{\Delta}(e_{a,b},\epsilon'_{a,b})/p_{a,b,\rm X}$. Also, we remind the reader that, for every $a,b\in{\rm A}$, $e_{a,b}$ is the observed number of bit errors in the set $\mathcal{X}^{a,b}$.\\

In what follows, given $S_{11,\rm X}^{\rm L}$ and $E_{11,\rm X}^{\rm U}$, we derive bounds on $n_{11,\rm Z}$ and $\phi_{11,\rm Z}$ (the quantities that enter the secret key length) via random sampling arguments. Let $N_{11,\rm Z}$ ($N_{11,\rm X}$) be the number of rounds where both Alice and Bob sent single photons and used the basis Z (X). Of course, $N_{11}=N_{11,\rm Z}+N_{11,\rm X}$ is the overall number of rounds where a basis match occurred and both parties sent single photons. In the absence of state preparation flaws, the quantum states sent by Alice and Bob that contain single photons on both sides are basis independent, meaning that Eve cannot distinguish in which basis they are prepared. As a consequence, the probability that Charles declares a successful BSM cannot depend on the basis choice. Thus, given the number $S_{11,\rm X}$ of rounds where both parties sent single photons in the basis X and Charles declared a successful BSM, one can estimate the corresponding number for the basis Z, $S_{11,\rm Z}$, via Serfling's inequality~\cite{Serfling}. Of course, this requires the knowledge of $N_{11,\rm Z}$  and $N_{11,\rm X}$ as well. Precisely,
\begin{equation}\label{Serfling}
P\left(S_{11,\rm Z}\leq{N_{11,\rm Z}\left(\frac{S_{11,\rm X}}{N_{11,\rm X}}\right)-\left(N_{11,\rm Z}+N_{11,\rm X}\right)\times\Upsilon\left(N_{11,\rm Z},N_{11,\rm X},\varepsilon\right)}\right)\leq{\varepsilon}
\end{equation}
holds for any $0<\varepsilon<{1}$ if we choose the deviation term $\Upsilon(N_{11,\rm Z},N_{11,\rm X},\varepsilon)$ to be defined by the function
\begin{equation}\label{upsilon}
\Upsilon(x,y,z)=\sqrt{(x+1)\ln(z^{-1})/(2y(x+y))}.
\end{equation}
For simplicity, we shall set a common error probability, $\varepsilon=\epsilon_{\rm S}$, for each usage of Serfling's inequality in this section.\\

Note that, as the quantities $N_{11,\rm Z}$, $N_{11,\rm X}$, and $S_{11,\rm X}$ are not known, one should derive statistical bounds on them and assume the worst-case scenario, \textit{i.e.}, the one that minimises the value of $S_{11,\rm Z}$. For the first two quantities one can use the standard Chernoff bound~\cite{Chernoff}, as their expected values are known to be $\mu_{11,\rm Z}=E\left[N_{11,\rm Z}\right]={N}q_{\rm Z}^{2}p_{1|\lambda}^{2}$ and $\mu_{11,\rm X}=E\left[N_{11,\rm X}\right]={N}q_{\rm X}^{2}\left(p_{\mu}p_{1|\mu}+p_{\nu}p_{1|\nu}+p_{\omega}p_{1|\omega}\right)^{2}$, where $p_{n|a}$ stands for the poissonian photon-number distribution with mean value $a$. Importantly, these expected values do not rely on the assumption of a particular channel model, but only on Alice's and Bob's state preparation process. Regarding, for instance, $N_{11,\rm Z}$, we have that $P\left(N_{11,\rm Z}>N_{11,\rm Z}^{\rm U}\right)<{\varepsilon'}$ and $P\left(N_{11,\rm Z}<N_{11,\rm Z}^{\rm L}\right)<{\varepsilon''}$ respectively hold for any $\varepsilon',\varepsilon''\in(0,1)$ if we set
\begin{equation}\label{Chernoff}
N_{11,\rm Z}^{\rm U}=\min\left\{\left\lceil{\mu_{\rm Z}+\Delta_{\rm U}(\mu_{\rm Z},\varepsilon')}\right\rceil,N\right\}\hspace{.5cm}\rm{and}\hspace{.5cm}N_{11,\rm Z}^{\rm L}=\max\left\{\left\lfloor{\mu_{\rm Z}-\Delta_{\rm L}(\mu_{\rm Z},\varepsilon'')}\right\rfloor,0\right\},
\end{equation}
where the deviation functions are given by~\cite{Chernoff}
\begin{equation}\label{Chernoff_intervals}
\Delta_{\rm U}(x,y)=\frac{\ln{y^{-1}}}{2}\left(1+\sqrt{1+\frac{8x}{\ln{y^{-1}}}}\right)\hspace{.5cm}\textrm{and}\hspace{.5cm}\Delta_{\rm L}(x,y)=\sqrt{2x\ln{y^{-1}}}.
\end{equation}
As usual, the superscript ``L" (``U") stands for ``lower" (``upper") bound, and the bounds on $N_{11,\rm X}$ are obtained substituting $\mu_{\rm Z}$ by $\mu_{\rm X}$ in Eq.~(\ref{Chernoff}). For simplicity, we shall set a common error probability, $\epsilon_{\rm C}$, for each usage of the Chernoff bound, as we already did for Serfling's inequality. In particular, we set $\varepsilon'=\varepsilon''=\epsilon_{\rm C}$.\\

Regarding $S_{11,\rm X}$, a lower bound $S_{11,\rm X}^{\rm L}$ was already derived in the first part of this note, and the corresponding error probability is denoted by $\epsilon_{11,\rm X}$. Coming next, we update the claim of Eq.~(\ref{Serfling}) by replacing $N_{11,\rm Z}$, $N_{11,\rm X}$ and $S_{11,\rm X}$ with the appropriate bounds minimising $S_{11,\rm Z}$, and by adding the corresponding error terms on the right-hand side. This yields $P\left(S_{11,\rm Z}\leq{S_{11,\rm Z}^{\rm L}}\right)\leq{\epsilon_{\rm S}+\epsilon_{11,\rm X}+2\epsilon_{\rm C}}$, for
\begin{equation}\label{Serfling_updated}
S_{11,\rm Z}^{\rm L}=\max\left\{\left\lfloor{N_{11,\rm Z}^{\rm L}\left(\frac{S_{11,\rm X}^{\rm L}}{N_{11,\rm X}^{\rm U}}\right)-\left(N_{11,\rm Z}^{\rm L}+N_{11,\rm X}^{\rm U}\right)\times\Upsilon\left(N_{11,\rm Z}^{\rm L},N_{11,\rm X}^{\rm U},\epsilon_{\rm S}\right)}\right\rfloor,0\right\}.
\end{equation}
Finally, using Serfling's inequality~\cite{Serfling} one can easily relate the lower bound on the number $n_{11,\rm Z}$ of single-photon successes in the random sample $\mathcal{Z'}\subset{\mathcal{Z}}$, with the lower bound on the number $S_{11,\rm Z}$ of single-photon successes in the original set $\mathcal{Z}$ (see the protocol description at the beginning of this note). Already incorporating Eq.~(\ref{Serfling_updated}), it follows that $P\left(n_{11,\rm Z}\leq{n_{11,\rm Z}^{\rm L}}\right)\leq{2\epsilon_{\rm S}+\epsilon_{11,\rm X}+2\epsilon_{\rm C}}$ for
\begin{equation}\label{Serfling_2}
n_{11,\rm Z}^{\rm L}=\max\left\{\left\lfloor{M\left(\frac{S_{11,\rm Z}^{\rm L}}{\left|\mathcal{Z}\right|}-\Lambda\left(\left|\mathcal{Z}\right|,M,\epsilon_{\rm S}\right)\right)}\right\rfloor,0\right\},
\end{equation}
where $\Lambda(x,y,z)=\sqrt{(x-y+1)\ln(z^{-1})/(2xy)}$ and $M$ is again the size of $\mathcal{Z'}$, which defines the post-processing block size (\textit{i.e.}, the size of the sifted keys).\\

In the derivation above, we used a basis indistinguishability argument to relate the ratio $S_{11,\rm Z}/N_{11,\rm Z}$ to the ratio $S_{11,\rm X}/N_{11,\rm X}$ via Serfling's inequality~\cite{Serfling}. The same argument also relates the ratio $e_{11,\rm Z}/n_{11,\rm Z}$ to the ratio $E_{11,\rm X}/S_{11,\rm X}$, where $e_{11,\rm Z}$ ($E_{11,\rm X}$) denotes the number of single-photon phase errors (bit errors) in the rounds indexed by $\mathcal{Z'}$ ($\mathcal{X}=\cup_{a,b}\mathcal{X}^{a,b}$). Precisely,
\begin{equation}\label{Serfling_3}
P\left(e_{11,\rm Z}\geq{n_{11,\rm Z}\left(\frac{E_{11,\rm X}}{S_{11,\rm X}}\right)+\left(S_{11,\rm X}+n_{11,\rm Z}\right)\times\Upsilon\left(n_{11,\rm Z},S_{11,\rm X},\epsilon_{\rm S}\right)}\right)\leq{\epsilon_{\rm S}}
\end{equation}
holds, where the deviation function $\Upsilon(x,y,z)$ is defined in Eq.~(\ref{Serfling}). Again, the quantities $n_{11,\rm Z}$, $S_{11,\rm X}$ and $E_{11,\rm X}$ are not known, in such a way that adequate bounds should be used instead. On the one side, a lower bound on $n_{11,\rm Z}$ was presented in the previous subsection, and the relevant bounds on $S_{11,\rm X}$ and $E_{11,\rm X}$ were derived in the first part of this note. Using these bounds and their respective error probabilities, one can update the claim of Eq.~(\ref{Serfling_3}) as $P\left(e_{11,\rm Z}\geq{e_{11,\rm Z}^{\rm U}}\right)\leq{3\epsilon_{\rm S}+\epsilon_{11,\rm X}+{\epsilon'}_{11,\rm X}+2\epsilon_{\rm C}}$, where
\begin{equation}\label{Serfling_3_updated}
e_{11,\rm Z}^{\rm U}=\min\left\{\left\lceil{{n_{11,\rm Z}^{\rm L}}\left(\frac{E_{11,\rm X}^{\rm U}}{S_{11,\rm X}^{\rm L}}\right)+\left(S_{11,\rm X}^{\rm L}+n_{11,\rm Z}^{\rm L}\right)\times\Upsilon\left(n_{11,\rm Z}^{\rm L},S_{11,\rm X}^{\rm L},\epsilon_{\rm S}\right)}\right\rceil,n_{11,\rm Z}^{\rm L}\right\}.
\end{equation}
To finish with, note that the single-photon phase error rate is, by definition, given by $\phi_{11,\rm Z}=e_{11,\rm Z}/n_{11,\rm Z}$. Thus, it follows that $P\left(\phi_{11,\rm Z}\geq{\phi_{11,\rm Z}^{\rm U}}\right)\leq{3\epsilon_{\rm S}+\epsilon_{11,\rm X}+{\epsilon'}_{11,\rm X}+2\epsilon_{\rm C}}$ for
\begin{equation}\label{upper_bound_phi_1z}
\phi_{11,\rm Z}^{\rm U}=\frac{e_{11,\rm Z}^{\rm U}}{n_{11,\rm Z}^{\rm L}},
\end{equation}
where $n_{11,\rm Z}^{\rm L}$ is given by Eq.~(\ref{Serfling_2}) and $e_{11,\rm Z}^{\rm U}$ is given by Eq.~(\ref{Serfling_3_updated}).

From the above PE procedure, it follows that the smooth parameter $\varepsilon$ (defined in Eq.~(\ref{uncertainty})) is upper-bounded as
\begin{equation}\label{smooth MDI-QKD}
\varepsilon\leq{}3\epsilon_{\rm S}+\epsilon_{11,\rm X}+{\epsilon'}_{11,\rm X}+2\epsilon_{\rm C}.
\end{equation}\\

\textbf{Authentication cost.} Following the authentication scheme presented in Supplementary Note 5, in order to quantify the total authentication cost of the lab-to-lab communications it suffices to specify the lengths of the different messages exchanged during the protocol. This is what we do next. According to the protocol description above in this note, we have
\begin{eqnarray}
\bigl|m_{\rm A}^{j}\bigr|&=&\left|a^{j}|_{c^{j}}\right|+\bigl|r_{\rm A}^{j}|_{c^{j},\rm X}\bigr|, \nonumber \\
|m_{\rm B}|&=&\left|s_{\mathcal{Z'}}\right|+\left|sy(s_{\rm B})\right|+\left|h_{\rm EV}(s_{\rm B})\right|+\left|h_{\rm EV}\ \rm description\right|+\left|h_{\rm PA}\ \rm description\right|,
\end{eqnarray}
where the expected sizes of $a^{j}|_{c^{j}}$ and $r_{\rm A}^{j}|_{c^{j},\rm X}$ for a typical channel model are given at the end of this note, $\left|s_{\mathcal{Z'}}\right|=\sum_{j=1}^{n_{\rm q}}\bigl|r_{\rm A}^{j}|_{c^{j},\rm Z}\bigr|$ (the expected sizes of all $r_{\rm A}^{j}|_{c^{j},\rm Z}$ being given at the end of the note too), the size of the syndrome $\left|sy(s_{\rm B})\right|$ depends on the EC protocol (and a typical model is given in the Results section of the main text), $\left|h_{\rm EV}(s_{\rm B})\right|=\lceil\log_{2}(2/\hat{\epsilon}_{\rm cor})\rceil$ bits, $\left|h_{\rm EV}\ \rm description\right|=2\lceil\log_{2}(2/\hat{\epsilon}_{\rm cor})\rceil$ bits and $\left|h_{\rm PA}\ \rm description\right|=Mn_{\rm q}+l-1$ bits, $l$ denoting the extractable secret key length, given by Eq.~(\ref{MDI-QKD AC}) (Eq.~(\ref{MDI-QKD PN})) in the AC, AN and PC corruption models (PN corruption model) of the QKD modules.\\

\textbf{Calculation of $N$ and $E_{\rm tol}$ for the simulations.} Here, we derive proper values for the number of transmission rounds per QKD pair, $N$, and for the threshold bit error rate of the EC protocol, $E_{\rm tol}$, based on respective restrictions on the abortion probabilities of the sifting step and the error verification step. The analysis relies on a typical channel model presented below in this note.

We calculate $N$ first. For this purpose, let us impose a common abortion probability $\gamma_{\rm sift}/n_{\rm q}$ for each sifting step ($n_{\rm q}$ of them in total). That is, we demand that $P(\left|\mathcal{Z}_{j}\right|<{M})\leq\gamma_{\rm sift}/n_{\rm q}$ for all $j=1,\ldots,n_{\rm q}$, where $\left|\mathcal{Z}_{j}\right|$ is the set of detection events when both parties use basis Z and $M$ is the pre-specified size of the sifted keys (\textit{i.e.}, the block size). Using the Chernoff's inequality~\cite{Chernoff}, this condition is met if we set the number of signals transmitted per module in each QKD pair to $N=\zeta(M,G_{\rm Z,Z}^{\lambda,\lambda},\gamma_{\rm sift}/n_{\rm q})$, where
\begin{equation}\label{transmitted Chernoff}
\zeta(x,y,z)=\left\lceil{\frac{x}{y}+\frac{\ln{(1/z)}}{y}\left[1+\sqrt{1+\frac{2x}{\ln{(1/z)}}}\right]}\right\rceil,
\end{equation}
and $G_{\rm Z,Z}^{\lambda,\lambda}$ is the probability that any given round of the QKD session between QKD$_{\mathrm{A}_{j}}$ and QKD$_{\mathrm{B}_{j}}$ contributes to $\mathcal{Z}_{j}$. An expression of $G_{\rm Z,Z}^{\lambda,\lambda}$ for a typical channel model is given at the end of this note.

Now, let us calculate $E_{\rm tol}$. Following the MDI-QKD protocol given at the beginning of this note, the reconciliation of $s_{\rm A}$ with $s_{\rm B}$ is performed separately on each $s_{\rm A}^{j}$. If, for simplicity, one assumes that the EC protocol corrects up to a fraction $E_{\rm tol}$ of bit errors (and no more) with certainty, either $E_{j}\leq{E_{\rm tol}}$ for all $j$ or the (single) EV step aborts, where $E_{j}$ denotes the actual error rate between $s_{\rm A}^{j}$ and $s_{\rm B}^{j}$. Thus, applying Chernoff's inequality~\cite{Chernoff}, $P\left(\rm{EV\hspace{.05cm}aborts}\right)\leq{\gamma_{\rm EC}}$ holds for any $\gamma_{\rm EC}\in(0,1)$ if
\begin{equation}\label{threshold QBER MDI-QKD}
E_{\rm tol}=\min\left\{1,E_{\rm Z,Z}^{\lambda,\lambda}+\frac{\Delta_{\rm U}(E_{\rm Z,Z}^{\lambda,\lambda}M,\gamma_{\rm EC}/n_{\rm q})}{{M}}\right\}
\end{equation}
where $E_{\rm Z,Z}^{\lambda,\lambda}$ is the expected bit error rate for the basis Z and the common intensity $\lambda$, and the deviation function $\Delta_{\rm U}(x,y)$ is defined in Eq.~(\ref{Chernoff_intervals}). An expression of $E_{\rm Z,Z}^{\lambda,\lambda}$ for a typical channel model is given at the end of this note.\\

\textbf{Channel model.} In this section, we derive expressions for the expected values of the observables of the protocol, considering the setup illustrated in Supplementary Figure~\ref{fig:setup}. To begin with, let us elaborate on the mathematical models we use.
\begin{enumerate}
	\item {Laser sources.} Alice's and Bob's photon sources emit PR-WCP of the form
	\begin{equation}
	\rho_{\tau}=\frac{1}{2\pi}\int_{0}^{2\pi}\ket{\tau}\bra{\tau}\hspace{.1cm}d\gamma,
	\end{equation}
	where $\ket{\tau}=\exp\left(\tau{a}^\dagger-\tau^*{a}\right)\ket{0}$ is a coherent state, with amplitude $\tau=|\tau|e^{i\gamma}\in\mathbb{C}$. Here, ${a}^\dagger$ ($a$) and $\ket{0}$ are the creation (annihilation) operator and the vacuum state for mode $a$, such that a Fock state with $n$ photons in this mode is given by $\ket{n}={{a}^{\dagger{n}}}/n!\ket{0}$.
	\item {Channel and detector loss.} An effective beam-splitter with transmittance $\eta=\eta_{\rm ch}\eta_{\rm det}$ is used to jointly model channel loss ($\eta_{\rm ch}$) and detector loss ($\eta_{\rm det}$) on each side. The transformation reads
	\begin{align}\label{loss}
	&p^{\dagger}\xrightarrow{}\sqrt{\eta}\hspace{.05cm}r^{\dagger}+\sqrt{1-\eta}\hspace{.05cm}s^{\dagger}, \nonumber \\
	&q^{\dagger}\xrightarrow{}\sqrt{\eta}\hspace{.05cm}s^{\dagger}-\sqrt{1-\eta}\hspace{.05cm}r^{\dagger},
	\end{align}
	where the quantum signal enters through the input port $p$, a vacuum state enters through the input port $q$, the output port $r$ leads to unit detection efficiency detectors, and the output port $s$ represents channel and detection loss. In turn, $\eta_{\rm ch}=10^{-\alpha_{\rm att}L/10}$, $\alpha_{\rm att}$ being the attenuation coefficient of the channel (in dB/km), and $L$ being the transmission length between each party and the central node (in km).
	\item {Basis choice and polarization misalignment.} Let $a_{\rm h}^{\dagger}$ ($a_{\rm v}^{\dagger}$) denote the creation operator of a photon with horizontal (vertical) polarization in a pre-fixed basis Z. For each party, the selection of the basis setting $\theta\in\left\{0,\pi/4\right\}$ and the occurrence of a polarization misalignment $\delta_{\rm mis}>0$ jointly transform $a_{\rm h}^{\dagger}$ and $a_{\rm v}^{\dagger}$ according to the following unitary operation:
	\begin{align}\label{misalignment}
	&a_{\rm h}^{\dagger}\xrightarrow{}\cos(\theta+\delta_{\rm mis})\hspace{.05cm}a_{\rm h}^{\dagger}+\sin(\theta+\delta_{\rm mis})\hspace{.05cm}a_{\rm v}^{\dagger}, \nonumber \\
	&a_{\rm v}^{\dagger}\xrightarrow{}\cos(\theta+\delta_{\rm mis})\hspace{.05cm}a_{\rm v}^{\dagger}-\sin(\theta+\delta_{\rm mis})\hspace{.05cm}a_{\rm h}^{\dagger}.
	\end{align}
	In short, for any given $\delta_{\rm mis}$, setting $\theta=0$ ($\theta=\pi/4$) in Eq.~(\ref{misalignment}) jointly models that the party selected basis Z (X) and a polarization misalignment $\delta_{\rm mis}$ occurred in the channel.
	\item {Photo-detectors.} Threshold detectors are considered, meaning that each of them is modeled with a POVM consisting of only two elements: $\left\{E_{\textrm{no click}},E_{\textrm{click}}\right\}$. As the detector loss is already accounted for in the channel model, the POVM here must describe unit efficiency photo-detectors, but having a non-zero dark count probability $p_{\rm d}$. That is,
	\begin{equation}\label{detectors}
	E_{\textrm{no click}}=(1-p_{\rm d})\ket{0}\bra{0},\hspace{.5cm}E_{\textrm{click}}=\unit-E_{\textrm{no click}}.
	\end{equation}
	The operator $\unit$ denotes the identitiy operator in the photon-number basis, \textit{i.e.}, $\unit=\sum_{n=0}^{\infty}\ket{n}\bra{n}$.\\
\end{enumerate}

\textbf{Relevant experimental parameters.} First of all, let us introduce some convenient notation. At every round of the protocol, $\theta_{\rm A}$ $(\theta_{\rm B})\in\{\rm{0,\pi/4}\}$ denotes Alice's (Bob's) basis setting, where, as usual, 0 ($\pi/4$) stands for basis Z (X). Similarly, $i$ and $j$ $\in\{1,2\}$ respectively denote Alice's and Bob's polarization states, such that, for basis Z (X), $1$ means ``h" (``+") and $2$ means ``v" (``-"). Regarding the photo-detectors, they are numbered by $w\in\{1,2,3,4\}$ as shown in Supplementary Figure~\ref{fig:setup}. Also, for each photo-detector $w$, it is convenient to introduce an ``arm index" $s_{w}\in\{1,2\}$ specifying whether it is on the right arm ($s_{w}=1$) or the left arm ($s_{w}=2$) of the detection scheme, and another ``polarization index" $k_{w}\in\{1,2\}$ specifying whether they detect the horizontal ($k_{w}=1$) or the vertical $(k_{w}=2)$ component of the pulses coming from the polarizing beam-splitters (tagged by the symbol ``$\otimes$") in Supplementary Figure~\ref{fig:setup}.\\
\begin{figure}[!htbp]
	\centering 
	\includegraphics[width=8.5cm]{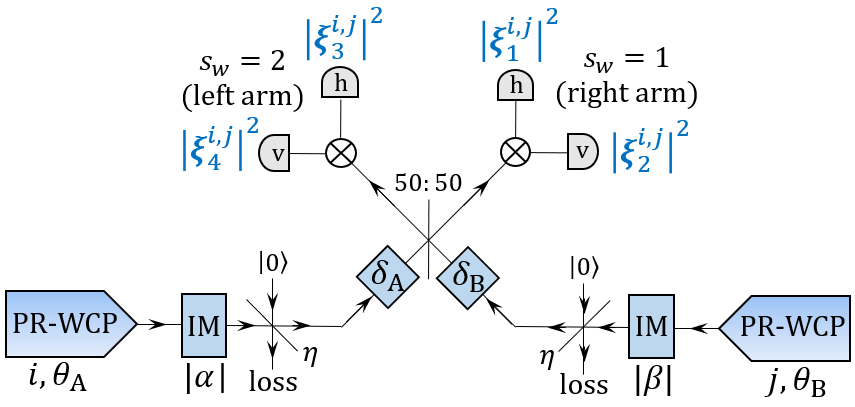} 
	\caption{Supplementary Figure 2. Schematic of the decoy-state MDI-QKD setup. Alice (Bob) holds a laser source that emits PR-WCPs in any of the four BB84 states, defined by a polarization setting $i$ $(j)\in\{1,2\}$ and a basis setting $\theta_{\rm A}$ $(\theta_{\rm B})\in\{\rm{0,\pi/4}\}$. An intensity modulator (IM) selects the amplitude $|\alpha|$ ($|\beta|$) of Alice's (Bob's) laser pulse. The overall one-sided efficiency is denoted by $\eta=\eta_{\rm ch}\eta_{\rm det}$, where $\eta_{\rm det}$ is the detector efficiency (set to a common value for all the photo-detectors) and $\eta_{\rm ch}=10^{-\alpha{L}/{10}}$ is the transmission efficiency, $\alpha$ (dB/km) being the attenuation coefficient of the channel and $L$ (km) being the common transmission length between each party and the central node. The angle $\delta_{\rm A}$ $(\delta_{\rm B})\geq{0}$ denotes the polarization misalignment occurring in the left (right) arm of the setup (denoted by $\delta_{\rm mis}$ in Eq.~(\ref{misalignment})) and the symbol ``$\otimes$" stands for polarizing beam-splitter (PBS). Blue color is used for the intensities $|\xi^{i,j}_{w}|^{2}$ that arrive at the detectors ($w\in\{1,2,3,4\}$). Each detector has an ``arm index" $s_{w}\in\{1,2\}$ that specifies whether it is on the right arm ($s_{w}=1$) or the left arm ($s_{w}=2$) of the detection scheme, and a polarization index $k_{w}\in\{1,2\}$ specifying whether it detects the horizontal ($k_{w}=1$) or the vertical $(k_{w}=2)$ component of the pulses coming from the PBSs. For simplicity, this last index is not shown in the figure.}
	\label{fig:setup}
\end{figure}

Let us assume for the moment that Alice's (Bob's) laser emits pure coherent states with complex amplitude $\alpha$ ($\beta$) in the BB84 state defined by $i$ and $\theta_{\rm A}$ ($j$ and $\theta_{\rm B}$). The quantum state at the input port of the detectors also factors as the product of four coherent states, $\ket{\phi_{\rm det}}=\ket{\xi^{i,j}_{1}}\ket{\xi^{i,j}_{2}}\ket{\xi^{i,j}_{3}}\ket{\xi^{i,j}_{4}}$, $\xi^{i,j}_{w}$ denoting the incoming amplitude to detector $w$ for settings $i$ and $j$ (the dependence on the intensity settings, $\alpha$ and $\beta$, and the basis settings, $\theta_{\rm A}$ and $\theta_{\rm B}$, is omitted for readability). Precisely, it can be shown that
\begin{equation}\label{amplitudes}
\xi^{i,j}_{w}=\sqrt{\frac{\eta}{2}}\biggl[\alpha\Theta_{\textrm{A},i,k_{w}}+(-1)^{s_{w}}\beta\Theta_{\textrm{B},j,k_{w}}\biggr],
\end{equation}
where $\eta=\eta_{\rm ch}\eta_{\rm det}$ is the overall one-sided efficiency (accounting for both the transmission efficiency of the channel, $\eta_{\rm ch}$, and the detection efficiency of Charles' detectors, $\eta_{\rm det}$), and $\Theta_{\textrm{A},l,m}$ $(\Theta_{\textrm{B},l,m})$ is the $(l,m)$-th element of the matrix $\Theta_{\rm A}$ ($\Theta_{\rm B}$), which incorporates Alice's (Bob's) measurement setting, $\theta_{\rm A}$ $(\theta_{\rm B})$, and the polarization misalignment occurring in her (his) side of the channel, $\delta_{\rm A}\geq{}0$ $(\delta_{\rm B}\geq{}0)$:
\begin{equation}
\Theta_{\rm A}=\begin{bmatrix} 
\hspace{.2cm}\cos(\theta_{\rm A}+\delta_{\rm A}) & \sin(\theta_{\rm A}+\delta_{\rm A}) \\
-\sin(\theta_{\rm A}+\delta_{\rm A}) & \cos(\theta_{\rm A}+\delta_{\rm A}) \\
\end{bmatrix}\hspace{.2cm}\textrm{and}\hspace{.2cm}
\Theta_{\rm B}=\begin{bmatrix} 
\hspace{.2cm}\cos(\theta_{\rm B}+\delta_{\rm B}) & \sin(\theta_{\rm B}+\delta_{\rm B}) \\
-\sin(\theta_{\rm B}+\delta_{\rm B}) & \cos(\theta_{\rm B}+\delta_{\rm B}) \\
\end{bmatrix}.
\end{equation}
The cases of interest are $\theta_{\rm A}=\theta_{\rm B}=0$ (basis Z match) and $\theta_{\rm A}=\theta_{\rm B}=\pi/4$ (basis X match). If, without loss of generality, we set $\alpha=|\alpha|$ and $\beta=|\beta|{e}^{i\gamma}$, the intensities (squared modulus of the amplitudes) at the detectors read
\begin{equation}\label{intensities_z}
{|\xi_{w}^{i,j}|}^{2}=\frac{\eta}{2}\biggl[|\alpha|^{2}\Theta_{\textrm{A},i,k_{w}}^{2}+|\beta|^{2}\Theta_{\textrm{B},j,k_{w}}^{2}+(-1)^{s_{w}}2|\alpha||\beta|\Theta_{\textrm{A},i,k_{w}}\Theta_{\textrm{B},j,k_{w}}\cos{\gamma}\biggr],\hspace{.2cm}w\in\{1,2,3,4\}.
\end{equation}
Since a success at the central node is heralded by the click of exactly two detectors referred to orthogonal polarizations, the set of possible successful events reads 
\begin{equation}\label{successes}
\Omega=\left\{(1,2),(3,4),(1,4),(2,3)\right\}.
\end{equation}
As an example, let us compute the probability ${P}_{(1,2)}^{i,j}$ of the successful event $(1,2)$. This probability factors as
\begin{equation}\label{factors}
{P}_{(1,2)}^{i,j}=p(\textrm{3\hspace{.1cm}and\hspace{.1cm}4\hspace{.1cm}do\hspace{.1cm}not\hspace{.1cm}click})\times{p}(\rm{1\hspace{.1cm}and\hspace{.1cm}2\hspace{.1cm}click}).
\end{equation}
Then, from our detector model and the poissonian statistics of coherent states, we have that
\begin{eqnarray}\label{factors2}
p(\textrm{3\hspace{.1cm}and\hspace{.1cm}4\hspace{.1cm}do\hspace{.1cm}not\hspace{.1cm}click})&=&(1-p_{\rm d})^{2}e^{-\left({|\xi^{i,j}_{3}|^{2}}+{|\xi^{i,j}_{4}|^{2}}\right)}, \nonumber \\
p(\textrm{1\hspace{.1cm}and\hspace{.1cm}2\hspace{.1cm}click})&=&
\left(1-e^{-|\xi^{i,j}_{1}|^{2}}\right)\left(1-e^{-|\xi^{i,j}_{2}|^{2}}\right)+ \nonumber \\
&+&p_{\rm d}\left[\left(1-e^{-|\xi^{i,j}_{1}|^{2}}\right)e^{-|\xi^{i,j}_{2}|^{2}}+\left(1-e^{-|\xi^{i,j}_{2}|^{2}}\right)e^{-|\xi^{i,j}_{1}|^{2}}\right]+p_{\rm d}^{2}e^{-\left({|\xi^{i,j}_{1}|^{2}}+{|\xi^{i,j}_{2}|^{2}}\right)}.
\end{eqnarray}
Putting both factors together and generalising the expression to an arbitrary successful event $(u,v)$, one obtains
\begin{equation}\label{success_1}
{P}_{(u,v)}^{i,j}=(1-p_{\rm d})^{2}\exp\biggl(-\sum_{w\neq{u,v}}{|\xi^{i,j}_{w}|^{2}}\biggr)\biggl[1-(1-p_{\rm d})\biggl(e^{-|\xi^{i,j}_{u}|^{2}}+e^{-|\xi^{i,j}_{v}|^{2}}\biggr)+(1-p_{\rm d})^{2}e^{-\left({|\xi^{i,j}_{u}|^{2}}+{|\xi^{i,j}_{v}|^{2}}\right)}\biggr],
\end{equation}
or, more conveniently,
\begin{equation}\label{success_2}
\frac{{P}_{(u,v)}^{i,j}}{(1-p_{\rm d})^{2}}=\exp\left(-\sum_{w\neq{u,v}}{|\xi^{i,j}_{w}|^{2}}\right)-(1-p_{\rm d})\left[\exp\left(-\sum_{w\neq{u}}{|\xi^{i,j}_{w}|^{2}}\right)+\exp\left(-\sum_{w\neq{v}}{|\xi^{i,j}_{w}|^{2}}\right)\right]+(1-p_{\rm d})^{2}\exp\left(-\sum_{w}{|\xi^{i,j}_{w}|^{2}}\right).
\end{equation}
Recalling that ${P}_{(u,v)}^{i,j}$ was computed assuming pure coherent states, one needs to average over phase values in order to derive the resulting probability for PR-WCPs, which we denote by ${p}_{(u,v),\alpha,\beta,\theta_{\rm A},\theta_{\rm B}}^{i,j}=\frac{1}{2\pi}\int_{0}^{2\pi}{P}_{(u,v)}^{i,j}\hspace{.1cm}d\gamma$. For convenience, this notation explicitly shows that, in any round of the protocol, the probability of a successful detection event $(u,v)\in\Omega$ depends on all the protocol settings. The explicit calculation of this integral yields
\begin{eqnarray}\label{success}
\frac{{p}_{(u,v),\alpha,\beta,\theta_{\rm A},\theta_{\rm B}}^{i,j}}{(1-p_{\rm d})^{2}}&=&\exp\left[-\frac{\eta}{2}\sum_{w\neq{u,v}}{\left(|\alpha|^{2}\Theta_{\mathrm{A},i,k_{w}}^{2}+|\beta|^{2}\Theta_{\mathrm{B},j,k_{w}}^{2}\right)}\right]I_{0,\rm sym}\left(\eta|\alpha||\beta|{\sum_{w\neq{u,v}}{(-1)^{s_{w}+1}\Theta_{\mathrm{A},i,k_{w}}\Theta_{\mathrm{B},j,k_{w}}}}\right) \nonumber \\
&-&(1-p_{\rm d})\exp\left[-\frac{\eta}{2}\sum_{w\neq{u}}{\left(|\alpha|^{2}\Theta_{\mathrm{A},i,k_{w}}^{2}+|\beta|^{2}\Theta_{\mathrm{B},j,k_{w}}^{2}\right)}\right]I_{0,\rm sym}\left(\eta|\alpha||\beta|{\sum_{w\neq{u}}{(-1)^{s_{w}+1}\Theta_{\mathrm{A},i,k_{w}}\Theta_{\mathrm{B},j,k_{w}}}}\right) \nonumber \\
&-&(1-p_{\rm d})\exp\left[-\frac{\eta}{2}\sum_{w\neq{v}}{\left(|\alpha|^{2}\Theta_{\mathrm{A},i,k_{w}}^{2}+|\beta|^{2}\Theta_{\mathrm{B},j,k_{w}}^{2}\right)}\right]I_{0,\rm sym}\left(\eta|\alpha||\beta|{\sum_{w\neq{v}}{(-1)^{s_{w}+1}\Theta_{\mathrm{A},i,k_{w}}\Theta_{\mathrm{B},j,k_{w}}}}\right) \nonumber \\
&+&(1-p_{\rm d})^{2}\exp\left[-\frac{\eta}{2}\sum_{w}{\left(|\alpha|^{2}\Theta_{\mathrm{A},i,k_{w}}^{2}+|\beta|^{2}\Theta_{\mathrm{B},j,k_{w}}^{2}\right)}\right]I_{0,\rm sym}\left(\eta|\alpha||\beta|{\sum_{w}{(-1)^{s_{w}+1}\Theta_{\mathrm{A},i,k_{w}}\Theta_{\mathrm{B},j,k_{w}}}}\right), \nonumber \\
\end{eqnarray}
where we have introduced the function $I_{0,\rm sym}(x)=[I_{0}(x)+I_{0}(-x)]/2=\frac{1}{2\pi}\int_{0}^{2\pi}e^{x\cos{\gamma}}d\gamma$, $I_{0}$ being the modified Bessel function of the first kind.

From Eq.~(\ref{success}), one can compute the probability $Q^{\alpha,\beta}_{\theta_{\rm A},\theta_{\rm B}}$ that any given round yields a successful BSM at the central node when Alice (Bob) uses basis $\theta_{\rm A}$ $(\theta_{\rm B})$ and intensity $|\alpha|^{2}$ $(|\beta|^{2})$,
\begin{equation}\label{P}
Q^{\alpha,\beta}_{\theta_{\rm A},\theta_{\rm B}}=\frac{1}{4}\sum_{(u,v)\in{\Omega}}\left[{p}_{(u,v),\alpha,\beta,\theta_{\rm A},\theta_{\rm B}}^{1,1}+{p}_{(u,v),\alpha,\beta,\theta_{\rm A},\theta_{\rm B}}^{\textrm{1,2}}+{p}_{(u,v),\alpha,\beta,\theta_{\rm A},\theta_{\rm B}}^{\textrm{2,1}}+{p}_{(u,v),\alpha,\beta,\theta_{\rm A},\theta_{\rm B}}^{\textrm{2,2}}\right],\hspace{.2cm}\theta_{\rm A},\theta_{\rm B}\in\{0,\pi/4\}.
\end{equation}
Similarly, one can compute the bit error rates, $E^{\alpha,\beta}_{0,0}$ and $E^{\alpha,\beta}_{\frac{\pi}{4},\frac{\pi}{4}}$, for the basis coincidences, given by
\begin{eqnarray}\label{Q}
&&Q^{\alpha,\beta}_{0,0}E^{\alpha,\beta}_{0,0}=\frac{1}{4}\sum_{(u,v)\in{\Omega}}\left[{p}_{(u,v),\alpha,\beta,0,0}^{\textrm{1,1}}+{p}_{(u,v),\alpha,\beta,0,0}^{\textrm{2,2}}\right]\hspace{.2cm}\textrm{and}\nonumber \\
&&Q^{\alpha,\beta}_{\frac{\pi}{4},\frac{\pi}{4}}E^{\alpha,\beta}_{\frac{\pi}{4},\frac{\pi}{4}}=\frac{1}{4}\left\{\sum_{(u,v)\in\Omega_{1}}\left[{p}_{(u,v),\alpha,\beta,\frac{\pi}{4},\frac{\pi}{4}}^{1,1}+{p}_{(u,v),\alpha,\beta,\frac{\pi}{4},\frac{\pi}{4}}^{2,2}\right]\hspace{.2cm}+\sum_{(u,v)\in\Omega_{2}}\left[{p}_{(u,v),\alpha,\beta,\frac{\pi}{4},\frac{\pi}{4}}^{1,2}+{p}_{(u,v),\alpha,\beta,\frac{\pi}{4},\frac{\pi}{4}}^{2,1}\right]\right\},
\end{eqnarray}
where $\Omega_{1}=\{(1,4),(2,3)\}$ and $\Omega_{2}=\{(1,2),(3,4)\}$.
Note that only the polarization settings h,h and v,v (\textit{i.e.}, $i=j=1$ and $i=j=2$) contribute to the basis Z bit error rate. This is so because, for these rounds, Bob flips his bit irrespectively of the successful outcome of the BSM. On the contrary, for the basis X bit error rate, the post-selection of $\ket{\psi^{-}}$ in the BSM (events $(1,4)$ and $(2,3)$) entails a bit flip, while the post-selection of $\ket{\psi^{+}}$ (events $(1,2)$ and $(3,4)$) does not, thus leading to the definition of the bit error rate given in Eq.~(\ref{Q}).\\

Finally, we write down the expected values of the observables required for the simulations, according to the channel model above. For this purpose, we introduce the quantities $G_{\rm Z,Z}^{\lambda,\lambda}=q_{\rm Z}^{2}Q^{\sqrt{\lambda},\sqrt{\lambda}}_{0,0}$, $G_{\rm X,X}^{a,b}=q_{\rm X}^{2}p_{a}p_{b}Q_{\frac{\pi}{4},\frac{\pi}{4}}^{\sqrt{a},\sqrt{b}}$, $G_{\rm Z,X}^{\lambda,b}=q_{\rm Z}q_{\rm X}p_{b}Q^{\sqrt{\lambda},\sqrt{b}}_{0,\frac{\pi}{4}}$, $G_{\rm X,Z}^{a,\lambda}=q_{\rm Z}q_{\rm X}p_{a}Q^{\sqrt{a},\sqrt{\lambda}}_{\frac{\pi}{4},0}$, $E_{\rm Z,Z}^{\lambda,\lambda}=E_{0,0}^{\sqrt{\lambda},\sqrt{\lambda}}$ and $E_{\rm X,X}^{a,b}=E_{\frac{\pi}{4},\frac{\pi}{4}}^{\sqrt{a},\sqrt{b}}$, with $a,b\in\mathrm{A}=\{\mu,\nu,\omega\}$. From these quantities, it follows that
\begin{eqnarray}\label{expected_observables}
&& E\bigl[\bigl|\mathcal{Z}_{j}|\bigr]=G_{\rm Z,Z}^{\lambda,\lambda}N, \nonumber \\
&& E\bigl[E_{j}\bigr]=E_{\rm Z,Z}^{\lambda,\lambda}, \nonumber \\
&& E\bigl[\bigl|\mathcal{X}_{j}^{a,b}\bigr|\bigr]=G_{\rm X,X}^{a,b}N, \nonumber \\
&& E\bigl[e_{a,b}^{j}\bigr]=E_{\rm X,X}^{a,b}G_{\rm X,X}^{a,b}N, \nonumber \\
&& E\bigl[|a^{j}|_{c^{j}}|\bigr]=\biggl(G_{\rm Z,Z}^{\lambda,\lambda}+\sum_{a,b\in\rm A}G_{\rm X,X}^{a,b}+\sum_{b\in\rm A}G_{\rm Z,X}^{\lambda,b}+\sum_{a\in\rm A}G_{\rm X,Z}^{a,\lambda}\biggr)2N\hspace{.1cm}\mathrm{bits}, \nonumber \\
&& E\bigl[\bigl|r_{\rm A}^{j}|_{c^{j},\rm X}\bigr|\bigr]=\biggl(\sum_{a,b\in\rm A}G_{\rm X,X}^{a,b}+\sum_{a\in\rm A}G_{\rm X,Z}^{a,\lambda}\biggr)N\hspace{.1cm}\mathrm{bits}\hspace{.2cm}\textrm{and} \nonumber \\
&& E\bigl[\bigl|r_{\rm A}^{j}|_{c^{j},\rm Z}\bigr|\bigr]=\biggl(G_{\rm Z,Z}^{\lambda,\lambda}+\sum_{b\in\rm A}G_{\rm Z,X}^{\lambda,b}\biggr)N\hspace{.1cm}\mathrm{bits}.
\end{eqnarray}
We recall that $N$ is the number of signals transmitted per module in each QKD pair.
\section*{Supplementary Note 7: decoy-state BB84}\label{BB84}
To further illustrate the applicability of the results in the main text, we combine Protocol with the standard decoy-state BB84 scheme~\cite{decoy2,decoy,decoy3,Lim} with three common decoy intensities per basis. Again, asterisks are omitted for readability throughout this note.\\

\textbf{QKD protocol.} Again, although the protocol description below assumes that the corrupted devices do not deviate from the protocol, the security against actively misbehaving corrupted devices is established in the main text.\\ 

For $j=1,\ldots,n_{\rm q}$, $\textrm{QKD}_{\textrm{A}_{j}}$ creates a trio of strings $(r_{\rm A}^{j},k_{\rm A}^{j},a^{j})$. The string $r_{\rm A}^{j}\in\left\{0,1\right\}^{N}$ is fully random (polarization bits string). For all $i=1,\ldots,N$, $k_{\rm A}^{j}\in\left\{\textrm{Z,X}\right\}^{N}$ verifies $P\bigl[k_{{\rm A}_{i}}^{j}=\zeta\bigr]=q_{\zeta}$ with $\zeta\in\left\{\textrm{Z,X}\right\}$ (basis string), and $a^{j}\in\left\{\mu,\nu,\omega\right\}^{N}$ (intensities string) verifies $P\bigl[a_{i}^{j}=a\bigr]=p_{a}$ for $a\in{}A$, with $\mathrm{A}=\left\{\mu,\nu,\omega\right\}$. Similarly, $\textrm{QKD}_{\textrm{B}_{j}}$ creates its basis string $k_{\rm B}^{j}\in\left\{\textrm{Z,X}\right\}^{N}$ verifying $P\bigl[k_{{\rm B}_{i}}^{j}=\zeta\bigr]=q_{\zeta}$ with $\zeta\in\left\{\textrm{Z,X}\right\}$ and $i=1,\ldots,N$.\\

Let us now focus on a single QKD pair, say, the $j$-th one. For $i$ ranging from $1$ to $N$, steps (i) to (iii) are repeated.
\begin{enumerate}
	\item[(i)]\textit{State preparation.} $\textrm{QKD}_{\textrm{A}_{j}}$ prepares a PR-WCP with intensity $a_{i}^{j}$ in the BB84 state defined by $k_{{\rm A}_{i}}^{j}$ and $r_{{\rm A}_{i}}^{j}$.
	\item[(ii)]\textit{Transmission.} $\textrm{QKD}_{\textrm{A}_{j}}$ sends the state to $\textrm{QKD}_{\textrm{B}_{j}}$ via the quantum channel. 
	\item[(iii)]\textit{Measurement.} $\textrm{QKD}_{\textrm{B}_{j}}$ performs a measurement in basis $k_{{\rm B}_{i}}^{j}$ and stores the outcome in a classical value $r_{{\rm B}_{i}}^{j}\in{\left\{0,1,\emptyset\right\}}$, where $\emptyset$ is the symbol produced when no signal is detected. If a multiple click takes place, Bob assigns a random bit to this event.\\
\end{enumerate}
After the above quantum communication phase, the distributed QKD post-processing starts. Again, we focus on a single QKD pair, say, the $j$-th one.
\begin{enumerate}
	\item[1.]\textit{Distribution of data.}
	Let $r_{\rm A}^{j}|_{\rm X}$ ($r_{\rm A}^{j}|_{\rm Z}$) be the sub-string of $r_{\rm A}^{j}$ where $\textrm{QKD}_{\textrm{A}_{j}}$ uses basis X (Z) for the encoding. $\textrm{QKD}_{\textrm{A}_{j}}$ uses the Share protocol of a conditional VSS scheme (which depends on the selected corruption model of the CP units via Proposition 1 in the main text) to distribute shares of $r_{\rm A}^{j}|_{\rm Z}$ among the $\textrm{CP}_{A_{l}}$. Particularly, let $\sigma_{i}^{\rm A}$ denote the set of units that receives the $i$-th share of $r_{\rm A}^{j}|_{\rm Z}$, which without loss of generality is common for all $j=1,\ldots,n_{\rm q}$. $\textrm{QKD}_{\textrm{A}_{j}}$ communicates the trio $(k_{\rm A}^{j},a^{j},r_{\rm A}^{j}|_{\rm X})$ to every $\textrm{CP}_{A_{l}}\in\sigma_{1}^{\rm A}$. Similarly, regarding Bob's side, let $c^{j}$ be the string of detector clicks held by $\textrm{QKD}_{\textrm{B}_{j}}$, such that $c_{i}^{j}=0$ if $r_{{\rm B}_{i}}^{j}={\emptyset}$ and $c_{i}^{j}=1$ otherwise, and let ${k_{\rm B}}^{j}|_{c^{j}}$ be the restriction of the basis string ${k_{\rm B}}^{j}$ to the non-zero entries of $c^{j}$. Also, let $r_{\rm B}^{j}|_{c^{j},\rm X}$ ($r_{\rm B}^{j}|_{c^{j},\rm Z}$) be the restriction of $r_{\rm B}^{j}$ to the non-zero entries of $c^{j}$ where $\textrm{QKD}_{\textrm{B}_{j}}$ uses basis X (Z) for the measurements.
	$\textrm{QKD}_{\textrm{B}_{j}}$ uses the Share protocol of a conditional VSS scheme to distribute shares of $r_{\rm B}^{j}|_{c^{j},\rm Z}$ among the $\textrm{CP}_{B_{l'}}$, such that $\sigma_{i}^{\rm B}$ denotes the set of units that receives the $i$-th share of $r_{\rm B}^{j}|_{c^{j},\rm Z}$. Again, one can impose that $\sigma_{i}^{\rm B}$ is common for all $j=1,\ldots,n_{\rm q}$ without loss of generality. Then, $\textrm{QKD}_{\textrm{B}_{j}}$ communicates the trio $({c}^{j},k_{\rm B}^{j}|_{c^{j}},r_{\rm B}^{j}|_{c^{j},\rm X})$ to every $\textrm{CP}_{B_{l'}}\in\sigma_{1}^{\rm B}$. Finally, all the $\textrm{CP}_{\textrm{A}_{l}}\in\sigma_{1}^{\rm A}$ ($\textrm{CP}_{\textrm{B}_{l'}}\in\sigma_{1}^{\rm B}$) perform a consistency test on $(k_{\rm A}^{j},a^{j},r_{\rm A}^{j}|_{\rm X})$ ($({c}^{j},k_{\rm B}^{j}|_{c^{j}},r_{\rm B}^{j}|_{c^{j},\rm X})$).
	\item[2.]\textit{Sifting.}
	Every $\textrm{CP}_{A_{l}}\in\sigma_{1}^{\rm A}$ sends $(k_{\rm A}^{j},a^{j},r_{\rm A}^{j}|_{\rm X})$ to every $\textrm{CP}_{B_{l'}}\in\sigma_{1}^{\rm B}$, which individually apply MV. Then, each $\textrm{CP}_{B_{l'}}\in\sigma_{1}^{\rm B}$ builds the index sets
	\begin{equation}\label{index_sets_BB84}
	\mathcal{Z}_{j}=\left\{i|c_{i}^{j}=1,k_{{\rm A}_{i}}^{j}=k_{{\rm B}_{i}}^{j}=\mathrm{Z}\right\}\hspace{.2cm}\textrm{and}\hspace{.2cm}\mathcal{X}_{j}^{a}=\left\{i|c_{i}^{j}=1,k_{{\rm A}_{i}}^{j}=k_{{\rm B}_{i}}^{j}=\mathrm{X},a_{i}^{j}=a\right\},
	\end{equation}
	for all $a\in{\rm A}$, and checks if the sifting condition $\bigl|\mathcal{Z}_{j}\bigr|\geq{}M$ is met, for a pre-established threshold value $M$. If it is not met, the $\textrm{CP}_{\textrm{B}_{l'}}\in{}\sigma_{1}^{\rm B}$ abort the protocol. In case of not aborting, the $\textrm{CP}_{B_{l'}}\in\sigma_{1}^{\rm B}$ units forward $\mathcal{Z}_{j}$ to the rest of the units, which apply MV. All together the $\textrm{CP}_{\textrm{B}_{l'}}$ perform a RBS protocol to select a random subset $\mathcal{Z'}_{j}\subseteq{\mathcal{Z}_{j}}$, of size $M$. Then, Bob's units locally perform the sifting. Precisely, every $\textrm{CP}_{\textrm{B}_{l'}}$ builds its shares of the sifted key $s_{\rm B}^{j}=r_{\rm B}^{j}|_{\mathcal{Z'}_{j}}$ from those of $r_{\rm B}^{j}|_{c^{j},\rm Z}$ (discarding the data external to $\mathcal{Z'}_{j}$).
	\item[3.]\textit{Parameter estimation.}
	For each $a\in{\rm A}$, every $\textrm{CP}_{B_{l'}}\in\sigma_{1}^{\rm B}$ unit builds the PE strings $r_{\rm B}^{j}|_{\mathcal{X}_{j}^{a}}$ and $r_{\rm A}^{j}|_{\mathcal{X}_{j}^{a}}$ from the respective strings $r_{\rm B}^{j}|_{c^{j},\rm X}$ and $r_{\rm A}^{j}|_{\rm X}$, discarding the data external to $\mathcal{X}_{j}^{a}$. Then, each of them computes the numbers of bit errors
	\begin{equation}\label{error_numbers_BB84}
	e_{a}^{j}=\sum_{k=1}^{\left|\mathcal{X}_{j}^{a}\right|}r_{{\rm A}_{k}}^{j}\bigr|_{\mathcal{X}_{j}^{a}}\oplus{r_{{\rm B}_{k}}^{j}\bigr|_{\mathcal{X}_{j}^{a}}},
	\end{equation}
	for $a\in{}A$, where $r_{{\rm A}_{k}}^{j}\bigr|_{\mathcal{X}_{j}^{a}}$ ($r_{{\rm B}_{k}}^{j}\bigr|_{\mathcal{X}_{j}^{a}}$) denotes the $k$-th bit of the corresponding string. Using $\left|\mathcal{Z}_{j}\right|$ and the different $\bigl|\mathcal{X}_{j}^{a}\bigr|$ and $e_{a}^{j}$ ($a\in{\rm A}$), every $\textrm{CP}_{\textrm{B}_{l'}}\in{}\sigma_{1}^{\rm B}$ computes a lower bound on the number $n_{1,\rm Z}^{j}$ of single-photon successes in $\mathcal{Z'}_{j}$ and an upper bound on the single-photon phase-error rate $\phi_{1,\rm Z}^{j}$ associated to the single-photon successes in $\mathcal{Z'}_{j}$.
\end{enumerate}
Although the rest of the post-processing is identical to that of the MDI-QKD protocol (except from the fact that no bit flips are required to correlate $s_{\rm A}$ and $s_{\rm B}$), we include it here for completeness. The above steps 1 to 3 are performed for all $j=1,\ldots,n_{\rm q}$. At this stage, every $\textrm{CP}_{\textrm{B}_{l'}}\in{}\sigma_{1}^{\rm B}$ derives a lower bound $l$ (given in the next section) on the secret key length that can be extracted from the concatenated sifted key $s_{\rm B}=s_{\rm B}^{1}\ldots{}s_{\rm B}^{n_{\rm q}}$ via PA. If a $\textrm{CP}_{\textrm{B}_{l'}}\in{}\sigma_{1}^{\rm B}$ finds $l\leq{0}$, it aborts the protocol.
\begin{enumerate}
	\item[4.]\textit{RBS generation.}
	If the protocol does not abort, every $\textrm{CP}_{\textrm{B}_{l'}}\in{}\sigma_{1}^{\rm B}$ forwards $l$ to the rest of Bob's units, which apply MV. All $\textrm{CP}_{\textrm{B}_{l'}}$ perform a RBS generation protocol to randomly select two 2-universal hash functions $h_{\rm EV}$ and $h_{\rm PA}$, respectively devoted to error verification (EV) and PA. Following~\cite{Fung}, if Toeplitz matrices are used for this purpose, $2\lceil\log_{2}(2/\hat{\epsilon}_{\rm cor})\rceil$ ($Mn_{\rm q}+l-1$) bits are required to specify $h_{\rm EV}$ ($h_{\rm PA}$).
	\item[5.]\textit{Information reconciliation.} Every $\textrm{CP}_{\textrm{B}_{l'}}$ computes its shares of (1) the concatenated syndromes string $sy_{\rm B}=sy(s_{\rm B}^{1})\ldots{}sy(s_{\rm B}^{n_{\rm q}})$ and (2) the EV tag $h_{\rm EV,B}=h_{\rm EV}(s_{\rm B})$. All together, the $\textrm{CP}_{\textrm{B}_{l'}}$ reconstruct $sy_{\rm B}$ and $h_{\rm EV,B}$ via the Reconstruct protocol of a conditional VSS scheme (see the Methods section in the main text). Each $\textrm{CP}_{\textrm{B}_{l'}}\in{}\sigma_{1}^{\rm B}$ sends the following items to every $\textrm{CP}_{\textrm{A}_{l}}\in{}\sigma_{1}^{\rm A}$:
	\begin{enumerate}
		\item[1.]{The string $s_{\mathcal{Z'}}=s_{\mathcal{Z'}_{1}}\ldots{}s_{\mathcal{Z'}_{n_{\rm q}}}$, where $s_{\mathcal{Z'}_{j}}$ specifies, say, the positions in $r_{\rm A}^{j}|_{\rm Z}$ that contribute to $\mathcal{Z'}_{j}$.}
		\item[2.]{The syndrome information $sy(s_{\rm B})$, together with the description of $h_{\rm EV}$ and the EV tag $h_{\rm EV}(s_{\rm B})$.}
		\item[3.]{The description of $h_{\rm PA}$.}
	\end{enumerate}
	Each $\textrm{CP}_{\textrm{A}_{l}}\in{}\sigma_{1}^{\rm A}$ decides on all three items via MV and communicate $s_{\mathcal{Z'}}$, $h_{\rm EV}$ and $h_{\rm PA}$ to the rest of Alice's units, which apply MV too. Then, they proceed as follows. Using $s_{\mathcal{Z'}}$, all $\textrm{CP}_{\textrm{A}_{l}}$ shrink their shares of $r_{\rm A}|_{\rm Z}=r_{\rm A}^{1}|_{\rm Z}\ldots{}r_{\rm A}^{n_{\rm q}}|_{\rm Z}$ into shares of $s_{\rm A}=s_{\rm A}^{1}\ldots{}s_{\rm A}^{n_{\rm q}}$, where $s_{\rm A}^{j}=r_{\rm A}^{j}|_{\mathcal{Z'}_{j}}$. All the $\textrm{CP}_{\textrm{A}_{l}}$ compute shares of $sy(s_{\rm A})$ from those of $s_{\rm A}$ and then perform the Reconstruct protocol of a conditional VSS scheme to agree on $sy(s_{\rm A})$. Coming next, the $\textrm{CP}_{\textrm{A}_{l}}\in{}\sigma_{1}^{\rm A}$ compute the error pattern $\hat{e}$ from $sy(s_{\rm B})$ and $sy(s_{\rm A})$ and update the first share of $s_{\rm A}$ XOR-ing it with $\hat{e}$ (\textit{i.e.}, key reconciliation is achieved by acting on a single share). We denote the corrected key by $\hat{s}_{\rm A}=s_{\rm A}\oplus{\hat{e}}$. Using $h_{\rm EV}$, all the $\textrm{CP}_{\textrm{A}_{l}}$ compute their shares of $h_{\rm EV}(\hat{s}_{\rm A})$ and reconstruct it via the Reconstruct protocol of a conditional VSS scheme. Then, each $\textrm{CP}_{\textrm{A}_{l}}\in{}\sigma_{1}^{\rm A}$ checks that $h_{\rm EV}(\hat{s}_{\rm A})=h_{\rm EV}(s_{\rm B})$. Otherwise, it aborts the protocol.
	\item[6.]\textit{Privacy amplification.}
	In case of not aborting, all the $\textrm{CP}_{\textrm{A}_{l}}$ compute their shares of Alice's final key $S_{\rm A}=h_{\rm PA}(\hat{s}_{\rm A})$. Similarly, if no abortion is notified, all the $\textrm{CP}_{\textrm{B}_{l'}}$ compute their shares of Bob's final key $S_{\rm B}=h_{\rm PA}(s_{\rm B})$.\\
\end{enumerate}

\textbf{Secret key length formula in the AC, AN and PC corruption models for the QKD modules.} Here, we particularize the secret key length formula of the AC model, that is, Eq.~(\ref{looser}), for the decoy-state BB84 protocol presented above. The formula is tight within the AC, AN and PC corruption models for the QKD modules, and the analysis is identical to the one for the MDI-QKD protocol given in Supplementary Note 6, so we omit the details here for simplicity. Precisely, the extractable key length is
\begin{equation}\label{BB84 AC}
l^{*}=\left\lfloor{\min_{j}\biggl\{n_{1,\rm Z}^{j,\rm L*}\left[1-h(\phi_{1,\rm Z}^{j,\rm U*})\right]-\bigl|{sy}^{*}(s_{\rm B}^{j*})\bigr|\biggr\}-\log_{2}\left(\frac{1}{\hat{\epsilon}_{\rm cor}\epsilon_{\rm PA}^{2}\delta}\right)}\right\rfloor,
\end{equation}
where $n_{1,\rm Z}^{\mathrm{h},\rm L*}$ ($\phi_{1,\rm Z}^{\mathrm{h},\rm U*}$) stands for a lower (upper) bound on $n_{1,\rm Z}^{\mathrm{h}*}$ ($\phi_{1,\rm Z}^{\mathrm{h}*}$), $h(\cdot)$ is the binary entropy function, $|{sy}^{*}(s_{\rm B}^{j*})|$ is the size of the $j$-th EC syndrome, $\hat{\epsilon}_{\rm cor}$ is the correctness parameter, $\epsilon_{\rm PA}$ is the error probability of the privacy amplification, and $\delta>0$. Also, as shown in Supplementary Note 1, the above key length is $\epsilon_{\rm sec}$-secret for all $\epsilon_{\rm sec}=\hat{\epsilon}_{\rm sec}+\epsilon_{\rm AU}$, with $\hat{\epsilon}_{\rm sec}\geq{2\varepsilon+\delta+\epsilon_{\rm PA}}$, where $\varepsilon$ is upper-bounded by the sum of the error probabilities of the estimates of $n_{1,\rm Z}^{\mathrm{h},\rm L*}$ and $\phi_{1,\rm Z}^{\mathrm{h},\rm U*}$, and $\epsilon_{\rm AU}$ is the pre-agreed total error probability of the authentication, which depends on the corruption model of the CP units (see the Results section in the main text). Explicit expressions of $n_{1,\rm Z}^{j,\rm L*}$ and $\phi_{1,\rm Z}^{j,\rm U*}$ in terms of the observables of the protocol are given in the next section, together with an upper bound on the smooth-parameter $\varepsilon$.\\

\textbf{Secret key length in the PN corruption model for the QKD modules.} Following Eq.~(\ref{looser PN}) of Supplementary Note 2, within the PC corruption model the following tighter key length formula holds,
\begin{equation}\label{BB84 PN}
l^{*}=\Biggl\lfloor\min_{v}\sum_{j\neq{v}}^{n_{\rm q}}\biggl\{n_{1,\rm Z}^{j,\rm L*}\left[1-h(\phi_{1,\rm Z}^{j,\rm U*})\right]-\bigl|{sy}\bigl(s_{\rm B}^{j*}\bigr)\bigr|\biggr\}-\log_{2}\left(\frac{1}{\hat{\epsilon}_{\rm cor}\epsilon_{\rm PA}^{2}\delta^{n_{\rm q}-1}}\right)\Biggr\rfloor,
\end{equation}
where $\epsilon_{\rm sec}=\hat{\epsilon}_{\rm sec}+\epsilon_{\rm AU}$ with $\hat{\epsilon}_{\rm sec}\geq{(n_{\rm q}-1)(2\varepsilon+\delta)+\epsilon_{\rm PA}}$ and $\epsilon_{\rm AU}$ is again pre-agreed by Alice and Bob.\\

\textbf{Parameter estimation.} Here, we give the analytical bounds $n_{1,\rm Z}^{j,\rm L*}$ and $\phi_{1,\rm Z}^{j,\rm U*}$ that enter Eq.~(\ref{BB84 AC}). These bounds were originally presented in~\cite{Lim} and we include them here for completeness. Since the analysis is common for all $j=1,\ldots,n_{\rm q}$, we drop the QKD pair index $j$ and refer to any of the QKD pairs.\\

The decoy-state bounds below require $\mu>\nu+\omega$ and $\nu>\omega\geq{0}$, where we recall that $\mathrm{A}=\left\{\mu,\nu,\omega\right\}$ is the set of intensity settings. In addition, let us introduce the decomposition $\mathcal{Z'}=\cup_{a\in{\mathrm{A}}}\mathcal{Z'}^{a}$, where $\mathcal{Z'}^{a}=\left\{i\in\mathcal{Z'}|a_{i}=a\right\}$. The observed sizes of the sets $\mathcal{Z'}^{a}$ determine $n_{1,\rm Z}^{\rm L*}$. Precisely, we have that for all $\epsilon_{\rm H}\in(0,1)$, $P\left(n_{1,\rm Z}<n_{1,\rm Z}^{\rm L}\right)<3\epsilon_{\rm H}$ holds for
\begin{equation}\label{single photon successes}
n_{1,\rm Z}^{\rm L}=\biggl\lfloor{\frac{\mu\tau_{1}}{\mu(\nu-\omega)-(\nu^{2}-\omega^{2})}\biggl\{\frac{e^{\nu}}{p_{\nu}}\biggl[\bigl|\mathcal{Z'}^{\nu}\bigr|-\delta(M,\epsilon_{\rm H})\biggr]-\frac{e^{\omega}}{p_{\omega}}\biggl[\bigl|\mathcal{Z'}^{\omega}\bigr|+\delta(M,\epsilon_{\rm H})\biggr]-\frac{\nu^{2}-\omega^{2}}{\mu^{2}}\frac{e^{\mu}}{p_{\mu}}\biggl[\bigl|\mathcal{Z'}^{\mu}\bigr|+\delta(M,\epsilon_{\rm H})\biggr]\biggr\}\biggr\rfloor},
\end{equation}
where $\tau_{1}=\mu{}e^{-\mu}p_{\mu}+\nu{}e^{-\nu}p_{\nu}+\omega{}e^{-\omega}p_{\omega}$ and $\delta(x,y)=\sqrt{(x/2)\ln{y^{-1}}}$ is the deviation term that follows from the use of Hoeffding's inequality~\cite{Hoeffding}. Such inequality is used three times in Eq.~(\ref{single photon successes}) (with a common error probability, $\epsilon_{\rm H}$) to obtain adequate one-sided bounds on the expected values of $\bigl|\mathcal{Z'}^{\mu}\bigr|$, $\bigl|\mathcal{Z'}^{\nu}\bigr|$ and $\bigl|\mathcal{Z'}^{\omega}\bigr|$, respectively, given their realisations. Note that, for this task, one could also apply the inverse Chernoff-bound given in Supplementary Note 8.\\

Similarly, $\bigl|\mathcal{X}^{\mu}\bigr|$, $\bigl|\mathcal{X}^{\nu}\bigr|$ and $\bigl|\mathcal{X}^{\omega}\bigr|$ determine a lower bound on the number $S_{1,\rm X}^{\rm L}$ of rounds in $\mathcal{X}=\cup_{a\in{A}}\mathcal{X}^{a}$ where Alice sent single photons. Precisely, $P\left(S_{1,\rm X}<S_{1,\rm X}^{\rm L}\right)<3\epsilon_{\rm H}$ holds for
\begin{eqnarray}\label{single photon successes X}
&&S_{1,\rm X}^{\rm L}=\nonumber \\
&&\biggl\lfloor{\frac{\mu\tau_{1}}{\mu(\nu-\omega)-(\nu^{2}-\omega^{2})}\biggl\{\frac{e^{\nu}}{p_{\nu}}\biggl[\bigl|\mathcal{{X}}^{\nu}\bigr|-\delta\left(\bigl|\mathcal{X}\bigr|,\epsilon_{\rm H}\right)\biggr]-\frac{e^{\omega}}{p_{\omega}}\biggl[\bigl|\mathcal{X}^{\omega}\bigr|+\delta\left(\bigl|\mathcal{X}\bigr|,\epsilon_{\rm H}\right)\biggr]-\frac{\nu^{2}-\omega^{2}}{\mu^{2}}\frac{e^{\mu}}{p_{\mu}}\biggl[\bigl|\mathcal{X}^{\mu}\bigr|+\delta\left(\bigl|\mathcal{X}\bigr|,\epsilon_{\rm H}\right)\biggr]\biggr\}\biggr\rfloor},\nonumber \\
\end{eqnarray}
where we assumed a common error probability, $\epsilon_{\rm H}$, for each usage of Hoeffding's inequality~\cite{Hoeffding} again.\\

Regarding the number $E_{1,\rm X}$ of single-photon errors in $\mathcal{X}$, it turns out that $P\left(E_{1,\rm X}>E_{1,\rm X}^{\rm U}\right)<2\epsilon_{\rm H}$ holds for
\begin{equation}\label{single photon errors X}
E_{1,\rm X}^{\rm U}=\biggl\lceil{\frac{\tau_{1}}{\nu-\omega}\biggl\{\frac{e^{\nu}}{p_{\nu}}\biggl[e_{\nu}+\delta(e,\epsilon_{\rm H})\biggr]-\frac{e^{\omega}}{p_{\omega}}\biggl[e_{\omega}-\delta({e},\epsilon_{\rm H})\biggr]\biggr\}\biggr\rceil},
\end{equation}
where we recall that $e_{a}$ is the observed number of errors in $\mathcal{X}^{a}$ ($a\in{A}$) and we defined $e=\sum_{a}e_{a}$. Also, the error probability $2\epsilon_{\rm H}$ follows from the composition of two usages of Hoeffding's inequality~\cite{Hoeffding}. Finally, as we did for the parameter estimation in the MDI-QKD protocol, we use Serfling's inequality~\cite{Serfling} to relate the number $e_{1,\rm Z}$ of single-photon errors in $\mathcal{Z'}$ with the number $E_{1,\rm X}$ of single-photon errors in $\mathcal{X}$. To be precise, it follows that $P\left(e_{1,\rm Z}>e_{1,\rm Z}^{\rm U}\right)<8\epsilon_{\rm H}+\epsilon_{\rm S}$ holds for
\begin{equation}\label{single photon errors Z}
e_{1,\rm Z}^{\rm U}=\min\left\{\left\lceil{{n_{1,\rm Z}^{\rm L}}\left(\frac{E_{1,\rm X}^{\rm U}}{S_{1,\rm X}^{\rm L}}\right)+\left(S_{1,\rm X}^{\rm L}+n_{1,\rm Z}^{\rm L}\right)\times\Upsilon\left(n_{1,\rm Z}^{\rm L},S_{1,\rm X}^{\rm L},\epsilon_{\rm S}\right)}\right\rceil,n_{1,\rm Z}^{\rm L}\right\},
\end{equation}
where the deviation function $\Upsilon(x,y,z)$ is given by Eq.~(\ref{upsilon}) and $\epsilon_{\rm S}$ is the error probability of Serfling's inequality~\cite{Serfling}. Equivalently, the single-photon phase error rate $\phi_{1,\rm Z}$ verifies $P\left(\phi_{1,\rm Z}\geq{\phi_{1,\rm Z}^{\rm U}}\right)\leq{8\epsilon_{\rm H}+\epsilon_{\rm S}}$ for
\begin{equation}
\phi_{1,\rm Z}^{\rm U}=\frac{e_{1,\rm Z}^{\rm U}}{n_{1,\rm Z}^{\rm L}},
\end{equation}
where $n_{1,\rm Z}^{\rm L}$ is given in Eq.~(\ref{single photon successes}) and $e_{1,\rm Z}^{\rm U}$ is given in Eq.~(\ref{single photon errors Z}).

From the above PE procedure it follows that the smooth parameter $\varepsilon$ (presented below Eq.~(\ref{BB84 AC})) is upper-bounded as
\begin{equation}\label{smooth BB84}
\varepsilon\leq{}8\epsilon_{\rm H}+\epsilon_{\rm S}.
\end{equation}\\

\textbf{Authentication cost.} Following Supplementary Note 5, in order to quantify the overall authentication cost it suffices to specify the lengths of the classical messages exchanged in the lab-to-lab communications, as we did for the MDI-QKD protocol. In particular, from the protocol description above, we have
\begin{eqnarray}
\bigl|m_{\rm A}^{j}\bigr|&=&\bigl|k_{\rm A}^{j}\bigr|+\bigl|a^{j}\bigr|+\bigl|r_{\rm A}^{j}|_{\rm X}\bigr|, \nonumber \\
|m_{\rm B}|&=&\left|s_{\mathcal{Z'}}\right|+\left|sy(s_{\rm B})\right|+\left|h_{\rm EV}(s_{\rm B})\right|+\left|h_{\rm EV}\ \rm description\right|+\left|h_{\rm PA}\ \rm description\right|.
\end{eqnarray}
In the previous equation, $k_{\rm A}^{j}$ is a string of N bits, $a^{j}$ is a string of $N$ trits (that can be accomodated with $2N$ bits), $r_{\rm A}^{j}|_{\rm X}$ is a string with an expected size of $E\bigl[\bigl|r_{\rm A}^{j}|_{\rm X}\bigr|\bigr]=q_{\rm X}N$ bits and $\left|s_{\mathcal{Z'}}\right|=\sum_{j=1}^{n_{\rm q}}\bigl|r_{\rm A}^{j}|_{\rm Z}\bigr|$, where $E\bigl[\bigl|r_{\rm A}^{j}|_{\rm Z}\bigr|\bigr]=q_{\rm Z}N$ bits for all $j=1,\ldots{},n_{\rm q}$. The size of the syndrome $\left|sy(s_{\rm B})\right|$ depends on the EC protocol (and a typical model is given in the Results section of the main text), $\left|h_{\rm EV}(s_{\rm B})\right|=\lceil\log_{2}(2/\hat{\epsilon}_{\rm cor})\rceil$ bits, $\left|h_{\rm EV}\ \rm description\right|=2\lceil\log_{2}(2/\hat{\epsilon}_{\rm cor})\rceil$ bits and $\left|h_{\rm PA}\ \rm description\right|=Mn_{\rm q}+l-1$ bits, $l$ denoting the extractable secret key length, given by Eq.~(\ref{BB84 AC}) (Eq.~(\ref{BB84 PN})) within the AC, AN and PC corruption models (PN corruption model) for the QKD modules.\\

\textbf{Calculation of $N$ and $E_{\rm tol}$ for the simulations.} Here, we give adequate values for the number $N$ of signals transmitted per $\mathrm{QKD}_{\mathrm{A}_{j}}$, $j=1,\ldots,n_{\rm q}$, and for the threshold bit error rate of the EC protocol, $E_{\rm tol}$, based on respective restrictions on the abortion probabilities of the sifting step and the error verification step. The analysis relies on a typical channel model described below in this note.

We calculate $N$ first. Let us impose a common abortion probability $\gamma_{\rm sift}/n_{\rm q}$ for each sifting step ($n_{\rm q}$ of them in total). That is, we demand that $P(\left|\mathcal{Z}_{j}\right|<{M})\leq\gamma_{\rm sift}/n_{\rm q}$ for all $j=1,\ldots,n_{\rm q}$, where $\left|\mathcal{Z}_{j}\right|$ is the set of detection events in which both parties use basis Z and $M$ is the pre-specified size of the sifted keys (\textit{i.e.}, the block size). Using the Chernoff's inequality~\cite{Chernoff}, this condition is met if we set the number of signals transmitted per $\mathrm{QKD}_{\mathrm{A}_{j}}$ to $N=\zeta(M,\sum_{a}G_{\rm Z,Z}^{a},\gamma_{\rm sift}/n_{\rm q})$, where $\zeta(x,y,z)$ is defined in Eq.~(\ref{transmitted Chernoff}) and $\sum_{a}G_{\rm Z,Z}^{a}$ is the probability that any given round contributes to $\mathcal{Z}_{j}$ (see the channel model below).

Now, let us compute $E_{\rm tol}$. Following the BB84 protocol at the beginning of this note, EC is applied separately on each pair of sifted keys $(s_{\rm A}^{j},s_{\rm B}^{j})$. Assuming, for simplicity, that the EC protocol certainly corrects up to a fraction $E_{\rm tol}$ of bit errors (and no more), either $E_{j}\leq{E_{\rm tol}}$ for all $j$ or the (single) EV step aborts, where $E_{j}$ denotes the actual error rate between $s_{\rm A}^{j}$ and $s_{\rm B}^{j}$. Thus, applying Chernoff's inequality~\cite{Chernoff}, $P\left(\rm{EV\hspace{.05cm}aborts}\right)\leq{\gamma_{\rm EC}}$ holds for any $\gamma_{\rm EC}\in(0,1)$ if
\begin{equation}\label{threshold QBER BB84}
E_{\rm tol}=\min\left\{1,E_{\rm Z}+\frac{\Delta_{\rm U}(E_{\rm Z}M,\gamma_{\rm EC}/n_{\rm q})}{{M}}\right\}
\end{equation}
where $E_{\rm Z}$ is the expected bit error rate for the basis Z, \textit{i.e.,} $E_{\rm Z}=E[E_{j}]$, and the deviation function $\Delta_{\rm U}(x,y)$ is defined in Eq.~(\ref{Chernoff_intervals}). An expression of $E_{\rm Z}$ for a typical channel model is given below.\\

\textbf{Channel model.} For the simulations, we adapt the typical channel model presented for the MDI-QKD setup (see Supplementary Note 6) to the decoy-state BB84 setup illustrated in Supplementary Figure~\ref{fig:setup_2}.
\begin{figure}[!htbp]
	\centering 
	\includegraphics[width=7cm,height=1.8cm]{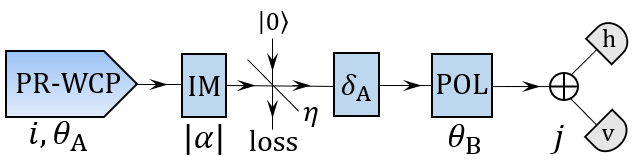} 
	\caption{Supplementary Figure 3. Schematic of the decoy-state BB84 setup.
		Alice holds a laser source that emits PR-WCPs in any of the four BB84 states, defined by a polarization setting $i\in\{1,2\}$ and a basis setting $\theta_{\rm A}\in\{\rm{0,\pi/4}\}$. An intensity modulator (IM) selects the amplitude $|\alpha|$ of Alice's laser pulse. The overall efficiency of the system is denoted by $\eta=\eta_{\rm ch}\eta_{\rm det}$, where $\eta_{\rm det}$ is the detector efficiency (set to a common value for both photo-detectors) and $\eta_{\rm ch}=10^{-\alpha{L}/{10}}$ is the transmission efficiency, $\alpha$ (dB/km) being the attenuation coefficient of the channel and $L$ (km) being the transmission length. The angle $\delta_{\rm A}\geq{}0$ denotes the polarization misalignment occurring in the channel.
		On the other hand, Bob holds a detection system that consists of a polarization modulator (POL), a polarizing beam-splitter (PBS) denoted by the symbol ``$\oplus$", and two single-photon detectors. POL selects Bob's measurement setting, $\theta_{\rm B}\in\{\rm{0,\pi/4}\}$, and the corresponding outcome is recorded in $j\in\{1,2,\emptyset\}$. Precisely, Bob sets $j=1$ ($j=2$) if a click is observed in the detector that detects the horizontal (vertical) component of the incoming pulse and $j=\emptyset$ if no click is observed. If both detectors click, the outcome is randomly assigned to $j=1$ or $j=2$.}
	\label{fig:setup_2}
\end{figure}

Most of the notation is common with the channel model in Supplementary Note 6: $\eta_{\rm det}$ denotes the detector efficiency and $\eta_{\rm ch}=10^{-\alpha_{\rm att}L}$ denotes the transmission efficiency, $\alpha_{\rm att}$ (dB/km) being the attenuation coefficient and $L$ (km) being the transmission distance between Alice and Bob. Similarly, $p_{\rm d}$ stands for the dark count probability of the photo-detectors and $\delta_{\rm A}$ stands for the polarization misalignment occurring in the channel. In this setup, the relevant experimental parameters are the detection probability and the probability of having a bit error with amplitude $|\alpha|$, given that both parties selected the same measurement setting (\textit{i.e.,} Z or X). We denote these parameters by $Q^{\alpha}$ and $E^{\alpha}$, respectively, and they are basis-independent in the considered channel model. In particular, explicit calculation of $Q^{\alpha}$ and $E^{\alpha}$ using this model yields
\begin{eqnarray}\label{model}
&&Q^{\alpha}=1-(1-p_{\rm d})^{2}e^{-\eta|\alpha|^{2}}, \nonumber \\
&&Q^{\alpha}E^{\alpha}=\frac{p_{\rm d}^{2}}{2}+p_{\rm d}(1-p_{\rm d})\bigl(1+h_{\eta,\alpha,\delta_{\rm A}}\bigr)+(1-p_{\rm d})^{2}\left(\frac{1}{2}+h_{\eta,\alpha,\delta_{\rm A}}-\frac{1}{2}e^{-\eta|\alpha|^{2}}\right) \nonumber \\
\end{eqnarray}
where $\eta=\eta_{\rm det}\eta_{\rm ch}$ and we defined $h_{\eta,\alpha,\delta_{\rm A}}=\bigl(e^{-\eta|\alpha|^{2}\cos^2(\delta_{\rm A})}-e^{-\eta|\alpha|^{2}\sin^2(\delta_{\rm A})}\bigr)/2$. These expressions account for the fact that multiple clicks are randomly assigned to a specific detection outcome (see the caption of Supplementary Figure~\ref{fig:setup_2} for more details).\\

Finally, we write down the expected values of the observables required for the simulations. For this purpose, we introduce the quantities $G_{\rm Z,Z}^{a}=q_{\rm Z}^{2}p_{a}Q^{\sqrt{a}}$, $G_{\rm X,X}^{a}=q_{\rm X}^{2}p_{a}Q^{\sqrt{a}}$, $\hat{E}^{a}=E^{\sqrt{a}}$ where $a\in{}A$ and $p_{\rm a}$ is the probability that Alice uses the intensity setting $a$. From these quantities, it follows that
\begin{eqnarray}\label{expected_observables BB84}
&& E\bigl[\bigl|\mathcal{Z'}^{a}_{j}|\bigr]=\frac{G_{\rm Z,Z}^{a}}{\sum_{a}G_{\rm Z,Z}^{a}}M, \nonumber \\
&& E\bigl[\bigl|\mathcal{X}_{j}^{a}\bigr|\bigr]=G_{\rm X,X}^{a}N,\hspace{.2cm}\textrm{and} \nonumber \\
&& E\bigl[e_{a}^{j}\bigr]=\hat{E}^{a}G_{\rm X,X}^{a}N,
\end{eqnarray}
where we recall that $N$ is the number of signals transmitted per QKD pair and $M$ is the size of each sifted key. Also note that, for each $j=1,\ldots,n_{\rm q}$ all three sets $\mathcal{Z'}^{a}_{j}$ contribute to the $j$-th sifted key. Thus, averaging over all three intensity settings, the expected QBER in the basis Z is 
\begin{equation}
E_{\rm Z}=\frac{\sum_{a}\hat{E}^{a}G_{\rm Z,Z}^{a}}{\sum_{a}G_{\rm Z,Z}^{a}}.
\end{equation}
Remarkably, the formula above corresponds to the \textit{a priori} expected bit error rate between any pair of sifted keys, \textit{i.e.}, the expected error rate without using the knowledge of the actual set sizes $\bigl|\mathcal{Z'}_{j}^{a}\bigr|$. The knowledge of the set sizes indeed provides slightly more accurate values of the expected bit error rates, but these would be different for each $j$. Thus, for simplicity, we use the common \textit{a priori} expected bit error rate for all $j$.\\

\textbf{Performance evaluation.}
\begin{figure}[!htbp]
	\centering 
	\includegraphics[width=17cm,height=7cm]{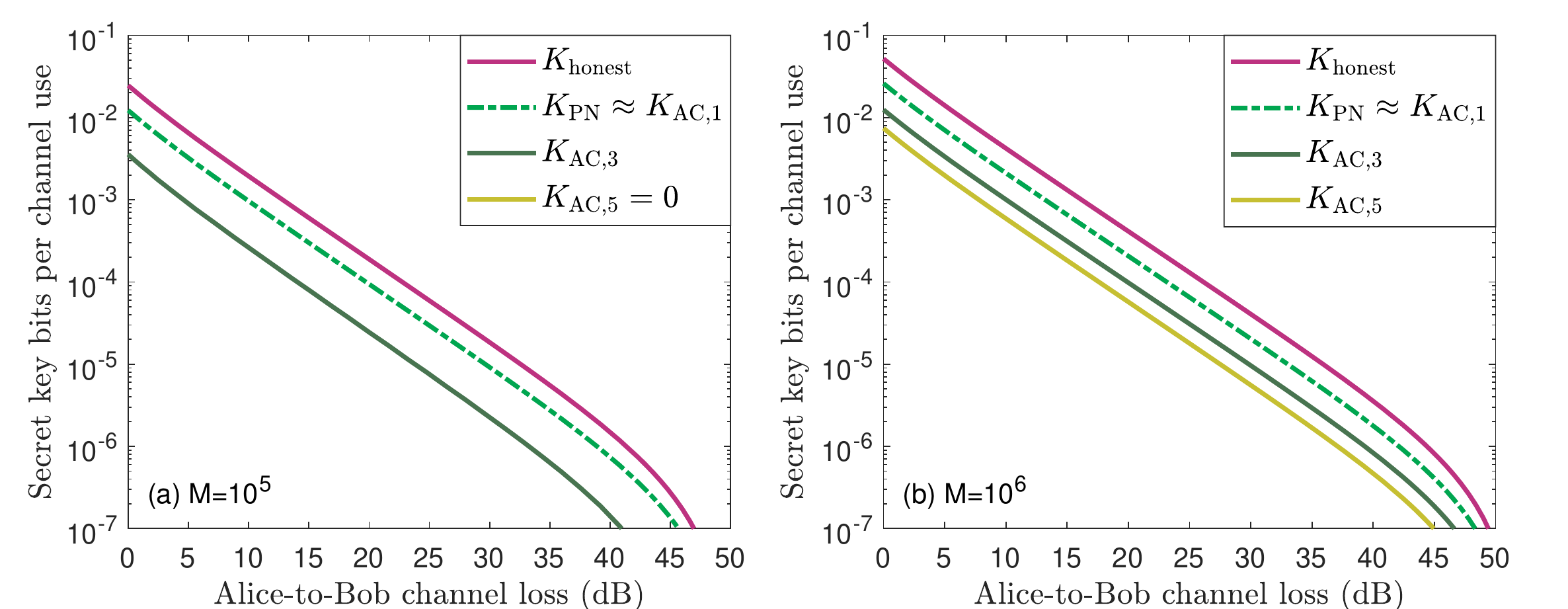}
	\caption{Supplementary Figure 4. Secret key rate, $K$, that results from the decoy-state BB84 protocol with redundant devices presented in Supplementary Note 6. Two distinct block-sizes are considered, \textbf{(a)} $M=10^{5}$ and \textbf{(b)} $M=10^{6}$. In each case, $K$ is plotted as a function of the channel loss between Alice and Bob for various adversarial scenarios with malicious devices. In both figures, the purple line is the secret key rate in the standard scenario where each party holds a trusted QKD module and a trusted classical post-processing (CP) unit. On the contrary, green lines are used for different adversarial scenarios. Precisely, the dashed-dotted phosphorescent line is the secret key rate assuming that the corrupted devices are passive and non-collaborative, which requires the use of two QKD pairs and two CP units per lab (all of them being possibly malicious) to provide security. Meanwhile, the solid non-phosphorescent green lines assume active and collaborative corrupted devices. These latter lines further assume the same number, say $t$, of malicious QKD pairs and malicious CP units per lab, which requires the use of at least $n_{\rm q}=t+1$ QKD pairs and $n_{\rm c}=3t+1$ CP units per party to provide security. Specifically, the dark (light) green line corresponds to $t=3$ ($t=5$).}
	\label{fig:performance_2}
\end{figure}
In Supplementary Figure~\ref{fig:performance_2}, we plot the secret key rate that one can extract combining Protocol with the decoy-state QKD scheme presented in this note, as a function of the channel loss between Alice and Bob. Both the security and the experimental parameters are set following the criteria described in the Results section of the main text, that is, they are common with the simulations of MDI-QKD presented in Fig. 2 there. As in that figure, two different block sizes are considered, (a) $M=10^{5}$ and (b) $M=10^{6}$, and various distinct adversarial scenarios are included. The reader is referred to the discussion of Fig. 2 in the main text for a comment on the results presented in Fig.~\ref{fig:performance_2} (such discussion is common for both figures).
\section*{Supplementary Note 8: inverse Chernoff bound}\label{inverse_Chernoff}
Here, we rephrase the statement of the inverse Chernoff-bound presented in~\cite{Zhang,Bahrani}.

Let $X_{1},\ldots,X_{N}$ be independent Bernouilli random variables such that $P[X_{i}=1]=p_{i}$, and let $X=\sum_{i=1}^{N}X_{i}$ and $\mu=E[X]=\sum_{i=1}^{N}p_{i}$, where $E[\cdot]$ denotes the expected value. Let $x$ be the observed outcome of $X$ for a given trial (that is, $x\in\mathbb{N}$). Then, $x$ satisfies
\begin{equation}
x=\mu+\delta
\end{equation}
except with probability at most $\epsilon_{\rm L}+\epsilon_{\rm U}$, where the parameter $\delta\in\left[-\Delta(x,\epsilon_{\rm L}),\hat{\Delta}(x,\epsilon_{\rm U})\right]$ and
\begin{eqnarray}\label{hat}
&&\hat{\Delta}(x,y)=x\left[W_{0}(-e^{-c_{x,y}})+1\right], \nonumber \\
&&{\Delta}(x,y)=\left\{
\begin{array}{ll}
-x\left[W_{-1}(-e^{-c_{x,y}})+1\right] & \textrm{if}\hspace{.2cm}x\neq{0}, \\
\ln{y^{-1}} & \textrm{if}\hspace{.2cm}x={0}. \\
\end{array} 
\right. \nonumber \\
\end{eqnarray}
Here, $W_{j}$ stands for the $j$-th branch of the $W$ Lambert function and $c_{x,y}$ is defined as $c_{x,y}=1+\ln{y^{-1}}/x$. Also, $\epsilon_{\rm L}$ and $\epsilon_{\rm U}$ are the one-sided error probabilities.
\section*{Supplementary Note 9: generalised chain rule for conditional smooth min-entropies}\label{chain_rule}
The first chain rule for conditional smooth min-entropies given in~\cite{Vitanov} can be restated as follows: for all $\epsilon_{2}',\epsilon_{1}\geq{0}$, $\epsilon_{2}>2\epsilon_{2}'+\epsilon_{1}$,
\begin{equation}\label{original_chain_rule}
H_{\rm min}^{\epsilon_{2}}(Z_{2}Z_{1}|E)\geq{H_{\rm min}^{\epsilon_{2}'}(Z_{2}|Z_{1}E)+H_{\rm min}^{\epsilon_{1}}(Z_{1}|E)-\log_{2}\left(\frac{1}{\epsilon_{2}-2\epsilon_{2}'-\epsilon_{1}}\right)}.
\end{equation}
Using mathematical induction, the previous claim is easily generalised to
\begin{equation}\label{generalised_chain_rule}
H_{\rm min}^{\epsilon_{n}}(Z_{n}\ldots{}Z_{1}|E)\geq{\sum_{j=2}^{n}\left[H_{\rm min}^{\epsilon'_{j}}(Z_{j}|Z_{j-1}\ldots{}Z_{1}E)-\log_{2}\left(\frac{1}{\epsilon_{j}-2\epsilon_{j}'-\epsilon_{j-1}}\right)\right]}+H_{\rm min}^{\epsilon_{1}}(Z_{1}|E),
\end{equation}
for all $\epsilon_{j},\epsilon'_{j}$ such that
\begin{equation}\label{epsilons}
\epsilon_{1}\geq{0},\hspace{.1cm}\left\{\epsilon'_{j}\geq{0},\hspace{.1cm}\epsilon_{j}>2\epsilon'_{j}+\epsilon_{j-1}\right\}_{j=2}^{n}\hspace{.1cm}
\end{equation}
and $n\geq{2}$. We prove it in what follows. First, the case $n=2$ trivially holds, as it reduces to Eq.~(\ref{original_chain_rule}). Let us now assume that the proposition holds for a specific $n=m$ larger than two and consider the case $n=m+1$. Again, from Eq.~(\ref{original_chain_rule})
\begin{equation}\label{induction}
H_{\rm min}^{\epsilon_{m+1}}(Z_{m+1}\ldots{}Z_{1}|E)\geq{H_{\rm min}^{\epsilon_{m+1}'}(Z_{m+1}|Z_{m}\ldots{}Z_{1}E)+H_{\rm min}^{\epsilon_{m}}(Z_{m}\ldots{}Z_{1}|E)}-\log_{2}\left(\frac{1}{\epsilon_{m+1}-2\epsilon_{m+1}'-\epsilon_{m}}\right),
\end{equation}
for all $\epsilon_{m+1}',\epsilon_{m}\geq{0}$, $\epsilon_{m+1}>2\epsilon_{m+1}'+\epsilon_{m}$. Note that the above equation is simply a recasting of Eq.~(\ref{original_chain_rule}). Precisely, $Z_{2}$ and $Z_{1}$ in the left-hand side of Eq.~(\ref{original_chain_rule}) are respectively replaced by  $Z_{m+1}$ and the multivariable $Z_{m}\ldots{Z_{1}}$. 

The second term in the right-hand side of Eq.~(\ref{induction}) can be lower-bounded using the induction hypothesis and the proposition for $n=m+1$ follows, which completes the proof.\\

In what follows, we deduce a restricted version of Eq.~(\ref{generalised_chain_rule}) used to derive the extractable key length in the PN corruption model for the QKD modules (Supplementary Note 2). Precisely, let us assume the following simplifications:
\begin{enumerate}
	\item{For $j=2,\ldots,{n}$, $H_{\rm min}^{\epsilon'_{j}}(Z_{j}|Z_{j-1}\ldots{}Z_{1}E)=H_{\rm min}^{\epsilon'_{j}}(Z_{j}|E)$ for all $\epsilon'_{j}$.}
	\item{$H_{\rm min}^{\epsilon_{1}}(Z_{1}|E)=0$ for all $\epsilon_{1}$.}
	\item{$\epsilon_{1}=0$ and $\epsilon_{j}=\epsilon_{j-1}+2\epsilon'_{j}+\delta$ for $j=2,\ldots{},n$ and $\delta>0$.}
	\item{$\epsilon'_{j}=\varepsilon$ for $j=2,\ldots{},n$ and $\varepsilon>0$.}
\end{enumerate}
Note that assumptions 3 and 4 are such that the conditions given in Eq.~(\ref{epsilons}) trivially hold and $\epsilon_{n}=(n-1)(2\varepsilon+\delta)$. From 1 to 4, Eq.~(\ref{generalised_chain_rule}) easily reduces to
\begin{equation}\label{simplified}
H_{\rm min}^{(n-1)(2\varepsilon+\delta)}(Z_{n}\ldots{}Z_{1}|E)\geq{\sum_{j=2}^{n}H_{\rm min}^{\varepsilon}(Z_{j}|E)-\log_{2}\left(\frac{1}{\delta^{n-1}}\right)},
\end{equation}
with $\varepsilon,\delta>0$ and $n\geq{}2$.
\section{References and Supplementary References}


\begin{thebibliography}{99}
	\bibitem{BB84}
	Bennett, C. H. \& Brassard, G. Quantum cryptography: public key distribution and coin tossing. \textit{In Proc. IEEE International Conference on Computers, Systems \& Signal Processing}, 175–179 (IEEE, NY, Bangalore, India, 1984).
	%
	\bibitem{review1}
	Scarani, V., Bechmann-Pasquinucci, H., Cerf, N. J., Dušek, M., Lütkenhaus, N. \& Peev, M. The security of practical quantum key distribution. \textit{Reviews of Modern Physics} \textbf{81,} 1301 (2009).
	%
	\bibitem{review2}
	Lo, H.-K., Curty, M. \& Tamaki, K. Secure quantum key distribution. \textit{Nature Photonics} \textbf{8,} 595 (2014).
	%
	\bibitem{Feihu}
	Xu F., Ma X., Zhang Q., Lo, H.-K., Pan J.-W. Secure quantum key distribution with realistic devices. \textit{Reviews of Modern Physics} \textbf{92,} 025002 (2020).
	%
	\bibitem{DHS}
	Diffie, W. \& Hellman, M. New directions in cryptography. \textit{IEEE Transactions on Information Theory} \textbf{22,} 644-654 (1976).
	%
	\bibitem{RSA}
	Rivest, R. L., Shamir, A. \& Adleman, L. A method for obtaining digital signatures and public-key cryptosystems. \textit{Communications of the ACM} \textbf{21,} 120-126 (1978).
	%
	\bibitem{Gligor}
	Gligor, V. D. A guide to understanding covert channel analysis of trusted systems, Vol. 30 (National Computer Security Center, 1994).
	%
	\bibitem{Zander}
	Zander, S., Armitage, G. \& Branch, P. A survey of covert channels and countermeasures in computer network protocols. \textit{IEEE Communications Surveys \& Tutorials} \textbf{9,} 44-57 (2007).
	%
	\bibitem{Prevelakis}
	Prevelakis, V. \& Spinellis, D. The Athens affair. \textit{IEEE Spectrum} \textbf{4,} 26-33 (2007).
	%
	\bibitem{Yang}
	Yang, K., Hicks, M., Dong, Q., Austin, T. \& Sylvester, D. A2: Analog malicious hardware. In \textit{IEEE symposium on security and privacy} 18-37 (IEEE, 2016).
	%
	\bibitem{Robertson}
	Robertson, J. \& Riley, M. The big hack: how China used a tiny chip to infiltrate US companies. \textit{Bloomberg Businessweek} \textbf{4} (2018).
	%
	\bibitem{Adee}
	Adee, S. The hunt for the kill switch. \textit{IEEE Spectrum} \textbf{45,} 34-39 (2008).
	%
	\bibitem{Becker}
	Becker, G. T., Regazzoni, F., Paar, C. \& Burleson, W. P. Stealthy dopant-level hardware trojans. In \textit{International Workshop on Cryptographic Hardware and Embedded Systems}, 197-214 (Springer, Berlin, Heidelberg, 2013).
	%
	\bibitem{Mayers}
	Mayers, D. \& Yao, A.~C.~C. Quantum cryptography with imperfect apparatus. In \textit{Proceedings of the 39th Annual Symposium on Foundations of Computer Science}, 503-509 (1998).
	%
	\bibitem{Acin}
	Acín, A., Brunner, N., Gisin, N., Massar, S., Pironio, S. \& Scarani, V. Device-independent security of quantum cryptography against collective attacks. \textit{Physical Review Letters} \textbf{98,} 230501 (2007).
	\bibitem{Vazirani}
	Vazirani, U. \& Vidick, T. Fully device-independent quantum key distribution. \textit{Physical Review Letters} \textbf{113,} 140501 (2014).
	%
	\bibitem{Rotem}
	Arnon-Friedman, R., Dupuis, F., Fawzi, O., Renner, R. \& Vidick, T. Practical device-independent quantum cryptography via entropy accumulation. \textit{Nature Communications} \textbf{9,} 459 (2018).
	%
	\bibitem{Miller}
	Miller, C. A. \& Shi, Y. Robust protocols for securely expanding randomness and distributing keys using untrusted quantum devices. \textit{Journal of the ACM} \textbf{63,} 33 (2016).
	%
	\bibitem{Barrett}
	Barrett, J., Colbeck, R. \& Kent, A. Memory attacks on device-independent quantum cryptography. \textit{Physical Review Letters} \textbf{110,} 010503 (2013).
	%
	\bibitem{Curty}
	Curty, M. \& Lo, H.-K. Foiling covert channels and malicious classical post-processing units in quantum key distribution. \textit{npj Quantum Information} \textbf{5,} 14 (2019).
	%
	\bibitem{Wei}
	Li, W. \emph{et~al.} Experimental quantum key distribution secure against malicious devices. Preprint at https://arxiv.org/abs/2006.12863 (2020).
	%
	\bibitem{VSS}
	Chor, B., Goldwasser, S., Micali, S. \& Awerbuch, B. Verifiable secret sharing and achieving simultaneity in the presence of faults. In \textit{Proc. of the 26th Annual Symposium on Foundations of Computer Science (FOCS'85)}, 383–395 (IEEE Computer Society, Los Alamitos, California, 1985).
	%
	\bibitem{MPC}
	Cramer, R., Damgård, I. B. \& Nielsen, J. B. \textit{Secure Multiparty Computation and Secret Sharing} (Cambridge University Press, New York, USA, 2015).
	%
	\bibitem{Ben-Or}
	Ben-Or, M., Goldwasser, S. \& Wigderson, A. Completeness theorems for non-cryptographic fault-tolerant distributed computation. In \textit{Proceedings of the twentieth annual ACM symposium on Theory of computing}, 1-10 (1988).
	%
	\bibitem{Chaum}
	Chaum, D., Crépeau, C. \& Damgard, I. Multiparty unconditionally secure protocols. In \textit{Proceedings of the twentieth annual ACM symposium on Theory of computing}, 11-19 (1988).
	%
	\bibitem{Maurer}
	Maurer, U. Secure multi-party computation made simple. \textit{Discrete Applied Mathematics} \textbf{154,} 370-381 (2006).
	%
	\bibitem{Shamir}
	Shamir, A. How to share a secret. \textit{Communications of the ACM} \textbf{22,} 612-613 (1979).
	%
	\bibitem{Blakley}
	Blakley, G. R. Safeguarding cryptographic keys. In \textit{Proc. of the AFIPS 1979 National Computer Conference (NCC'79)}, 313–317 (AFIPS Press, New Jersey, 1979).
	%
	\bibitem{Mitra}
	Mitra, S., Wong, H.,-S. P. \& Wong, S. Stopping Hardware Trojans in Their Tracks. \textit{IEEE Spectrum} \textbf{20} (2015).
	%
	\bibitem{PA}
	Bennett, C. H., Brassard, G. \& Robert, J. M. Privacy amplification by public discussion. \textit{SIAM journal on Computing}, \textbf{17,} 210-229 (1988).
	%
	\bibitem{Tomamichel1}
	Tomamichel, M., Schaffner, C., Smith, A. \& Renner, R. Leftover hashing against quantum side information. \textit{IEEE Transactions on Information Theory} \textbf{57,} 5524-5535 (2011).
	%
	\bibitem{Lamport}
	Lamport, L., Shostak, R., \& Pease, M. The Byzantine Generals Problem. \textit{Transactions on Programming Languages and Systems} \textbf{4}, 382-401 (1982).
	%
	\bibitem{LFSR}
	Krawczyk, H. LFSR-based hashing and authentication. In \textit{Advances in Cryptology - CRYPTO’94, Lecture Notes in Computer Science} (Springer-Verlag, 1994) \textbf{893,} 129–139.
	%
	\bibitem{X.B.Wang}
	Zhou, Y.~H., Yu, Z.~W. \& Wang, X.-B. Making the decoy-state measurement-device-independent quantum key distribution practically useful. \textit{Physical Review A} \textbf{93,} 042324 (2016).
	%
	\bibitem{Lim}
	Lim, C. C. W., Curty, M., Walenta, N., Xu, F. \& Zbinden, H. Concise security bounds for practical decoy-state quantum key distribution. \textit{Physical Review A} \textbf{89,} 022307 (2014).
	%
	\bibitem{Yin}
	Yin, H. L. \emph{et~al.} Measurement-device-independent quantum key distribution over a 404 km optical fiber. \textit{Physical Review Letters} \textbf{117,} 190501 (2016).
	%
	\bibitem{Thales}
	nShield Solo HSMs, Thales Group.
	https://www.thalesesecurity.com/products/general-purpose-hsms/nshield-solo.
	%
	\bibitem{Gemalto}
	Hardware Security Modules, Gemalto.
	https://safenet.gemalto.com/dataencryption/hardware-security-modules-hsms/.
	%
	\bibitem{Amazon}
	AWS CloudHSM, Amazon Web Services.
	https://aws.amazon.com/cloudhsm/.
	%
	\bibitem{Salvail}
	L. Salvail \emph{et~al.} Security of trusted repeater quantum key distribution networks. \textit{Journal of Computer Security} \textbf{18,} 61-87 (2010).
	%
	\bibitem{Peev}
	Peev, M., \emph{et~al.} The SECOQC quantum key distribution network in Vienna. \textit{New Journal of Physics} \textbf{11,} 075001 (2009).
	%
	\bibitem{Sasaki}
	Sasaki, M., \emph{et~al.} Field test of quantum key distribution in the Tokyo QKD Network. \textit{Optics Express} \textbf{19,} 10387-10409 (2011).
	%
	\bibitem{Renner}
	Renner, R. Security of quantum key distribution. \textit{International Journal of Quantum Information} \textbf{6,} 1-127 (2008).
	%
	\bibitem{Vitanov}
	Vitanov, A., Dupuis, F., Tomamichel, M. \& Renner, R. Chain rules for smooth min-and max-entropies. \textit{IEEE Transactions on Information Theory} \textbf{59,} 2603-2612 (2013).
	%
	\bibitem{Fung}
	Fung, C.~H.~F., Ma, X. \& Chau, H.-F. Practical issues in quantum-key-distribution postprocessing. \textit{Physical Review A} \textbf{81,} 012318 (2010).
	%
	\bibitem{Curty1}
	Lo, H.-K., Curty, M. \& Qi, B. Measurement-device-independent quantum key distribution. \textit{Physical Review Letters} \textbf{108,} 130503 (2012).
	%
	\bibitem{Curty2}
	Curty, M., Xu, F., Cui, W., Lim, C. C. W., Tamaki, K. \& Lo, H.-K. Finite-key analysis for measurement-device-independent quantum key distribution. \textit{Nature Communications} \textbf{5,} 1-7 (2014).
	%
	\bibitem{Tomamichel2}
	Tomamichel, M., Lim, C.~C.~W., Gisin, N. \& Renner, R. Tight finite-key analysis for quantum cryptography. \textit{Nature Communications} \textbf{3,} 634 (2012).
	%
	\bibitem{Serfling}
	Serfling, R.~J. Probability inequalities for the sum in sampling without replacement. \textit{The Annals of Statistics}, 39-48 (1974).
	%
	\bibitem{Chernoff}
	Mitzenmacher, M. \& Upfal, E. \textit{Probability and computing: Randomization and probabilistic techniques in algorithms and data analysis} (Cambridge University Press, 2017).
	%
	\bibitem{decoy2}
	Hwang, W. Y. Quantum key distribution with high loss: toward global secure communication. \textit{Physical Review Letters} \textbf{91,} 057901 (2003).
	%
	\bibitem{decoy}
	Lo, H.-K., Ma, X. \& Chen, K. Decoy state quantum key distribution. \textit{Physical Review Letters} \textbf{94,} 230504 (2005).
	%
	\bibitem{decoy3}
	Wang, X. B. Beating the photon-number-splitting attack in practical quantum cryptography. \textit{Physical Review Letters} \textbf{94,} 230503 (2005).
	%
	\bibitem{Lim}
	Lim, C. C. W., Curty, M., Walenta, N., Xu, F. \& Zbinden, H. Concise security bounds for practical decoy-state quantum key distribution. \textit{Physical Review A} \textbf{89,} 022307 (2014).
	%
	\bibitem{Hoeffding}
	Hoeffding, W. Probability inequalities for sums of bounded random variables. \textit{Journal of the American Statistical Association} \textbf{58,} 13–30 (1963).
	%
	\bibitem{Zhang}
	Zhang, Z., Zhao, Q., Razavi, M. \& Ma, X. Improved key-rate bounds for practical decoy-state quantum-key-distribution systems. \textit{Physical Review A} \textbf{95,} 012333 (2017).
	%
	\bibitem{Bahrani}
	Bahrani, S., Elmabrok, O., Lorenzo, G. C. \& Razavi, M. Wavelength assignment in quantum access networks with hybrid wireless-fiber links. \textit{Journal of the Optical Society of America B} \textbf{36,} 99-108 (2019).
	%
\end{thebibliography}
\end{document}